%% file: arxiv-main.tex
\documentclass[11pt]{article}
\usepackage{fullpage}

\usepackage[numbers]{natbib}
\usepackage[utf8]{inputenc} %
\usepackage[T1]{fontenc}    %
\usepackage[hidelinks]{hyperref} %
\usepackage{url}            %
\usepackage{booktabs}       %
\usepackage{amsfonts}       %
\usepackage{nicefrac}       %
\usepackage{microtype}      %
\usepackage{xcolor}         %
\usepackage{enumitem}
\usepackage[FIGTOPCAP]{subfigure}
\usepackage{graphicx}
\usepackage{floatrow}
\usepackage{multirow}
\newfloatcommand{capbtabbox}{table}[][\FBwidth]

\usepackage{amsmath,amsthm, amssymb, mathrsfs}
\usepackage{tikz}
\usepackage{color}
\usepackage{hyperref}
\usepackage{diagbox}
\usepackage[font = small,labelfont = bf,tableposition = bottom]{caption}
\usepackage{array,xcolor,colortbl}
\usepackage{tabularray}
\usepackage{wrapfig}
\usepackage{placeins}

\usepackage[ruled,vlined]{algorithm2e}
\SetKwInOut{Input}{Input}
\SetKwRepeat{Do}{do}{while}
\SetProcNameSty{texttt}

\theoremstyle{plain}
\newtheorem{MainResult}{Main Result}
\newtheorem{theorem}{Theorem}
\newtheorem{lemma}{Lemma}
\newtheorem{proposition}{Proposition}
\newtheorem{corollary}{Corollary}
\SetKwInOut{Parameter}{Parameters}

\input{macro}

\theoremstyle{definition}
\newtheorem{assumption}{Assumption}
\newtheorem{definition}{Definition}
\newtheorem{example}{Example}
\theoremstyle{remark}
\newtheorem{remark}{Remark}

\title{An Isotonic Mechanism for Overlapping Ownership}
\author{
Yifan Guo\thanks{Graduate School of Business, Stanford University; Email: \texttt{yifanguo@stanford.edu}.}
\and Weijie Su\thanks{Department of Statistics and Data Science, The Wharton School, University of Pennsylvania; Email: \texttt{suw@wharton.upenn.edu}.}
\and Jibang Wu\thanks{New York University, Shanghai; Email: \texttt{wujibang@nyu.edu}.}
\and Haifeng Xu\thanks{Department of Computer Science, University of Chicago; Email: \texttt{haifengxu@uchicago.edu}.}
}
\date{}

\begin{document}

\maketitle

\input{abstract}

\input{intro}
\input{nash}
\input{partition}
\input{exp}
\input{discussion}

\section*{Acknowledgments}

This work was supported in part by NSF through CCF-2303372 and CCF-1934876, ARO through W911NF-23-1-0030, Analytics at Wharton, and Wharton AI and Analytics for Business.

\bibliographystyle{plainnat}
\bibliography{refer}

\newpage
\appendix
\input{proofs}
\input{exp-append}

\end{document}

%% file: macro.tex
\RequirePackage{bm}

\newtheorem{thm}{Theorem}[section]

\newcommand{\by}{\times}
\newcommand{\abs}[1]{\left| #1 \right|}

\newcommand{\union}{\cup}
\newcommand{\intersect}{\cap}

\renewcommand{\hat}{\widehat}
\renewcommand{\tilde}{\widetilde}
\renewcommand{\bar}{\overline}

\newcommand{\norm}[1]{\|#1\|}

\DeclareMathOperator{\obj}{obj}

\DeclareMathOperator{\OPT}{OPT}
\DeclareMathOperator{\ALG}{ALG}

\def\min{\qopname\relax n{min}}
\def\max{\qopname\relax n{max}}

\def\argmin{\qopname\relax n{argmin}}
\def\argmax{\qopname\relax n{argmax}}

\def\Pr{\qopname\relax n{\mathbf{Pr}}}
\def\Ex{\qopname\relax n{\mathbf{E}}}

\newcommand{\ba}{\bm{a}}
\newcommand{\bb}{\bm{b}}

\newcommand{\bp}{\bm{p}}

\newcommand{\br}{\bm{r}}

\newcommand{\bx}{\bm{x}}
\newcommand{\bz}{\bm{z}}
\renewcommand{\by}{\bm{y}}

\newcommand{\bI}{\bm{I}}

\newcommand{\bP}{\bm{P}}

\newcommand{\bR}{\bm{R}}

\newcommand{\cA}{\mathcal{A}}

\newcommand{\cE}{\mathcal{E}}

\newcommand{\cG}{\mathcal{G}}

\newcommand{\cI}{\mathcal{I}}
\newcommand{\cJ}{\mathcal{J}}

\newcommand{\cM}{\mathcal{M}}
\newcommand{\cN}{\mathcal{N}}
\newcommand{\cO}{\mathcal{O}}
\newcommand{\cP}{\mathcal{P}}

\newcommand{\cS}{{\mathcal{S}}}
\newcommand{\cT}{{\mathcal{T}}}

\newcommand{\cV}{\mathcal{V}}
\newcommand{\cW}{\mathcal{W}}

\newcommand{\RR}{\mathbb{R}}

\newcommand{\E}{{\mathbb E}}

\newcommand{\bbeta}{\bm{\beta}}

\newcommand{\bepsilon}{\bm{\epsilon}}
\newcommand{\bvarepsilon}{\bm{\varepsilon}}

\newcommand{\bmu}{\bm{\mu}}

\newenvironment{lp*}{\begin{equation*}  \begin{array}{lll}}{\end{array}\end{equation*}}

%% file: abstract.tex
\begin{abstract}
 Motivated by the problem of improving peer review at large scientific conferences, this paper studies how to elicit self-evaluations to improve review scores in a natural many-to-many owner-item (e.g., author-paper) setting with overlapping ownership. We design a simple, efficient, and truthful mechanism to elicit self-evaluations from item owners and calibrate noisy review scores in the existing evaluation process (e.g., papers' review scores from peers).

Our approach starts by partitioning the owner-item relation structure into disjoint blocks, each sharing a common set of co-owners. We then elicit the ranking of items from each owner and employ isotonic regression to produce adjusted item scores, aligning with both the reported rankings and raw item review scores. We prove that truth-telling by all owners is a payoff-dominant Nash equilibrium for any valid partition of the overlapping ownership sets under natural conditions. Moreover, truthfulness depends on eliciting rankings independently within each block, making block-partition optimization crucial for improving statistical efficiency. Despite the computational intractability of the general optimization problem, we develop a nearly linear-time greedy algorithm that finds a high-quality block partition with robust approximation guarantees. Experiments on both synthetic data and real-world conference review data demonstrate the effectiveness of our mechanism in a pressing real-world problem.
\end{abstract}

%% file: intro.tex
\section{Introduction}

\begin{figure}[t]
    \centering
    \includegraphics[width=0.42\textwidth]{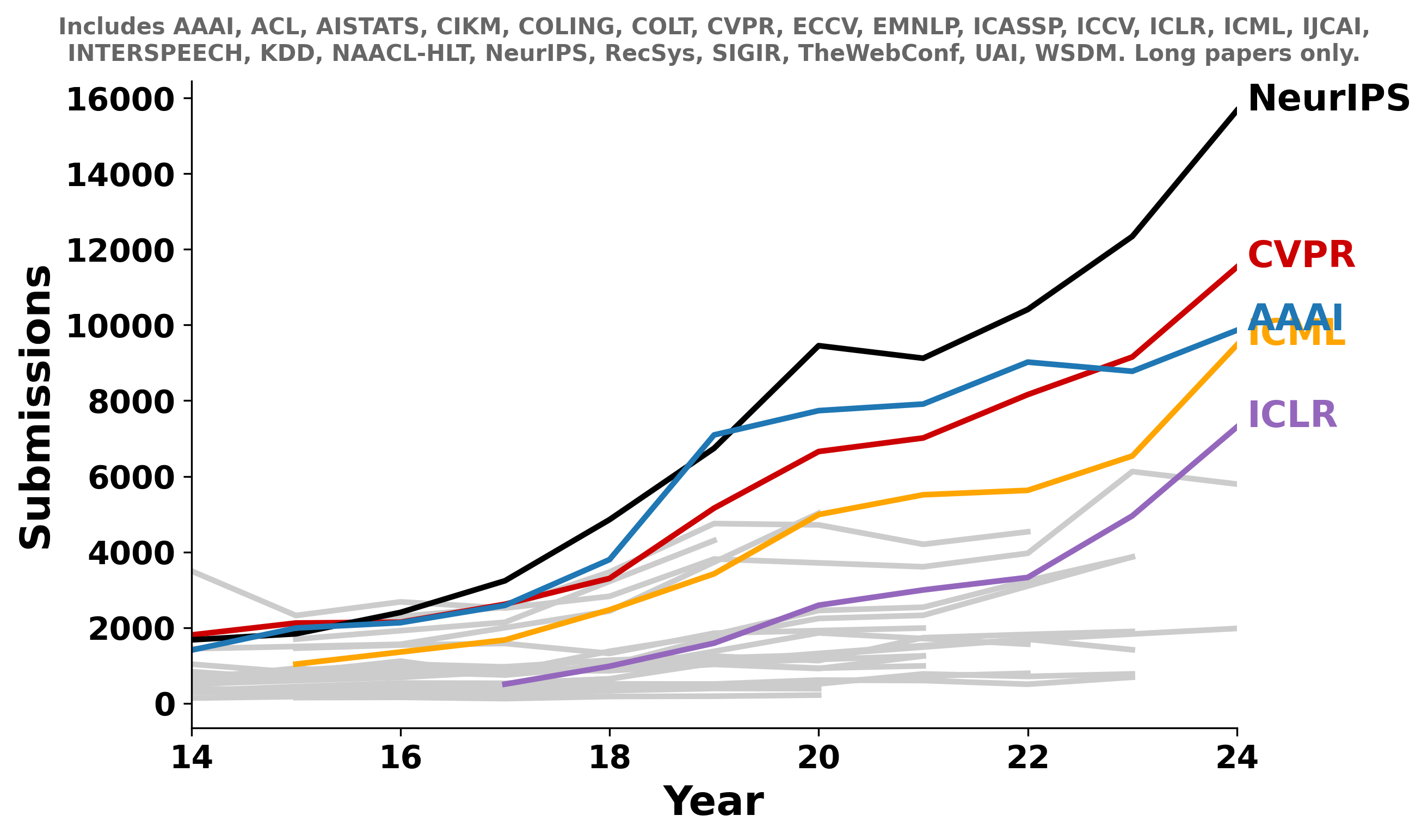}
    \caption{The number of paper submissions in major machine learning and related conferences from 2014 to 2024.}
    \label{fig:conference-size}
\end{figure}

In recent years, major AI and machine learning conferences such as NeurIPS, ICML, and ICLR have faced a concerning decline in the quality of peer review, posing a significant challenge to the global machine learning community~\citep{langford2015arbitrariness,brezis2020arbitrariness,tomkins2017reviewer,lipton2019troubling}. In particular, the NeurIPS 2014 experiment showed that $49.5\%$ of the papers accepted by one committee would be rejected by another~\citep{nipsexp,franccois2015arbitrariness,langford2015arbitrariness}. This inconsistency probability was $50.6\%$ for NeurIPS 2021~\citep{cortes2021inconsistency}. This troubling trend can be largely attributed to a structural imbalance between the surge in submission volumes and lagged growth of the number of qualified reviewers~\citep{shah2022challenges}.
As demonstrated by Figure~\ref{fig:conference-size}, many of these conferences have handled around 10,000 full paper submissions recently, reflecting an almost ten-fold growth over the past decade, whereas a large portion of the reviewers are inexperienced researchers (e.g., junior graduate students)~\citep{shah2018design, stelmakh2021novice, russo2021some}.
Similar trends are seen across various research communities~\citep{mccook2006peer, lajtha2010should, gropp2017peer, checco2021ai}, highlighting a systemic issue in today's peer review systems. To mitigate this issue, there have been progressive research efforts to improve peer review processes.
Some approaches employ machine learning and optimization techniques to improve reviewer assignments, reduce bias, and automate review procedures, while others adopt an economic approach modeling the incentives of participants to encourage high-quality reviews (see, e.g., a survey by \cite{shah2022challenges}).

{Motivated by pressing concerns as mentioned above, this paper proposes a mechanism that blends together statistical and economic design, and can effectively incentivize  authors to share evaluations of their own papers to calibrate peer review scores.} Conceptually, the motivation is natural: if the reviewer pool cannot supply sufficient information for all the submissions,  conference organizers could resort to other sources --- in our case, the authors themselves.
However, while self-criticism is a merit in science, self-evaluation has been largely overlooked due to the obvious conflicts of interest. That is, authors may be reluctant to provide honest assessments, if revealing a mediocre rating of their own work puts them at disadvantage.
Consequently, estimations from manipulated data may be hardly useful. This presents a fundamental design challenge:
\smallskip
\begin{quote}
\emph{How can statistical estimation processes effectively elicit and use data from strategic actors while accounting for both statistical efficiency and incentive compatibility?}
\end{quote}
\smallskip

{
Examining the above question within a clean and stylized model of overlapping ownership, this paper embarks on a path to bring the mechanism design perspective to statistical estimation and proposes an approach, termed the Isotonic Mechanism, that demonstrates a definitive solution to this design challenge.
In particular, while most learning-algorithm designs have focused on improving statistical efficiency, this work takes one step further by tracing the source of the data. Specifically, by jointly accounting for statistical efficiency and incentive compatibility, we design a mechanism that enables the elicitation and use of information traditionally deemed inaccessible.
We view the demonstration of this feasibility in a high-stakes application as an important conceptual contribution, which we hope to spark a broader research agenda at the \emph{interplay} of statistical and incentive design. Furthermore, given the prevalence of many-to-many owner-item relationships with overlapping ownership structures in real-world settings, we anticipate that our framework can be similarly applied to refine evaluation and award selection across a variety of related batch portfolio-review domains, which we describe in more detail below.
}
{
\interfootnotelinepenalty=10000
The conference application is more than a motivating analogy.
\citet{su2024analysis} analyzed a ranking experiment at ICML~2023. At ICML~2026, authors with multiple submissions were asked to provide self-rankings. After pre-rebuttal reviews, projected scores from the Isotonic Mechanism were used as discrepancy signals to flag submissions that might require additional area-chair attention or emergency reviewing.\footnote{\raggedright See the ICML~2026 self-ranking policy announcement, \url{https://blog.icml.cc/2026/01/12/introducing-icml-2026-policy-for-self-ranking-in-reviews/}, and the ICML~2026 Call for Papers, \url{https://icml.cc/Conferences/2026/CallForPapers}.\par}
This deployment illustrates that author rankings can supplement reviewer scores and help allocate scarce reviewing attention without directly determining acceptance decisions.
Beyond conference peer review, the same owner-assisted calibration structure also arises in internal company project or proposal review, where employees may belong to multiple project teams, and in internal screening of jointly produced creative works such as songs, demos, scripts, or film projects, where teams of creators repeatedly assemble across many productions~\citep{demere2019role,o2011multiple,guimera2005team}.
}

To elicit an owner's evaluation about her owned items,  a key challenge to begin with, even putting incentive issues aside, is to determine what kind of data could be elicited from owners  with reasonable accuracy.
One approach might be to ask for highly fine-grained data, such as the owners’ own evaluation values (called \emph{item scores} henceforth) for their items. However, even if owners  are completely honest, they may lack the precise knowledge needed to provide reliable scores. Alternatively, one could request more general information (e.g., asking owners to identify their favorite item), but this may not offer sufficient value for improving review scores. The challenge lies in finding an effective middle ground—eliciting data that is both accurate and useful. To this end, we propose focusing on the \emph{ranking} of an owner's items (e.g., an author's papers). Such relative information tends to be less noisy than absolute measures and has been successfully applied in various fields, such as learning from human preferences~\citep{yue2009interactively,bai2022training}.

Next comes our core scientific question: is it possible to truthfully elicit ranking data from owners to improve the statistical efficiency of any pre-given noisy review scores (e.g., papers' original peer review scores)? Our starting point is an earlier work~\citep{su2021you} which showcased the possibility in a much simplified situation with only a single owner, say, Alice. For this stylized case, \citet{su2021you} considered a natural statistical model for review score estimation and proved that truthful ranking elicitation is possible under natural owner utility assumptions.
To formally describe this method, suppose Alice is the   owner of $n$ items (e.g., a single author for $n$ papers).   Letting $y_1, \ldots, y_n$ denote any pre-given noisy review scores of the $n$ items (e.g., average reviewer scores for Alice's single-authored $n$ papers). The mechanism elicits from Alice a ranking $\pi$ in the form of a permutation of $1, 2, \cdots, n$ that sorts her items in descending order of quality, and then use this elicited ranking to compute adjusted review scores that are the solution to the following convex optimization program:
\begin{align*}
\min_{\br \in \RR^{n}}\ & \sum_{i=1}^n (y_i - r_i)^2 \\
 \text{ s.t. } & r_{\pi(1)} \geq r_{\pi(2)} \geq \dots \geq r_{\pi(n)}.
\end{align*}
This optimization yields the well-known isotonic regression~\citep{barlow1972isotonic}, and its solution ensures consistency with the owner-provided ranking while remaining as close as possible to the original review scores in the least squares sense.
Under natural statistical modeling of the pre-given review scores, this approach provably improves items' score estimation. Moreover,  it also guarantees truthful elicitation of Alice's ranking, under standard utility and prior knowledge assumptions. Notably, the latter is a non-trivial property. Specifically, it crucially hinges on geometric properties of isotonic regressions, and would not hold for other estimation methods (e.g., switching from $l_2$-norm above to $l_1$-norm). This illustrates the importance  of designing appropriate statistical estimation methods that can align with incentives.

\begin{figure}[tbh]
  \begin{center}
    \includegraphics[width=0.35\textwidth]{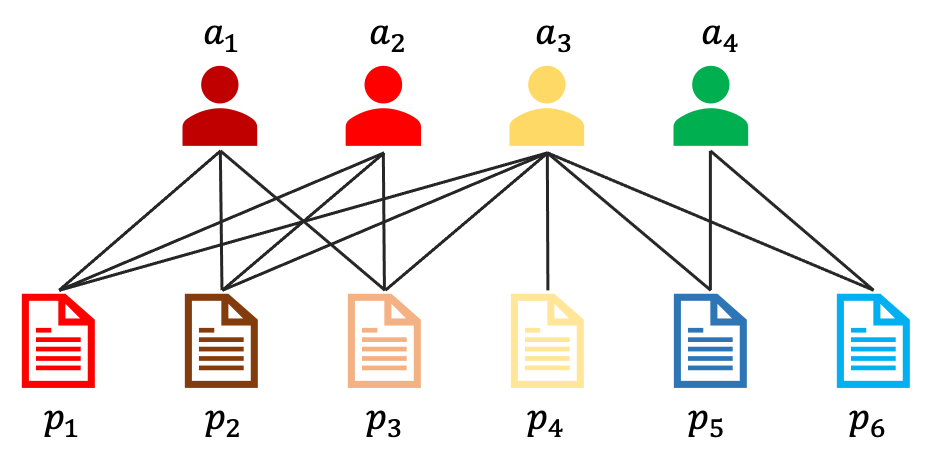}
  \end{center}
  \caption{An example of an owner-item ownership set shown as a bipartite graph.
  An edge between an owner and an item indicates that this individual owns the item.}
  \label{fig:bipartite}
\end{figure}

This work seeks to address the more general situation in which many individuals own many items, with overlapping ownership (e.g., overlapping paper authorship).
This strict generalization exhibits multiple new challenges.
First, since Alice's co-owners will also submit rankings for (some of) Alice's items,  Alice would have to account for the effect of her co-owners' data on the utility of Alice. This intricacy is due to the fundamental difference between single-agent  and  multi-agent decision making setup, and it leads to our different solution concepts, generalizing from Bayesian optimal decision to Bayes Nash equilibrium.
Second, an even more challenging issue is that  different items can \textit{partially} overlap in ownership, as illustrated in Figure~\ref{fig:bipartite}. This is an especially common situation in today's popular machine learning conferences.  In such cases, Alice's utility may be affected by other co-owners' rankings on items Alice does not own.   Naive adaptation of the previous Isotonic Mechanism designed for single-owner situations can lead to serious incentive issues. For instance, one natural approach is to apply the Isotonic Mechanism for every owner's items and output final item scores by averaging their adjusted review scores. Unfortunately, it is easy to find undesirable examples with untruthful behaviors under this mechanism (see Section \ref{sec3:non-truthful})  because some co-owners would misreport their items' ranking information in order to improve their own utilities. Therefore, a more principled ``truth serum'' is needed to effectively incorporate self-evaluations from a complex network of overlapping ownership into the reviewing mechanism.

To address the aforementioned challenges, this paper develops an approach that advances the earlier design using novel techniques from both algorithm design and mechanism design. Specifically, the mechanism first resorts to an algorithmic preprocessing procedure that partitions the owner-item ownership structure into blocks of items such that each block shares a common set of owners.
It then uses the ranking information elicited from the common owners in each block to calibrate the pre-given review scores of their items.
Under certain conditions, this proposed mechanism is marked by the following key characteristic:
\begin{MainResult}
Truthfully reporting rankings by all owners forms a Nash equilibrium under the proposed mechanism.
\end{MainResult}
Along with other results, the main result of truthfulness property is formally presented in Section~\ref{sec:ne}.\footnote{We make a remark on practical implementation of our mechanism. To guarantee truthfulness, it suffices for the designer to commit to only use the partial ranking information within the partitioned blocks. Thus, to ease the practical implementation, the designer can elicit every owner's full ranking of all her items but commit to use only part of the ranking information for the mechanism.
In paper review applications, the mechanism designer is typically the organizing committee of large conferences. They do have such power of commitment, since their decisions are usually audited by steering committees as well.}
This result significantly extends the previous truthfulness guarantee of the earlier work~\citep{su2021you} to cases with overlapping ownership and shows that truthful reporting is a best response for each owner when all other owners also report truthfully. In particular, our proof of this new result involves a novel technique concerning majorization in the ordering of item scores with respect to both the pre-given review noise and other owners' reports.
Moreover, the mechanism encourages truthful behavior in several ways. From a game-theoretic perspective, the truthful Nash equilibrium is payoff dominant, as any other Nash equilibrium gives no owner higher utility. Empirical studies of common-interest games also find that participants often favor payoff-dominant outcomes.

{
A potential criticism of the above global-partition approach is that we have to give up eliciting ranking information across blocks. One may naturally wonder whether the Isotonic Mechanism can be carefully tailored to elicit additional comparison information among items across partitioned blocks, so long as such information can still be truthfully elicited. A natural enlarged design space is to allow each owner to use her own personalized partition of items, which leads to an Isotonic Mechanism with personalized partition. Unfortunately, our second main result gives a negative answer.

\begin{MainResult}
Within the isotonic-regression mechanism class of Mechanism~\ref{algo:linear-isotonic} with personalized partitions, every mechanism that is truthful for every input admits an equivalent global-partition representation (Mechanism \ref{algo:deterministic-partition}). Consequently, within this class, truthful mechanisms cannot exploit ordering information across partition blocks.
In a Gaussian equal-score benchmark, we also quantify how splitting a block affects isotonic-regression risk.
\end{MainResult}

This result is formally presented in Section~\ref{subsec:partition-necessity}.
It should be interpreted as a within-class reduction result. Among isotonic-regression mechanisms with personalized partitions, any truthful design can be replaced by a global-partition mechanism without losing truthfully elicited ranking information.
In particular, the result does not rule out truthful mechanisms outside this class.
We further show that, once attention is restricted to global partitions, statistical efficiency is shaped by a tradeoff between informativeness and robustness. Proposition~\ref{prop:coarser-better} gives an exact informativeness result in a Gaussian equal-score benchmark, while Proposition~\ref{prop:more-owners} formalizes the aggregation benefit of additional noisy owner rankings.
Therefore, within the studied isotonic-regression framework, improving statistical efficiency while retaining truthfulness reduces to optimizing the choice of the global partition.
}
As is conceivable, the choice of partition affects the statistical performance of our mechanism.
For example, a partition for the instance in  Figure~\ref{fig:bipartite} can be simply $\{p_1, p_2, p_3, p_4, p_5\}$, which share a single common owner $\{a_3\}$, or can be $\{p_1, p_2, p_3\}, \{p_4\}$, and $\{p_5, p_6\}$, which share $\{a_1, a_2, a_3\}, \{a_3\}$, and $\{a_3, a_4\}$, respectively. To find high-quality owner-item ownership partitions, we define a natural \emph{class} of criteria to assess the quality of a partition. Within this class of criteria, we obtain the following result on this proposed mechanism:
\begin{MainResult}
A simple nearly linear-time greedy algorithm   can obtain  a   partition of the owner-item ownership structure for the proposed mechanism  that is near-optimal simultaneously for \emph{every} criterion in the class.
\end{MainResult}

This result is formally stated in Section \ref{sec:greedy-charm}. We also show that it is NP-hard to find the exactly optimal partition even under some simple criterion.
In contrast, our greedy algorithm is computationally efficient while achieving constant-ratio approximation \emph{simultaneously} for all the criteria in our considered class.  Such simultaneous approximation is a surprisingly nice property of our problem and is a rare phenomenon generally.~\footnote{We are aware of only one other situation in approximated majorization for minimizing symmetric convex objectives~\citep{goel2006simultaneous}, where there is a simultaneous approximation guarantee but their optimization problem is very different from ours and their ratio is logarithmic in general. This powerful result has recently been used by \citet{banerjee2023fair} for the welfare guarantee in information design.}

Finally, to validate the empirical performance of our mechanism, we apply it to conference peer review and conduct a series of experiments on both synthetic data and ICLR review data. Our experiments show that the proposed mechanism improves estimation performance relative to the original review scores across a variety of settings.

\section{Problem Formulation}
\label{sec:model}

This paper considers the review score calibration problem under a network of overlapping ownership.\footnote{Besides the motivating example of academic conference review, similar problems with overlapping ownership also exist in {internal company project or proposal review and in internal screening of jointly produced creative works such as songs, demos, scripts, and film projects~\citep{demere2019role,o2011multiple,guimera2005team}.} Hence, our technical sections study an abstract model with generic terms ``owner/item''.}
There are $m$ owners and $n$ items. The $j$-th owner owns a set of items $\cI^j \subseteq [n]=\{1,2,\dots,n\}$, with $|\cI^j|=n_j$.~\footnote{The ownership relation forms a bipartite graph (see, e.g., Figure~\ref{fig:bipartite}), and we formalize this structure in the proofs in Appendix~\ref{append:greedy-approximation}.}
For each item $i \in [n]$, let $R_i$ be its \emph{ground-truth score} and
$$
y_i = R_i + z_i
$$
be the \emph{raw review score} given by its reviewers with noise $z_i$.
As a friendly reminder of our notation, $i$ and $\cI$ are used to denote \emph{item} index and \emph{item} set, respectively, whereas $j$ is used to index the \emph{owner}.   Denote by $\bz = (z_1, z_2, \dots, z_n), \bR=(R_1, R_2, \dots, R_n)$ and $ \by =(y_1, y_2, \dots, y_n) $, respectively, the vector of noise, the ground-truth scores, and the raw review scores.

In applications such as conference paper reviews, the \emph{ground-truth scores} $\bR$ can be interpreted as the mean evaluation scores of the papers perceived across all experts in the community. This item evaluation score is mathematically well defined and always exists,\footnote{We remark that a paper's evaluation score should be carefully distinguished from its ``true merit,'' which is a vague concept whose existence might be arguable. Unlike merit, a paper's true evaluation score is well defined and is what peer review procedures and statistical methods like ours seek to estimate.} but it is difficult to observe accurately during the review phase because of limited reviewing resources and insufficient time to observe the items' impact. This is essentially the motivation for peer review, during which the ground-truth score is estimated by a few experts sampled from the community. The average review scores from these experts form the \emph{raw review scores} $\by$ --- noisy estimates of the ground-truth scores $\bR$.
Therefore, the high-level objective of a calibration mechanism $\cM$ is to output \emph{adjusted review scores} $\hat{\bR}$ that provide more accurate estimates of the ground-truth scores $\bR$.
An advantage of such a calibration mechanism for estimating the community's expected evaluation of papers is that it is naturally compatible with  existing peer review systems.

To tackle the review score calibration problem with the limited supply of qualified reviewers, we resort to the design of \emph{owner-assisted calibration mechanisms} that elicit and utilize self-evaluation data from the item owners.
We abstract this mechanism design problem into the construction of an estimator $\hat{\mathrm{R}}_{\cM}$ for $\bR$ that combines the raw scores $\by$ with the  information elicited from owners $\{\pi^j \}_{j=1}^{m} $ to improve the estimates of $\bR$.
In this paper, we focus on eliciting each owner's ranking of her items, where $\pi^j$ is a permutation of $1, 2, \ldots, n_j$ that specifies an ordering of the $j$-th owner's items in $\cI^j$. The benefits of this design choice and possible relaxations are discussed in Section~\ref{sec:discussion}.
The owner-assisted calibration mechanism then proceeds in the following stages.
{
\begin{enumerate}
    \item \textit{Before Review:} Each owner $j$ observes information about her items $\cI^j$, is informed of the calibration mechanism, and is then asked to report a ranking $\pi^j$ of her items.~\footnote{If an owner does not report a ranking, a uniformly random ranking will be used. As will be shown later, this design ensures that rational owners participate in the mechanism. It is also without loss of generality to assume that owners report the full ranking of their items; see Remark~\ref{rm:implementation}.
}
    \item \textit{After Review:} Once the raw review score $\by$ is realized, the mechanism accordingly estimates and outputs the adjusted review scores as $\hat{\bR} = \hat{\mathrm{R}}_{\cM}(\{\pi^j\}_{j=1}^{m}; \by)$.
\end{enumerate}
}

The key challenge of this design problem is to account jointly for statistical efficiency and incentive compatibility.
To formalize this problem, we adopt the mechanism design approach in modeling the strategic incentives of item owners on the calibration results.
That is, we assume each owner has a utility function on their items' adjusted review scores. Let owner $j$'s utility function be  $\mathrm{U}^j: \RR^{n_j} \to \RR$ such that each owner $j$ derives utility $\mathrm{U}^j\left([\hat{R}_i]_{i\in \cI^j}\right)$ from adjusted review scores $\hat{\bR}$, output by the calibration mechanism.
{We discuss how this  utility function based on review scores can be translated to the one based on the final outcomes in Section~\ref{sec:discussion}.}
We employ the Bayesian game framework~\citep{harsanyi1967games} to study the owners' strategic decision-making at the information elicitation stage during submission, where the realization of the raw review scores~$\by$ is uncertain to the owners (e.g., due to the unknown reviewer assignments and noise).
We say a profile of owners' reports $\{\pi^j\}_{j=1}^{m}$ forms a Bayes-Nash equilibrium (BNE)~\footnote{It suffices to consider pure-strategy NEs in most of this paper; we defer the notation for the more general definition of mixed-strategy NEs to Appendix~\ref{append:mixed-ne}.} under mechanism~$\cM$
if, for any owner $j \in [m]$, given the other owners' reports $\pi^{-j} = \{\pi^{j'}\}_{j'\neq j}$, owner $j$'s expected utility from reporting $\pi^j$ is no worse than that from reporting any other ranking $\tilde{\pi}^j$ or, mathematically,
\begin{equation*}\tag{\text{Equilibrium Condition}}
    \quad \Ex_{\by} \left[\mathrm{U}^{j}\left( \hat{\mathrm{R}}_{\cM}(\pi^j, \pi^{-j}; \by) \right)\right] \geq \Ex_{\by} \left[\mathrm{U}^{j}\left( \hat{\mathrm{R}}_{\cM}(\tilde{\pi}^j, \pi^{-j}; \by) \right) \right],\quad \forall j \in [m].
\end{equation*}
We say a mechanism $\cM$ is \emph{truthful} if every owner $j$ reporting the true ranking $\pi^j$ of her items forms a Bayes-Nash equilibrium under $\cM$. Through a standard revelation principle argument~\citep{nisan1999algorithmic}, it is without loss of generality to restrict the design space to truthful mechanisms. {This equilibrium condition guarantees truthfulness with respect to unilateral deviations by individual owners. We discuss potential extensions to address group manipulation in Section~\ref{sec:discussion}.}

To demonstrate the theoretical guarantees of our mechanism, we make the following  assumptions that are natural in peer review applications.
In Section~\ref{sec:exp}, we also conduct empirical studies that demonstrate the performance of our mechanism in settings beyond these assumptions.

{
\begin{assumption}[Informed Owners] \label{assum:informed-owners}
    For each $j$, the $j$-th owner knows the relative ranking of her own items in $\cI^j$, according to their ground-truth scores.
\end{assumption}
}

\begin{assumption}[Exchangeable Noise Distribution] \label{assum:exchangeable-noise}
The review noise vector $\bz = (z_1, \dots , z_n)$ follows an exchangeable distribution in the sense that $(z_1, \dots , z_n)$ has the same probability distribution in $\RR^n$ as in $\rho \circ \bz := (z_{\rho(1)}, \dots , z_{\rho(n)})$ for any permutation $\rho$ of $1,2,\dots, n$.
\end{assumption}

\begin{assumption}[Convex Utility] \label{assum:convex-utility}
For each $j$, the $j$-th owner's utility function takes the form of $\mathrm{U}^j(\hat{\bR}) = \sum_{i\in \cI^j} {U}^j(\hat{R}_i)$,
where ${U}^j: \RR \to \RR$ is a non-decreasing convex function.
\end{assumption}

{Assumption~\ref{assum:informed-owners} is an idealized benchmark that requires each owner to know the ranking of her own items under the ground-truth scores.\footnote{The assumption that owners know ranking information also appears in machine learning applications that learn parameters from human rankings or comparisons. Such rankings are often posited to be more accurate than absolute values \citep{yue2009interactively,bai2022training}.} This ranking may not follow from only the scores that the owner herself would assign. The assumption yields exact truthfulness. Section~\ref{sec:discussion} gives a single-owner relaxation based on the owner's perceived scores and an explicit condition on her predictive beliefs.
Assumption~\ref{assum:exchangeable-noise} imposes symmetry on review noises, and
Assumption~\ref{assum:convex-utility} is a tractable sufficient condition for our truthfulness analysis.
It is motivated by high-risk-high-reward environments, but Section~\ref{sec:discussion} also discusses its limitations and empirical relevance.
}

%% file: nash.tex
\section{An Isotonic Mechanism for Completely Overlapping Ownership}
\label{sec:ne}

We begin with a rank-calibrated score estimator $\hat{\mathrm{R}}(\pi ; \by ) $ that employs isotonic regression on raw review scores $\by$ and any reported ranking $\pi$ from the owner, via the following convex program,
\begin{align*}
\hat{\mathrm{R}}(\pi ; \by )  = \argmin_{\br \in \RR^{n}}\ & \norm{\by- \br}^2 \\
 \text{ s.t. } & r_{\pi(1)} \geq r_{\pi(2)} \geq \dots \geq r_{\pi(n)},
\end{align*}
where $\norm{\cdot}$ denotes the $\ell_2$ norm throughout this paper.

In the general setup with overlapping ownership, we would like to design a mechanism that aggregates information from as many owners as possible, while preserving the desirable properties of truthfulness and statistical efficiency. The most natural design is perhaps to use the (weighted) {averaged} scores from the estimates based on each owner's reported ranking.
The mechanism takes as input a problem instance specified by review scores $\by \in \RR^n$ and owner credentials $\{\alpha^{j}\}_{j=1}^{m}\in [0,1]^m$, and it outputs adjusted review scores~$\hat{\bR}$ using rankings elicited from all owners.
The owner credentials are weights $\{\alpha^{j}\}_{j=1}^{m}$ that specify the owners' levels of influence. We view them as part of the problem instance, reflecting each owner's expertise, reputation, and track record in that instance.
In practice, one could set $\alpha^j = 1$ to weight all reported rankings equally or choose personalized $\alpha^j$ to account for differences in expertise. We leave this choice to the practitioner's discretion.
We formally describe these procedures in Mechanism~\ref{algo:multi-unweighted}. The rest of this section is to analyze the situations when it does or does not work.

\begin{algorithm}[h]
\SetAlgorithmName{Mechanism}{mechanism}{List of Mechanisms}
        \caption{Isotonic Mechanism under Completely Overlapping Ownership}
        \label{algo:multi-unweighted}
        \KwIn{Review scores $\by \in \RR^n$, {owner} credentials $\{\alpha^{j}\}_{j=1}^{m}$. }
        \For{every $j \in [m]=\{1,2,\dots,m\}$ }{
        Elicit ranking $\pi^j$ from owner $j$. \\
        Solve for ranking-calibrated scores $ \hat{\bR}^j \gets \hat{\mathrm{R}}(\pi^j ; \by ) .$
        }
        Set $\hat{\bR} \gets  \sum_{ j = 1 }^{m} \alpha^{j} \hat{\bR}^j / \sum_{ j = 1 }^{m} \alpha^{j} $. \\
\Return $\hat{\bR}$.
\end{algorithm}

\subsection{Truthfulness under Completely Overlapping Ownership}
We start by considering a useful special case in which every submission is owned by every owner, referred to as the \emph{completely overlapping ownership} setting. Truth-telling forms a Nash equilibrium in this setting under Mechanism~\ref{algo:multi-unweighted}.~\footnote{This result can be related to the ex-post incentive compatibility of Mechanism~\ref{algo:multi-unweighted}, a common goal of truthful mechanism design.}
Moreover, this equilibrium is \emph{payoff-dominant}~\citep{harsanyi1988general}, i.e., Pareto superior to all other Nash equilibria in the game. Put simply, all owners prefer this equilibrium because it gives every owner the highest utility among all equilibrium outcomes.
This nice property makes it more plausible to expect agents' truthful behaviors, despite the potential existence of multiple equilibria in the game; more discussions about the truthfulness of this mechanism from the perspective of behavioral theory can be found in Remark~\ref{rm:truthful-nash}.

\begin{theorem}\label{thm:ne-same-item}
Under Assumptions \ref{assum:informed-owners}, \ref{assum:exchangeable-noise}, and \ref{assum:convex-utility}, if the ownership is completely overlapping,
\begin{enumerate}[leftmargin=*]
    \item Truthful reporting by all owners forms a Bayes-Nash equilibrium in Mechanism~\ref{algo:multi-unweighted}.
    \item This truthful-reporting equilibrium is payoff-dominant.
\end{enumerate}
\end{theorem}

 We defer the proof of Theorem~\ref{thm:ne-same-item} to Appendix~\ref{append:ne-same-item} and conclude this subsection with additional evidence from behavioral game theory on the truthfulness of the Isotonic Mechanism.

\begin{remark}[Truthfulness of Isotonic Mechanism from a behavioral angle]
\label{rm:truthful-nash}
Mechanism \ref{algo:multi-unweighted} induces the strategic game with special structures known as the \emph{common interest games}~\citep{harsanyi1988general,bacharach2018beyond}. We briefly highlight the connections here and discuss evidence from both empirical human preferences and behavioral theory on why truthful behavior should be expected in such games.
In its simplest form, the payoff matrix of a common interest game typically takes one of the three structures illustrated in Table \ref{tab:payoff-isotonic} (only orders of the payoff values matter).
Namely, there is an action profile (the upper left cell of each game) that represents the common interest of both players; this corresponds to the truthful action profile $(\pi^\star, \pi^\star)$ under the Isotonic Mechanism which forms a payoff-dominant Nash equilibrium, according to Theorem \ref{thm:ne-same-item}.
Meanwhile, the three variants of common interest games capture the different possible orders of payoff values in the lower right cell and those in the off-diagonal cells. Any variant of these payoff matrices may be realized under the Isotonic Mechanism, since our proof using Jensen's inequality only offers upper bounds on the utilities of these non-truthful action profiles (see e.g., Equation \eqref{eq:jensen}).

\begin{table}[h]
    \centering
\renewcommand*{\arraystretch}{1.2}
\begin{tabular}{c|c|c|}
   \multicolumn{1}{c}{}  & \multicolumn{1}{c}{$\pi^\star$}  & \multicolumn{1}{c}{$\pi^\circ$}\\ \cline{2-3}
  $\pi^\star$ & $10, 11$ & $4, 5$ \\ \cline{2-3}
  $\pi^\circ$ & $4, 5$ & $3, 4$ \\ \cline{2-3}
\end{tabular}
\qquad
\begin{tabular}{c|c|c|}
   \multicolumn{1}{c}{}  & \multicolumn{1}{c}{$\pi^\star$}  & \multicolumn{1}{c}{$\pi^\circ$}\\ \cline{2-3}
  $\pi^\star$ & $10, 11$ & $1, 1$ \\ \cline{2-3}
  $\pi^\circ$ & $1, 1$ & $3, 4$ \\ \cline{2-3}
\end{tabular}
\qquad
\begin{tabular}{c|c|c|}
   \multicolumn{1}{c}{}  & \multicolumn{1}{c}{$\pi^\star$}  & \multicolumn{1}{c}{$\pi^\circ$}\\ \cline{2-3}
  $\pi^\star$ & $10, 11$ & $1, 5$ \\ \cline{2-3}
  $\pi^\circ$ & $5, 1$ & $3, 4$ \\ \cline{2-3}
\end{tabular}
\caption{An illustration of three typical common interest game payoff matrices.}
    \label{tab:payoff-isotonic}
\end{table}

In experiments and real-life settings, people often choose the upper-left cell --- the \emph{payoff-dominant} outcome --- in the three game structures above~\citep{bacharach2018beyond}. This behavior is expected under the first payoff matrix because $(\pi^\star, \pi^\star)$ is the unique Nash equilibrium there.
It is somewhat reasonable under the second payoff matrix, as argued by \citet{harsanyi1988general} for its ``self-reinforcing property'' --- everyone thinks the other has no reason to prefer the payoff-dominated equilibrium outcomes. However, realizing the payoff dominant outcome becomes less clear under the third payoff matrix, because $(\pi^\circ, \pi^\circ)$ forms a ``risk-dominant'' Nash equilibrium --- i.e., compared to $\pi^\star$,  $\pi^\circ$ is  the less risky action under  uncertainty of the opponent's choice~\citep{harsanyi1988general} (specifically, under $(\pi^\circ, \pi^\circ)$ equilibrium,   row player gets $5$ instead of $3$ if the opponent deviates from the equilibrium action $\pi^\circ$ to $\pi^\star$).  Interestingly, standard game theory cannot account for this phenomenon of coordination on the payoff dominant outcome under the second and third payoff matrices, i.e., the Hi-Lo Paradox. One major theory that explains this phenomenon is
 the \emph{team reasoning} by
\citet{bacharach2018beyond} as a non-game-theoretic rationale, i.e., ``What do we want? And what should I do to play my part in achieving this?'' If there is common knowledge that both players adopt the team-reasoning mode of choosing their strategies, then both would choose the payoff-dominant equilibrium. These observations provide behavioral evidence for truthful reporting in the Isotonic Mechanism.

\end{remark}

\subsection{Non-truthfulness Beyond Completely Overlapping Ownership}\label{sec3:non-truthful}
Despite the properties of the Nash equilibrium shown in Theorem~\ref{thm:ne-same-item}, its assumption of completely overlapping ownership is not satisfied in many applications --- different papers are often written by different sets of authors. To extend Mechanism~\ref{algo:multi-unweighted} to partially overlapping ownership, one natural choice is to average the scores according to ownership, $\hat{R}_i \gets { \sum_{ j = 1 }^{m} e^{j}_i \hat{R}^j_i} / { \sum_{ j' = 1 }^{m} e^{j'}_i }$, where the binary ownership indicator $e^{j}_i = 1$ if and only if owner $j$ owns item $i$.
The question is whether truthful reporting still forms a Nash equilibrium. Unfortunately, the answer turns out to be ``No''.
Specifically, owners may gain from misreporting. One possible manipulation is to use the reputation of strong items to promote a weak item: this may hurt a strong item only slightly while helping the weak item substantially. We demonstrate this manipulation in the example below.

\begin{example}[Non-truthfulness under Partially Overlapping Ownership]\label{ex:non-truthful}
Consider a case of $m$ owners and $n=3$ items. The items have ground-truth scores $R_1 = 9, R_2=8, R_3 = 4$ --- one weak and two strong. The ownership is not completely overlapping. As illustrated in Table~\ref{tab:non-truthful}, the first $m-1$ owners jointly own the two strong items $1$ and $2$, while the $m$-th owner owns items $2$ and $3$.
For each $j$, let the $j$-th owner's utility be $\mathrm{U}^j(\hat{\bR}) = \sum_{i\in \cI^j} \max\{ \hat{R}_i - 5, 0 \}$.

\begin{table}[h]
\begin{center}
\begin{tabular}{>{\centering\arraybackslash}m{2cm}>{\centering\arraybackslash}m{1cm}>{\centering\arraybackslash}m{1cm}>{\centering\arraybackslash}m{1cm}>{\centering\arraybackslash}m{1cm}>{\centering\arraybackslash}m{1cm}}
\diagbox[width=20mm]{Item}{Owner}  & $1$ & $2$ & $\cdots$ & $m-1$  & $m$     \\
1 & 1 & 1 & $\cdots$ & 1 & 0   \\
2 & 1 & 1 & $\cdots$ & 1 & 1    \\
3 & 0 &  0 & $\cdots$ &  0 & 1  \\
\end{tabular}
\,
\begin{tabular}{c}
{True Scores} \rule[0ex]{0pt}{12mm} \\
$R_1 = 9$ \\
$R_2 = 8$ \\
$R_3 = 4$ \\
\end{tabular}
\end{center}
\caption{An illustration of the ownership matrix $E = (e_{i}^{j})_{n\times m}$ and ground-truth scores in Example~\ref{ex:non-truthful}.}
\label{tab:non-truthful}
\end{table}

For simplicity, suppose the reviews are noiseless so that, if all owners report their rankings truthfully, the calibration is perfect, i.e., $\hat{\bR} = \by = \bR$. However, the $m$-th owner does not have an incentive to report the ranking of items $2$ and $3$ truthfully when the first $m-1$ owners report the true ranking. Under the flipped ranking {$\tilde{R}_2^{m} \leq \tilde{R}_3^{m}$}, Mechanism~\ref{algo:multi-unweighted} gives $\tilde{R}_2^{m} = \tilde{R}_3^{m} = 6$ and adjusted scores $\tilde{R}_1 = 9$, $\tilde{R}_2 = \frac{1}{m}(\tilde{R}_2^{m} + \sum_{j=1}^{m-1} R_2^j) = 8 - \frac{2}{m}$, and $\tilde{R}_3 = \tilde{R}_3^{m} = 6$. The $m$-th owner's utility under truthful reporting is $3$, whereas the flipped ranking gives $3-\frac{2}{m}+1>3$ for any $m\geq 3$.
\end{example}

At a high level, such strategic behavior arises from an owner's uneven influence on the average scores of her different items --- we formalize this intuition in Theorem~\ref{thm:partition-necessary}. One might try to resolve this problem by carefully reweighting each owner's influence. This approach also fails. Specifically, in the example above, the $m$-th owner, as the sole owner of item 3, always fully determines its score. Yet this owner can never fully determine the score of item 2 unless we ignore the other owners' opinions or assign some weight to the raw review scores, which weakens the role of owner information in the calibration. Hence, we seek a different approach to aggregate the owners' information truthfully.

%% file: partition.tex
\section{Restoring Truthfulness via Partitioning}
\label{sec:partition}

\begin{figure}[tbh]
\centering
    \includegraphics[width=0.35\textwidth]{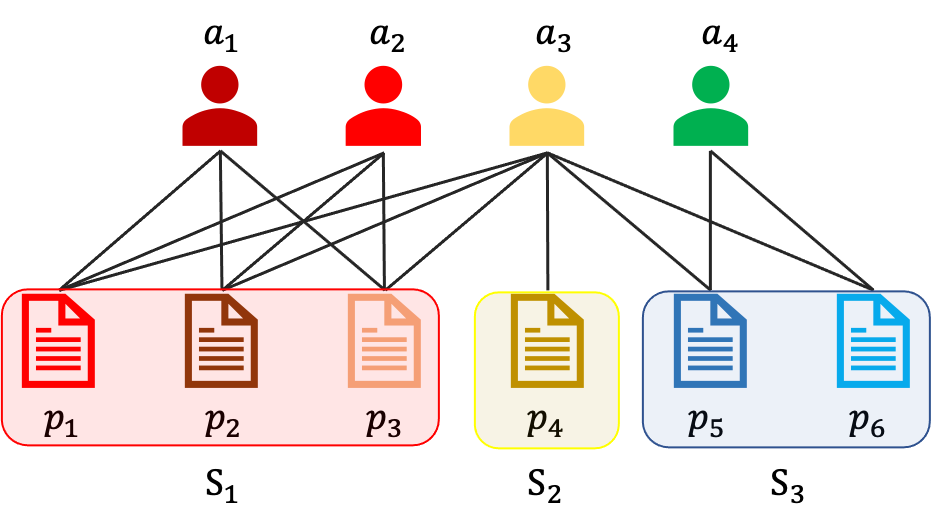}
\caption{A partition of partially overlapping ownership.}
\label{fig:partition}
\end{figure}

To ensure the truthfulness beyond complete ownership, this section studies a partition-based approach. Our main idea is to partition the ownership sets into multiple blocks, ensuring complete overlap within each block by some owners. We then individually apply Mechanism~\ref{algo:multi-unweighted} to each block, eliciting truthful rankings only from the owners who completely own that block, as illustrated in Figure~\ref{fig:partition}. Finally, we use the elicited rankings from these blocks to estimate the ground-truth scores. We formally describe this procedure in Mechanism~\ref{algo:deterministic-partition}.

\begin{algorithm}[h]
    \SetAlgorithmName{Mechanism}{mechanism}{List of Mechanisms}
        \caption{Partition-based Isotonic Mechanism for Partially Overlapping Ownership}
        \label{algo:deterministic-partition}
        \Parameter{Item set partition $\mathfrak{S} = \{\cS_1, \ldots, \cS_K\}$ based on ownership relations $\{  \cI^j \}_{j = 1}^{m}$.}
        \KwIn{Review score $\by \in \RR^n$, {owner} credentials $\{ \alpha^j \}_{j=1}^{m}$.}
\For{every $\cS_k \in \mathfrak{S}$}{
        Find all owners with complete ownership of $\cS_k$, i.e., $\cT_k \gets \{ j \in [m]:  \cS_k \subseteq \cI^j \}$. \\
        \If{$|\cT_k| = 0$  or  $|\cS_k| = 1$ }{Do not elicit any ranking information.}
        \Else{
        Apply Mechanism~\ref{algo:multi-unweighted}
        to item set $\cS_k$ with owner set $\cT_k$ and weights $\{ \alpha^j \}_{j\in \cT_k}$ to estimate the ground-truth score of items in $\cS_k$, denoted by $\hat{\bR}[\cS_k]$.
        }
        }
        \Return $\hat{\bR} = \{ \hat{\bR}[\cS_k]  \}_{k=1}^K$.
\end{algorithm}

Compared to Mechanism \ref{algo:multi-unweighted}, Mechanism \ref{algo:deterministic-partition} additionally determines a set of parameters which include a partition of item set  $\mathfrak{S} = \{\cS_1, \ldots, \cS_K\}$ that satisfies $\bigcup_{k=1}^{K} \cS_k = [n], \cS_k \cap \cS_{k'} = \varnothing$. We ask the parameters to only depend on the ownership relations for practical considerations: since the mechanism may be truthful only under some proper choice of parameters, the owners should be informed of the parameters during submission, when the ownership relations are formed but the review scores are not realized.
One can observe that in Mechanism \ref{algo:deterministic-partition}, each owner $j \in \cT_k$ has the complete ownership of the items in $\cS_k$ by construction.
As a corollary of Theorem \ref{thm:ne-same-item}, it is easy to show that Mechanism \ref{algo:deterministic-partition} is truthful under any partition  $\mathfrak{S}  = \{\cS_1, \ldots, \cS_K\}$ and any weights $\{ \{ \beta^j_k\}_{k=1}^{K} \}_{j=1}^{m}$, because we only elicit rankings of items within each block from those who completely own the block. Thus, all owners $j \in \cT_k$  will report truthful ranking over items in $\cS_k \subseteq \cI^{j}$.
Owing to the independence of score estimates for items between different partition blocks, the overall mechanism is truthful; we formalize it in the following corollary.

\begin{corollary} \label{prop:truthfulness-partition}
Mechanism~\ref{algo:deterministic-partition} is truthful for any input instance and parameter choice in the following sense: it forms a Bayes-Nash equilibrium for each owner to truthfully report the ranking of their items within each block specified by the partition.
\end{corollary}

{Moreover, Mechanism~\ref{algo:deterministic-partition} can be implemented in practice with a simple elicitation procedure. We discuss participation fairness separately in Section~\ref{sec:discussion}.}

\begin{remark}[Practical Implementation]\label{rm:implementation}
While Mechanism \ref{algo:deterministic-partition} only elicits the ranking of items within each block $\cS_k$, the mechanism designer may implement the mechanism by simply asking owners to rank all their items with the commitment to only use part of the ranking of each owner as specified by the mechanism. Under such a committed mechanism, any owner will be indifferent about the order of any two items in any two different partition sets $\cS_k$ and $\cS_{k'}$ since their order will never be used by the mechanism. Thus, the owner can be assumed to break ties in favor of the designer and reveal their full ranking truthfully.\footnote{There is also a formal  justification for such tie-breaking by adding a negligible amount of randomness for picking an arbitrarily different partition, which consequently creates a negligible amount of incentive for any owner to report other rankings truthfully. Such equilibrium refinement via randomization has appeared in the literature on mechanism design \citep{nisan1999algorithmic}.  } Notably, however, the designer's commitment to not using any order information beyond what the mechanism specified (despite such information being elicited) is important, since otherwise, truthfulness will not hold. In reality, the mechanism designer as a trusted authority (e.g., the organizers of a large ML conference) usually has such commitment power.

\end{remark}

\subsection{Reduction from Personalized Partition to Global Partition}\label{subsec:partition-necessity}

\begin{table}[tbh]
    \centering
    \begin{tblr}{colspec = {ccccc},
    cell{2}{2} = {red!30},
    cell{3}{2} = {red!30},
    cell{4}{3} = {yellow!30},
    cell{4}{4} = {yellow!30},
}
    \diagbox[width=20mm]{Item}{Owner}  & $1$ & $2$ & $3$       \\
         $1$ & 1 & 0 & 1   \\
         $2$ & 1 & 1 & 0     \\
         $3$ & 0 & 1 & 1   \\
\end{tblr}
\qquad
\begin{tblr}{colspec = {ccccc},
    cell{2}{2} = {red!30},
    cell{3}{2} = {red!30},
    cell{3}{3} = {yellow!30},
    cell{4}{3} = {yellow!30},
    cell{2}{4} = {cyan!20},
    cell{4}{4} = {cyan!20},
}
    \diagbox[width=20mm]{Item}{Owner}  & $1$ & $2$ & $3$       \\
         $1$ & 1 & 0 & 1   \\
         $2$ & 1 & 1 & 0     \\
         $3$ & 0 & 1 & 1   \\
\end{tblr}
    \caption{An illustration of partition-based mechanism (left) and a more general mechanism (right). In the partition-based mechanism, each color denotes a partitioned item set  block. In the more general mechanism, different owners can have different item set partitions, as marked by different colors.  }
    \label{tab:alternative-mechanism}
\end{table}

At this point, one may wonder whether restricting ranking elicitation to partitioned item sets   is necessary, as it inevitably forces the mechanism to ignore some of the {owners'} revealed information across the partition blocks.
For example, consider an instance with $m=n=3, \cI^1 =\{1,2\}, \cI^2 =\{2,3\}, \cI^3 =\{1,3\}  $ --- i.e., each owner owns two items, and each item has two co-owners (see  the illustration in Table \ref{tab:alternative-mechanism}). We can see that, with any partition of the item set,  Mechanism~\ref{algo:deterministic-partition}  can elicit ranking information from at most one owner. However, a more aggressive design could be eliciting  the ranking information from every owner $j$ about her items in $\cI^j$ and then applying a generalized version of Mechanism \ref{algo:multi-unweighted} by using the elicited (partial) ranking of $j$'s items. This is certainly a statistically more efficient design, but the key question is whether such a mechanism can still be guaranteed to be truthful.
{
Our next result gives a negative answer: within the isotonic-regression mechanism class with personalized partitions, every truthful mechanism can be reduced to an equivalent global-partition mechanism in terms of truthfully elicited information. Thus, within this class, one may need to give up cross-block comparisons to retain incentive guarantees. This identifies a limitation on truthfully elicitable information within the studied isotonic-regression framework, while leaving open the possibility of truthful mechanisms outside this class.
}

We start by generalizing Mechanism \ref{algo:deterministic-partition} to a broader class of Isotonic Mechanisms, as described in Mechanism~\ref{algo:linear-isotonic}.
In this more generalized Isotonic Mechanism, the calibrated score of any item $i\in [n]$ is now allowed to depend on the information elicited from all its owners, denoted by $\cJ^i \subseteq [m]$.
Concretely, the input to Mechanism   \ref{algo:deterministic-partition} and Mechanism~\ref{algo:linear-isotonic} is the same, but the parameter choices in Mechanism~\ref{algo:linear-isotonic} are strictly more general: (1) the item partition $\mathfrak{S}^j = \{\cS^{j}_1, \cS^{j}_2, \dots \cS^{j}_{K^j} \}$, satisfying  $\bigcup_{k=1}^{K^j} \cS_k^j = \cI^j, \cS_k^j \cap \cS_{k'}^j = \varnothing$, is now allowed to be different across owners; (2)  additionally, a  weight vector $\bbeta^j = [\beta_1^j, \beta_2^j, \dots, \beta_n^j] \in \RR_{\geq 0}^n$ is introduced to allow owner $j$'s more fine-grained influence on her items with  itemized weights. This is much more powerful than $j$'s influence   in Mechanism~\ref{algo:deterministic-partition} that is restricted to be the same among items. The parameters are also determined from the input problem instance, specifically the ownership relation $\{  \cI^j \}_{j = 1}^{m}$.
The adjusted score for each item is then similarly determined by a weighted linear combination of the rank-calibrated scores from each owner. Notably, parameter $\beta_i^j$  for any $i\not\in \cI^j$ is never used by Mechanism \ref{algo:linear-isotonic} hence can be arbitrary. We nevertheless defined $\bbeta^j $ as a vector in $\RR_{\geq 0}^n$ mainly for notational convenience.

\begin{algorithm}[t]
    \SetAlgorithmName{Mechanism}{mechanism}{List of Mechanisms}
        \caption{{Isotonic Mechanism with Personalized Partition}}
        \label{algo:linear-isotonic}
        \Parameter{Personalized item  partition $\mathfrak{S}^j = \{\cS_1^j, \ldots, \cS_{K^j}^j\}$ and itemized weights $\bbeta^j $   for each owner $j\in[m]$ based on ownership relations $\{  \cI^j \}_{j = 1}^{m}$.
}
        \KwIn{Review scores $\by \in \RR^n$, {owner} credentials $\{ \alpha^j \}_{j=1}^{m}$. }
        \For{every $j \in [m]$}{
        Apply Mechanism~\ref{algo:multi-unweighted} to owner $j$ with partition $\mathfrak{S}^j $ to estimate the ground-truth score, denoted by $\hat{R}^j_i$, for each item $i$ in $\cI^j$.
}
        \Return $\big\{ \hat{R}_i = \sum_{j\in \cJ^i}  \alpha^j \beta^j_i \hat{R}^j_i / \sum_{{j'}\in \cJ^i} \alpha^{j'} \beta^{j'}_{i} \big\}_{i=1}^n$.
\end{algorithm}

It is easy to see that the class of Mechanism \ref{algo:linear-isotonic} strictly contains the class of Mechanism \ref{algo:deterministic-partition}.
On the one hand, it can be immediately verified that  Mechanism~\ref{algo:deterministic-partition} with any item set partition $\mathfrak{S} = \{\cS_1, \ldots, \cS_K\}$ can be formulated as Mechanism~\ref{algo:linear-isotonic} with some parameter $\big\{\mathfrak{S}^j , \bbeta^j \big\}_{j=1}^{m}$ (see the proof of Theorem \ref{thm:partition-necessary} for the details)
On the other hand, Mechanism \ref{algo:linear-isotonic} does have strictly more expressivity than Mechanism~\ref{algo:deterministic-partition}. For example, the non-partition-based mechanism illustrated above in Table \ref{tab:alternative-mechanism} can be captured by the following elicitation rule:  $\mathfrak{S}^1=\{\{1,2\}\}, \bbeta^1=[1,1,0], \mathfrak{S}^2=\{\{2,3\}\}, \bbeta^2=[0,1,1]$ and so on.
One may now wonder whether this strictly more general Mechanism \ref{algo:linear-isotonic} is also strictly more powerful. Our next result shows that the answer is ``NO'' if the mechanism is truthful.

{
\begin{theorem}[Reduction from Personalized Partition to Global Partition] \label{thm:partition-necessary}
  Under Assumptions \ref{assum:informed-owners}, \ref{assum:exchangeable-noise}, and \ref{assum:convex-utility}, for any $\cM$ in the form specified in Mechanism~\ref{algo:linear-isotonic} that is truthful for every input, there exists a $\cM'$ in the form specified in Mechanism~\ref{algo:deterministic-partition} that elicits at least as much ranking information. Formally, if the order of any two items $i, i'$ is truthfully elicited by $\cM$, then their order is also truthfully elicited by $\cM'$.
\end{theorem}
}
{
Theorem~\ref{thm:partition-necessary} is a characterization result within the mechanism class of Mechanism~\ref{algo:linear-isotonic}. In particular, it shows that any truthful mechanism with personalized partitions in this isotonic-regression framework can be replaced by a truthful global-partition mechanism that preserves all truthfully elicited pairwise comparisons. It does not rule out truthful mechanisms outside this class.
}
{
The formal proof is deferred to Appendix~\ref{append:necessity-proof}. At a high level, our proof is structured with two main steps. First, we characterize the necessary conditions under which a mechanism of the form specified in Mechanism~\ref{algo:linear-isotonic} is truthful. Second, we show that any truthful mechanism in this personalized-partition class can be converted into a global-partition mechanism that elicits at least as much truthfully usable ranking information.
}

\begin{lemma}\label{lm:truthful-equivalence}
Under Assumptions \ref{assum:informed-owners}, \ref{assum:exchangeable-noise}, and \ref{assum:convex-utility}, if a Mechanism~\ref{algo:linear-isotonic} with parameters $\big\{\mathfrak{S}^j , \bbeta^j \big\}_{j=1}^{m}$ is truthful, then  the following two conditions must both hold:
\begin{enumerate}
\item[\textup{(I)}] Each owner has balanced influence on the items within each  partition blocks for any input, i.e.,
$$\text{for any } j\in [m], \cS\in \mathfrak{S}^j, i,i'\in \cS, \qquad \text{we have } \,  \omega^j_i = \omega^j_{i'},$$
where $\omega_i^j =  \alpha^j \beta^j_i / \sum_{j'\in \cJ^i} \alpha^{j'} \beta^{j'}_{i}$ denotes owner $j$'s relative influence on the   score of item $i$.
\item[\textup{(II)}] The parameters $\big\{\mathfrak{S}^j , \bbeta^j \big\}_{j=1}^{m}$  has a valid partition structure in the following sense, \vspace{-2mm}
\begin{equation}\label{eq:partition-almost}
  \text{for any } j,j'\in [m], \cS \in \mathfrak{S}^j, \cS' \in \mathfrak{S}^{j'},\text{ if }\cS\cap \cS' \neq \varnothing, \text{ then } \beta^j_i = \beta^j_{i'}, \text{ for all } i,i'\in \cS\cup \cS'.\footnote{We remark that  $\beta_i^j$ is set to $0$ for any  $ i\not\in \cI^j$ in this statement.}
\end{equation}
\end{enumerate}
Moreover, these two conditions are equivalent to each other.
\end{lemma}

\subsection{  Partition Optimization and Its Hardness }\label{subsec:partition-hard}
{
As illustrated above, any global partition makes Mechanism~\ref{algo:deterministic-partition} truthful. Moreover, within the mechanism class of Mechanism~\ref{algo:linear-isotonic}, every truthful mechanism with personalized partitions is equivalent to a truthful global-partition mechanism in terms of truthfully elicited ranking information. Hence, searching for an optimal truthful mechanism in this class reduces to searching for an optimal global partition.
}
{
These observations point to the last key piece of our design: identifying the item partition for optimal statistical efficiency. Since any partition yields a truthful mechanism (Corollary~\ref{prop:truthfulness-partition}), the remaining degree of freedom is the choice of partition for statistical performance. Throughout, we measure this performance by the \emph{risk} of the calibrated scores, i.e., the expected squared estimation error $\Ex\norm{\hat{\bR}-\bR}^2$; a lower risk means a higher \emph{statistical efficiency}, so the two notions move in opposite directions. Two structural properties are relevant: (i) \emph{block size} --- how many items share a block, and (ii) \emph{number of owners per block} --- how many owners can supply ranking information for each block. We formalize their statistical roles in two tractable settings that motivate our subsequent partition optimization.

\begin{proposition}[Informativeness Under Gaussian Equal Scores]
\label{prop:coarser-better}
Suppose $\bz \sim \mathcal{N}(0,\sigma^2 \bI_n)$ and the true scores are equal within each block of a partition $\mathfrak{S}$. Then the isotonic risk of a block of size $s$ is $\sigma^2 H_s$ with $H_s = \sum_{k=1}^{s} 1/k$, so a partition with block sizes $(s_1, \ldots, s_K)$ has total risk $\Ex\norm{\hat{\bR}^{(\mathfrak{S})} - \bR}^2 = \sigma^2 \sum_{k=1}^{K} H_{s_k}$ and total wellness $\sum_{k=1}^{K} w(s_k)$ for the induced wellness function
\[
w(0) := 0, \qquad w(s) := \sigma^2 s - \Ex\norm{\hat{\bR}^{([s])} - \bR}^2 = \sigma^2(s - H_s) \quad (s \geq 1),
\]
where $\hat{\bR}^{([s])}$ is the isotonic regression estimate on a single block of size $s$. This $w$ satisfies $w(0) = w(1) = 0$, is strictly increasing for positive integer block sizes, and is strictly convex on the non-negative integers.
\end{proposition}
The proof is in Appendix~\ref{append:coarser-better}.
Proposition~\ref{prop:coarser-better} gives an exact output-risk foundation for the informativeness objective in a canonical benchmark. Within this benchmark, splitting an equal-score block into smaller blocks strictly increases expected risk.
}

{
\begin{proposition}[Robustness: More Owners Reduce Risk Under Noisy Perceptions]\label{prop:more-owners}
Consider a single partition block of $s \geq 2$ items with true scores $\bR \in \RR^s$, $R_1 \geq \cdots \geq R_s$, and review scores $\by = \bR + \bz$.
Suppose each of $L$ owners independently observes perceived scores $\bP^j = \bR + \bepsilon^j$, where $\{\bepsilon^j\}_{j=1}^L$ are i.i.d.\ and independent of $\bz$, and reports the ranking $\pi^j$ induced by $\bP^j$.
Define the aggregated estimator
$
\hat{\bR}_L = \frac{1}{L}\sum_{j=1}^{L} \hat{\mathrm{R}}(\pi^j ; \by).
$
Then
\begin{equation}\label{eq:owners-bv}
\Ex\norm{ \hat{\bR}_L - \bR }^2 \;=\; {\Ex_{\bz}\norm{ \bmu(\by) - \bR }^2} \;+\; \frac{1}{L}{\Ex_{\bz}\left[\Ex_{\pi}\norm{ \hat{\mathrm{R}}(\pi;\by) - \bmu(\by) }^2\right]},
\end{equation}
where $\bmu(\by) := \Ex_{\pi}\left[\hat{\mathrm{R}}(\pi;\by)\right]$ is the expected isotonic regression estimate over the random ranking $\pi$ (induced by perception noise $\bepsilon$), conditional on $\by$.
In particular:
\begin{enumerate}[leftmargin=*]
\item $\Ex\norm{ \hat{\bR}_L - \bR }^2$ is non-increasing in $L$, with strict decrease whenever the variance term in~\eqref{eq:owners-bv} is positive, equivalently when perception noise yields distinct isotonic estimates with positive probability.
\item Under Assumption~\ref{assum:informed-owners} (no perception noise), every owner reports the true ranking $\pi^\star$ and the variance term vanishes, so the risk is the same for all $L \geq 1$.
\end{enumerate}
\end{proposition}

The proof is in Appendix~\ref{append:more-owners}; it is a direct application of the bias-variance decomposition for i.i.d.\ averages.
Proposition~\ref{prop:more-owners} addresses a key point: under Assumption~\ref{assum:informed-owners}, all owners report the same ranking, so adding owners does not reduce risk. The benefit of having multiple owners arises when Assumption~\ref{assum:informed-owners} is relaxed and different owners may hold conflicting private rankings.
}

{
These local statistical properties motivate two proxy principles for evaluating partition quality \emph{within the isotonic-regression mechanism class}: (1) \emph{Informativeness} --- prefer larger blocks, as justified by Proposition~\ref{prop:coarser-better} in the Gaussian equal-score benchmark; (2) \emph{Robustness} --- prefer more owners per block, justified by Proposition~\ref{prop:more-owners}.
However, there is an inherent tension between the two principles, as larger blocks usually have fewer joint owners, since multiple owners rarely jointly own many items. Pursuing larger blocks can therefore prevent us from eliciting rankings from multiple owners. To address this tradeoff, we formalize the informativeness principle as an optimization objective and the robustness principle as a constraint. We then study the computational complexity of the resulting partition optimization, whose objective serves as a proxy for statistical efficiency.\footnote{We emphasize that the proxy objectives defined below measure how much ranking information the isotonic mechanism can exploit, conditioned on using isotonic regression to produce calibrated scores. A trivial mechanism that outputs raw scores without adjustment could elicit the same rankings while remaining truthful (by owner indifference), but would not use the elicited information for calibration. Our proxy objectives are meaningful within the mechanism class studied in this paper.}
}

\vspace{2mm}
\noindent\textbf{Informativeness (as optimization objective). }
In regard to the performance, we aim to identify partitions that are most informative for the review score calibration. On one hand,  larger-sized blocks generally contain more ranking information, whereas in the extreme case if a block contains only one item,   the mechanism cannot elicit any ranking information within the block.
On the other hand, selecting a large block may render other blocks small, creating a complex tradeoff among partition choices. To systematically evaluate the quality of a partition $\mathfrak{S}$, we introduce an \emph{informativeness} objective function, \vspace{-2mm}
$$\obj(\mathfrak{S} ) = \sum_{k=1}^{K} w(|\cS_k|),$$
determined by some \emph{wellness function} $w$ on the size of each partition block. For example, consider the following two choices of the wellness function $w(\cdot)$. One is $w(x)=x^2$, a simple quadratic function of block size that relates to the number of pairwise comparisons within a block:
\begin{flalign} \label{eq:obj-pairwise}
 \textbf{comparison-focused objective:} \hspace{4mm}  \obj(\mathfrak{S} ) = \sum_{k=1}^{K} |\cS_k|^2. &&
\end{flalign}

\noindent Another choice is $w(x) =  \max \{   x - 1,  0 \} $, which is called \emph{size-focused} objective, defined below:
\begin{flalign} \label{eq:obj-size-focused}
 \textbf{size-focused objective:}   \hspace{17mm} \obj(\mathfrak{S} ) =  \sum_{k=1}^{K}     \max \{  |\cS_k| - 1 , 0 \} =    \sum_{k=1}^{K}     \left[ |\cS_k| - 1 \right]   =   n - K, &&
\end{flalign}
where the second equation is because any block in the partition has at least $1$ item (i.e., $|\cS_k| \geq 1$). Since $n$ is a constant, maximizing the size-focused objective is equivalent to minimizing the size of the partition $K$ (thus the name ``size-focused'').

Generally, it is reasonable to require $w(\cdot)$ to be monotonically \emph{increasing} and \emph{convex}. Together with $w(0)=0$, convexity implies that an original block is weakly preferred to any split: $w(l) \geq w(l_1)+w(l_2)$ whenever $l=l_1+l_2$. The normalization $w(0)=0$ also prevents empty blocks from contributing to partition wellness. Both functions above satisfy these properties, as does any monomial $w(x)=x^p$ for $p\geq1$. Beyond these properties, however, the exact form of $w(\cdot)$ is generally unknown. We therefore seek algorithms in Subsection~\ref{sec:greedy-charm} that perform well for \emph{every} wellness function satisfying these properties.

\vspace{2mm}
\noindent\textbf{Robustness (as optimization constraint). }
The informativeness objective above is the quantity we want the partition to maximize. Robustness is instead a constraint we wish the partition to satisfy. In regard to robustness, aggregating information from multiple owners can help mitigate errors in each owner's perceived ranking when Assumption~\ref{assum:informed-owners} does not hold. Thus, the minimum number of owners for each partition block matters. Formally, we say a partition $\mathfrak{S}  = \{\cS_1, \ldots, \cS_K\}$ is $L$-strong if at least $L$ owners share all items in each non-singleton block, i.e., $\left| \cT_k \right| \geq L$ for every $\cS_k \in \mathfrak{S}$ with $|\cS_k| > 1$. A larger $L$ makes the Isotonic Mechanism less sensitive to errors in any one owner's perceived ranking. Notably, $L \geq 1$ is the minimum requirement for Mechanism~\ref{algo:deterministic-partition} to work. Next we show that generating a feasible $L$-strong partition reduces to generating a feasible $1$-strong partition for a different ownership instance.  This reduction  helps us convert any $L$-strongness constraint to a certain $1$-strongness constraint  during our design of the partition optimization algorithm.

\begin{proposition}[Reduction to $1$-Strong Partition]\label{prop:simple-graph}
For any ownership instance $\cO = \{  \cI^j \}_{j \in [m]}$,  we can construct in $O\left( m^Ln \right)$ time a different ownership instance $\cO' = \{  \bar{\cI}^j \}_{j \in [m]}$ with the same owner and item set such that any  partition $\cS$ is $L$-strong in $\cO$ if and only if $\cS$ is $1$-strong in $\cO'$.
\end{proposition}

 Proposition \ref{prop:simple-graph}  shows that to optimize the informativeness objective, it is without loss of generality to design algorithms to maximize the objective under $1$-strongness, for a transformed ownership set instance.  We defer its formal proof to Appendix~\ref{append:simple-graph} but provide a simple example below to illustrate the main idea of the reduction: any $2$-strong partition in the ownership instance $\cO = \{ \{1,2\}, \{1,2,3\}, \{2,3\} \}$ is a $1$-strong partition in the ownership instance $\cO' = \{ \{1, 2\}, \{2\}, \{2, 3\} \}$, and vice versa.
 Essentially, the instance $\cO'$ is constructed by {using} one owner to represent each subset of $L$ owners in the original instance $\cO$, whose common items are owned by this owner.  This is a polynomial-time reduction for small constant $L$, which is often the case in academic conference review due to the small number of co-authors on a paper.
Notably, this reduction does not fundamentally simplify the problem due to the intrinsic hardness of maximizing the informativeness objective, as shown in the following proposition; see Appendix~\ref{append:hardness} for its formal proof.
\begin{proposition}[Hardness of Partition Optimization]\label{prop:hardness}
 It is NP-hard to find the optimal partition $\mathfrak{S}  = \{\cS_1, \ldots, \cS_K\}$ that maximizes the size-focused objective in Eq.  \eqref{eq:obj-size-focused} subject to  $1$-strongness.
\end{proposition}

\subsection{Fast Greedy Partition with Robust Approximation Guarantees}  \label{sec:greedy-charm}

\begin{algorithm}[h]
    \caption{A Greedy Algorithm for $1$-Strong Partition}
    \label{algo:greedy}
    \KwIn{Ownership sets $\{\cI^j \}_{j=1}^{m}$.}
    Initialize the partition as $\mathfrak{S}  = \{ \}$ and set of selected items $\bar{\cI} = \varnothing$.\\
    \While{ $\bar{\cI} \subset [n] $}{
         Determine the largest residual {item} set $\cS^* =    \cI^{j^*}  \setminus \bar{\cI} $ where  $j^*  = \argmax_{j \in [m] } |\cI^{j}  \setminus \bar{\cI} |$. \\
        Update $\bar{\cI} \gets \bar{\cI} \union \cS^* $ and $\mathfrak{S}  \gets \mathfrak{S}  \union \{ \cS^* \} $.
    }
    \Return $\mathfrak{S}$.
\end{algorithm}

Despite the computational hardness for the optimal partition, we show that a natural  greedy algorithm presented in Algorithm \ref{algo:greedy}  can efficiently find $1$-strong partitions with a provably good approximation guarantee. In words, this greedy algorithm just  iteratively selects the largest residual {item} set owned by some owner. This algorithm can be implemented in time almost linear in the input size $\sum_{j\in[m]}{|\cI^j|}$, up to a $\log(m)$ factor,  thus is essentially the fastest algorithm one could hope to design.

The highlight of this greedy algorithm is that, though the algorithm itself is fully agnostic to any partition objectives,  its output is a robustly high-quality partition for \emph{every} natural wellness function, i.e., monotone and convex wellness functions, as formalized in the following Theorem~\ref{thm:greedy-approximation}.
In particular, consider the following hypothesis class $\cW$ of  wellness functions
$$ \cW := \{ w: \RR_{\geq 0} \to \RR_{\geq 0} | w(0) = 0, w \text{ is convex, non-decreasing} \}.$$

Algorithm \ref{algo:greedy} enjoys the following strong approximation guarantees for every $w$ within   $\cW$, whose proof is deferred to Appendix \ref{append:greedy-approximation}. In Remark~\ref{rm:special-case}, we give the exact approximation ratio for some optimization objectives of special interests and show its tightness. It is also worth mentioning that while Algorithm \ref{algo:greedy} only finds a partition satisfying $1$-strongness, the algorithm as well as its constant approximation guarantees can be readily extended to find the optimal $L$-strong partition in $O\left( m^Ln \right)$ time, using the reduction in Proposition \ref{prop:simple-graph}.

\begin{theorem}[Robust Approximation of Greedy] \label{thm:greedy-approximation}
 For any ownership instance $\{  \cI^j \}_{j \in [m]}$ with $N =    \sum_{j\in[m]}{|\cI^j|} $  total number of {owner-item} pairs, Algorithm \ref{algo:greedy} runs in $O( N \log m)  $ time and  outputs a 1-strong partition that is simultaneously a $c(w)$-approximation of the optimal objective for every $w \in \cW$, where
$$c(w) = \inf \Big\{ \frac{w(x)}{ w'_{-}(x) x} \Big| w'_{-}(x) > 0, x\geq 2 \Big\}.$$
Above, $w'_{-}(a) = \lim_{x\to a^{-}} \frac{w(x)-w(a)}{x-a}$ denotes the left derivative of $w$ at $a$.
\end{theorem}

\begin{remark}\label{rm:special-case}
Theorem~\ref{thm:greedy-approximation} readily implies that the partition found by the greedy algorithm  \emph{simultaneously} approximates at least $1/2$ of both the optimal comparison-focused objective and size-focused objective.
More generally, for $\alpha$-th degree polynomial functions $w(x) = x^{\alpha}$, the greedy algorithm simultaneously guarantees $(1/\alpha)$-fraction of the optimal objective since $w(x)/[w'(a) x] = 1/\alpha$ for every $x$.  Such objective-agnostic property of Algorithm \ref{algo:greedy} is  especially desirable in our application scenarios where the concrete form of the partition objective is difficult (if not impossible) to know.
Moreover,  the approximation ratio  in Theorem \ref{thm:greedy-approximation} for the greedy  algorithm is provably tight under every monomial objective $f(\mathfrak{S} ) = \sum_{k=1}^{K} |\cS_k|^\alpha$, $\alpha > 1$, and the construction of lower bound instances can be found in Appendix~\ref{sec:tight}.
\end{remark}

%% file: exp.tex
\section{{Empirical Evaluation in the Context of ML Conference Review}} \label{sec:exp}

\begin{figure}[tbh]
  \begin{center}
    \includegraphics[width=0.66\textwidth]{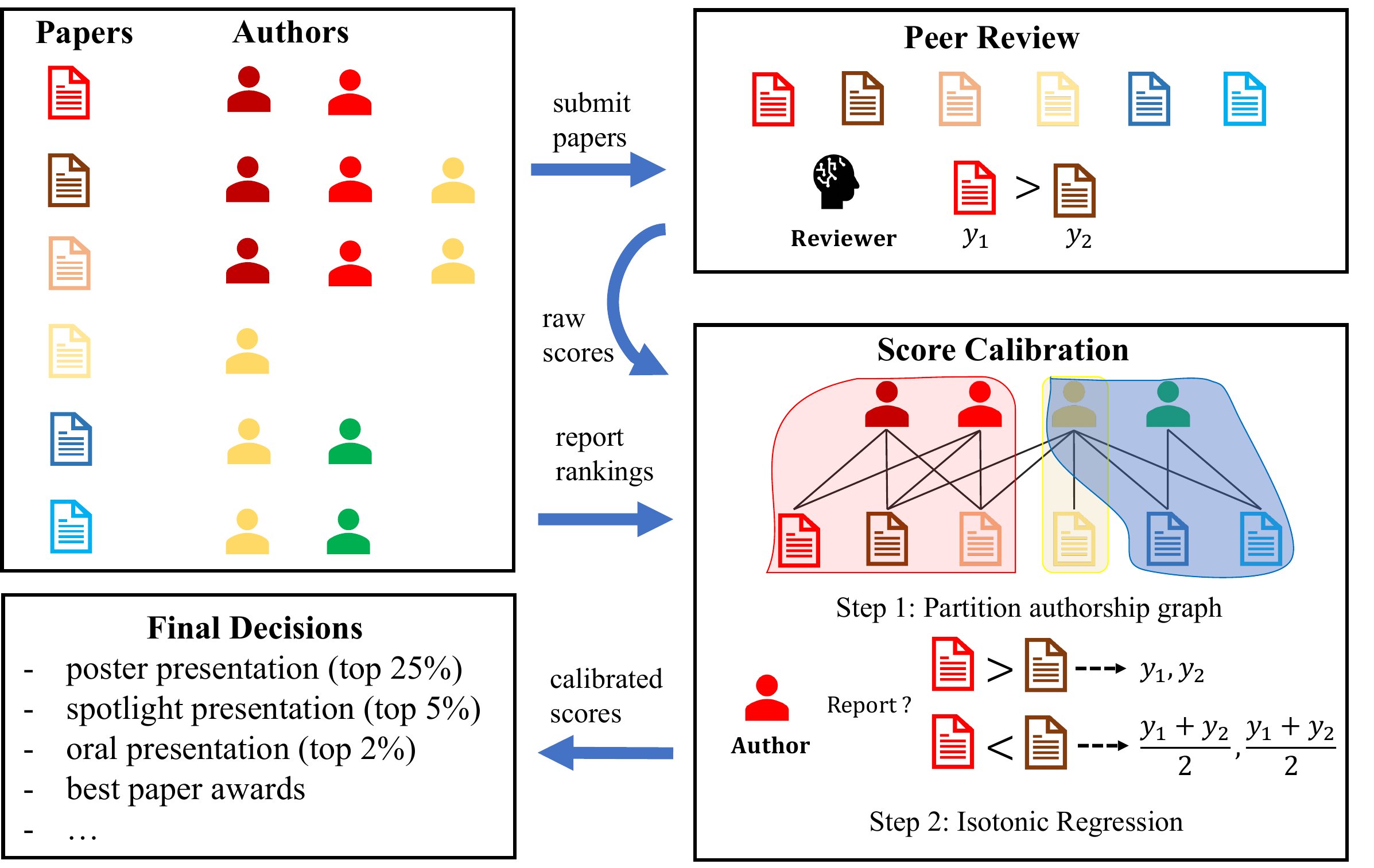}
  \end{center}
  \caption{An illustration of applying our proposed Isotonic Mechanism to the conference reviewing procedure.}
  \label{fig:isotonic-basics}
\end{figure}

 In this section, we study the empirical performance of our proposed mechanism {in our primary motivating domain of academic conference peer review. Figure~\ref{fig:isotonic-basics} illustrates how the mechanism can be applied in this setting.} One major evaluation challenge is that the performance measures depend on ground-truth scores, which are unavailable in most applications. Indeed, if the ground-truth scores were known, peer review would be unnecessary. We therefore use observed data from ICLR and ICML together with clearly specified surrogate targets.
Below, we describe our experimental setups and results.

\subsection{Experiment Setups: Datasets, Baselines and Metrics}
We start by describing the preprocessing procedures and characteristics of the two real-world datasets used in our experiments.

\begin{enumerate}[leftmargin=*]
    \item \textbf{ICLR 2021--2023:} {We collect ownership relations and review scores from ICLR 2021--2023 \citep{DBLP:conf/iclr/2021,DBLP:conf/iclr/2022,DBLP:conf/iclr/2023}, made publicly available through \href{https://openreview.net/group?id=ICLR.cc}{OpenReview.net}. For each paper, we use the mean of all available ratings as a surrogate ground-truth score $R_i$. In each of 100 independent replications, we generate the raw score as
    \[
    y_i=\operatorname{clip}(R_i+z_i,1,10),\qquad z_i\sim\cN(0,\sigma^2),
    \]
    for $\sigma\in\{1,1.5,2,2.5,3,3.5,4\}$. The clipping keeps every simulated score on the observed ICLR rating scale, so negative or above-range scores do not enter the mechanism.}

    \item \textbf{ICML 2023:} We collect ownership relations, review scores, and authors' reported rankings of their papers from ICML 2023 through a survey conducted by \href{https://openrank.cc/}{OpenRank.cc}. We retain only authors who participated in the survey and provided a ranking, together with their papers. We use these observed rankings in the separate one-shot evaluation described in Q3.
\end{enumerate}

{Neither dataset reveals a literal ground-truth score. We therefore use \emph{surrogate ground truth} consistently to denote the target against which MSE is evaluated. ICLR is a replicated controlled-noise simulation, whereas ICML is a separate one-shot evaluation using observed author rankings. Their numerical results are therefore not directly comparable.}

{We use Mechanism~\ref{algo:deterministic-partition} with uniform owner credentials ($\alpha^j=1$ for all $j$) to determine the adjusted scores $\hat{\bR}$ from review scores $\by$ and the elicited ranking information.}
{For comparison, we consider four approaches in the ICLR experiments, where all rankings are generated from the surrogate ground truth. The \emph{baseline} directly uses the raw scores. The \emph{randomized} approach uses a feasible randomized partition, while the \emph{greedy} approach uses the partition returned by Algorithm~\ref{algo:greedy}. We also include an \emph{oracle all-owner} benchmark, which assumes that every eligible owner truthfully ranks all of her papers. It applies isotonic regression separately to each eligible owner's papers and averages the estimates for papers with multiple eligible owners. Papers with no eligible owner retain their raw scores. This benchmark uses all available owner rankings but is not incentive compatible under partial overlap, so we treat it as an information-rich benchmark rather than a formal upper bound.}

{We evaluate the approaches by MSE, $\frac{1}{n}\norm{\bR-\hat{\bR}}^2$, and top-$30\%$ accuracy, $|\hat{\cP}_{@30}\cap\cP_{@30}|/|\cP_{@30}|$. Here, $\cP_{@30}$ and $\hat{\cP}_{@30}$ contain the top papers under the surrogate ground truth and adjusted scores, respectively. We also measure the share of baseline false rejections recovered by the greedy method:
\[
\frac{|(T_{30}\setminus B_{30})\cap G_{30}|}{|T_{30}\setminus B_{30}|},
\]
where $T_{30}$, $B_{30}$, and $G_{30}$ are the top-$30\%$ sets under the surrogate ground truth, raw scores, and greedy estimates.}
{Because clipping and averaged ratings can produce ties, each replication uses a score-independent random ordering to break ties at percentile cutoffs. The same ordering is used for the surrogate target and every method within that replication, so the comparison does not depend on paper input order.}

{Table~\ref{tab:iclr-stats} reports the filtered ownership graphs and the greedy partitions used in the ICLR experiments. An owner is eligible if she owns at least two papers. A ranking provider is an eligible owner who completely owns at least one multi-item block and is therefore asked for a ranking.}

 \begin{table}[tbh]
\centering
\small
\textit{Panel A: Filtered ownership graphs}\\[2pt]
\begin{tabular}{lrrr}
\toprule
Dataset & Papers & Owners & Eligible owners \\
\midrule
ICLR 2021 & 2,964 & 8,875 & 1,863 \\
ICLR 2022 & 3,261 & 10,411 & 2,209 \\
ICLR 2023 & 3,813 & 12,587 & 2,719 \\
\bottomrule
\end{tabular}

\vspace{2mm}
\textit{Panel B: Greedy partitions}\\[2pt]
\begin{tabular}{lcrrrr}
\toprule
Dataset & $L$ & Multi-item blocks & Calibrated papers & Providers & Pairwise comparisons \\
\midrule
ICLR 2021 & 1 & 571 & 1,936 & 743 & 4,274 \\
          & 2 & 440 & 1,086 & 976 & 1,045 \\
ICLR 2022 & 1 & 631 & 2,170 & 802 & 4,849 \\
          & 2 & 466 & 1,203 & 1,034 & 1,406 \\
ICLR 2023 & 1 & 753 & 2,584 & 966 & 5,434 \\
          & 2 & 601 & 1,488 & 1,307 & 1,448 \\
\bottomrule
\end{tabular}
      \caption{Statistics of the filtered ICLR ownership graphs and their greedy partitions. A provider completely owns at least one multi-item block; an $L=2$ block may therefore have multiple providers. The oracle all-owner benchmark uses all eligible owners in Panel A.}
      \label{tab:iclr-stats}
\end{table}

{We use $L=1$ for the real-data performance comparison. As Table~\ref{tab:iclr-stats} shows, requiring $L=2$ leaves substantially fewer calibrated papers and pairwise comparisons. We therefore vary $L$ only in the controlled synthetic experiment below.}

\subsection{Experiment Results}
Using these two real-world peer review datasets, we study three questions about the performance of our proposed mechanism.

\begin{figure}[tbh]
    \centering
    \subfigure[MSE on ICLR 2021 ]{\includegraphics[width=0.32\textwidth]{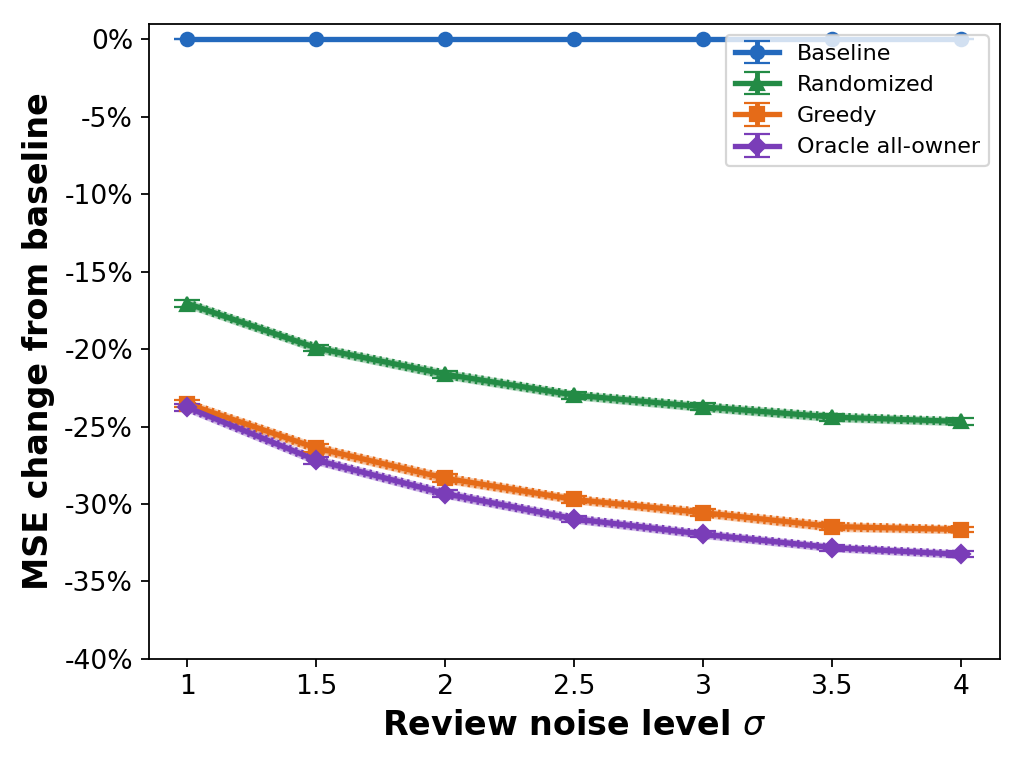}}
    \subfigure[MSE on  ICLR 2022 ]{\includegraphics[width=0.32\textwidth]{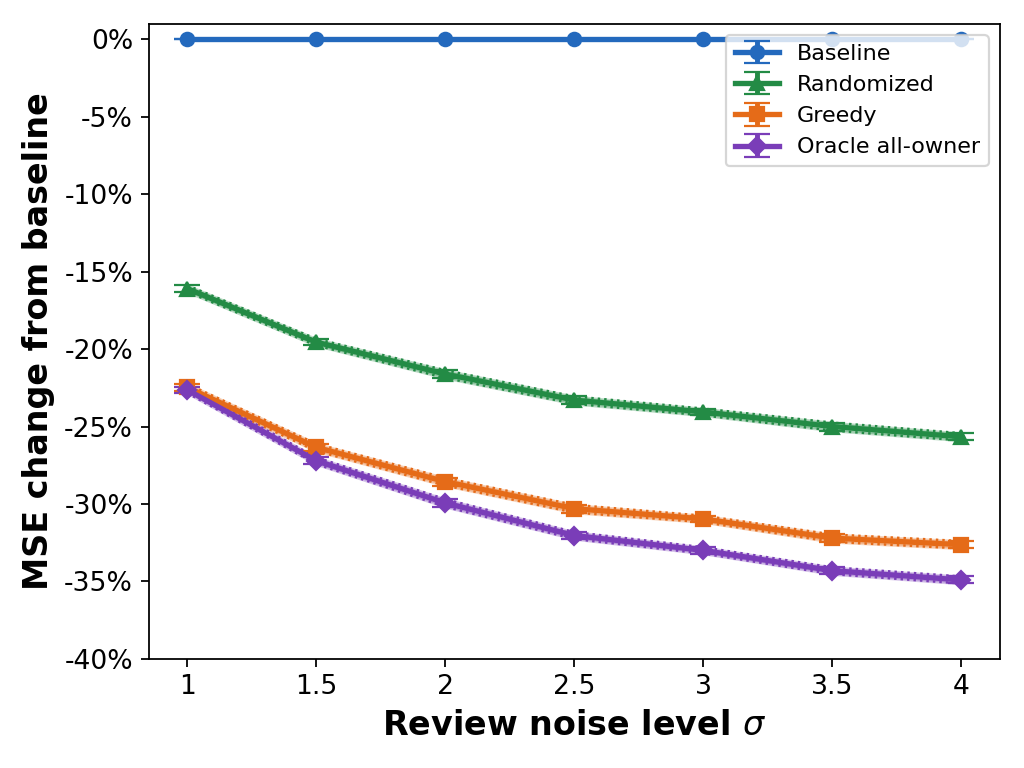}}
    \subfigure[MSE on  ICLR 2023 ]{\includegraphics[width=0.32\textwidth]{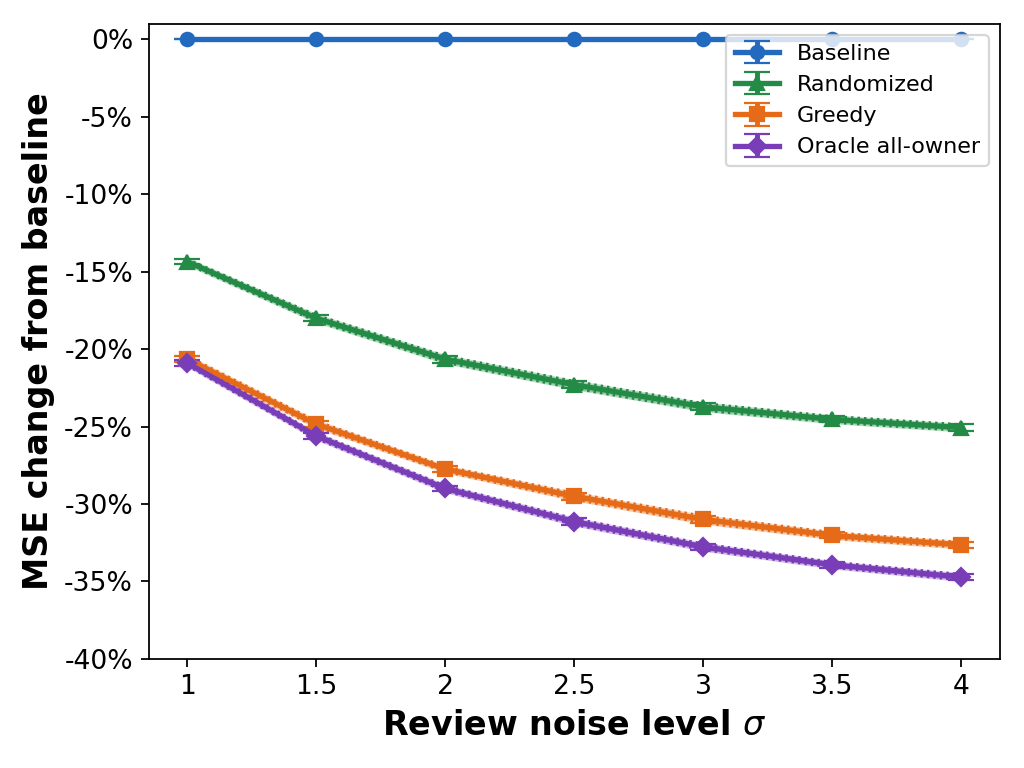}}
    \subfigure[Top-$30\%$, ICLR 2021]{\includegraphics[width=0.32\textwidth]{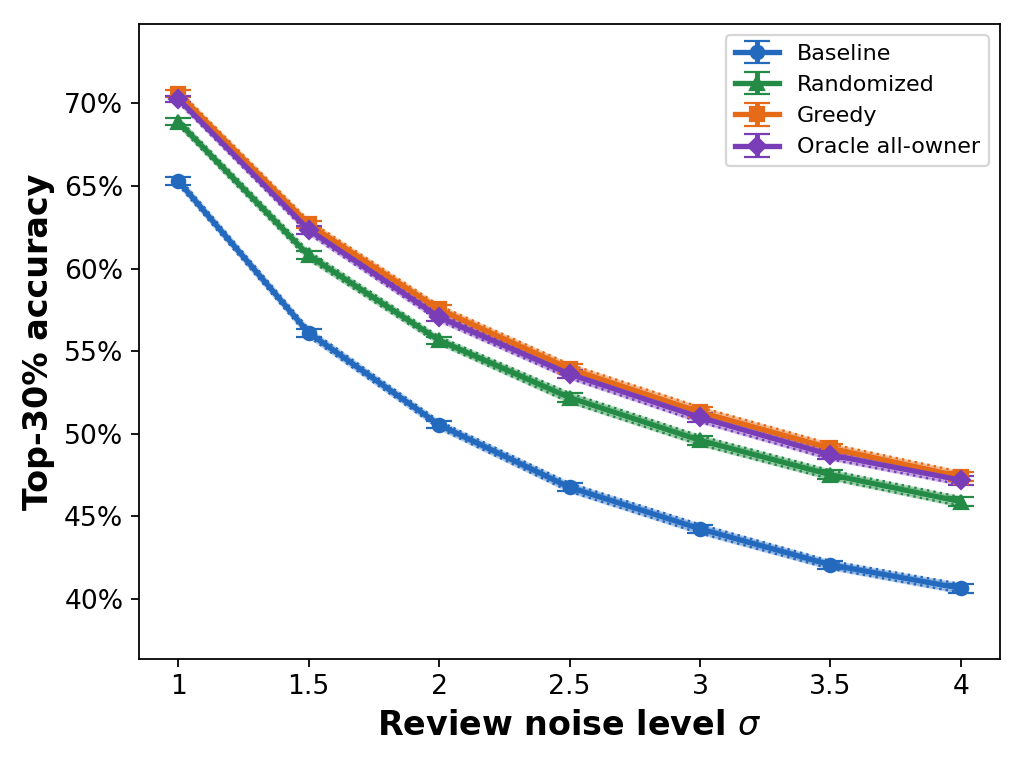}}
    \subfigure[Top-$30\%$, ICLR 2022]{\includegraphics[width=0.32\textwidth]{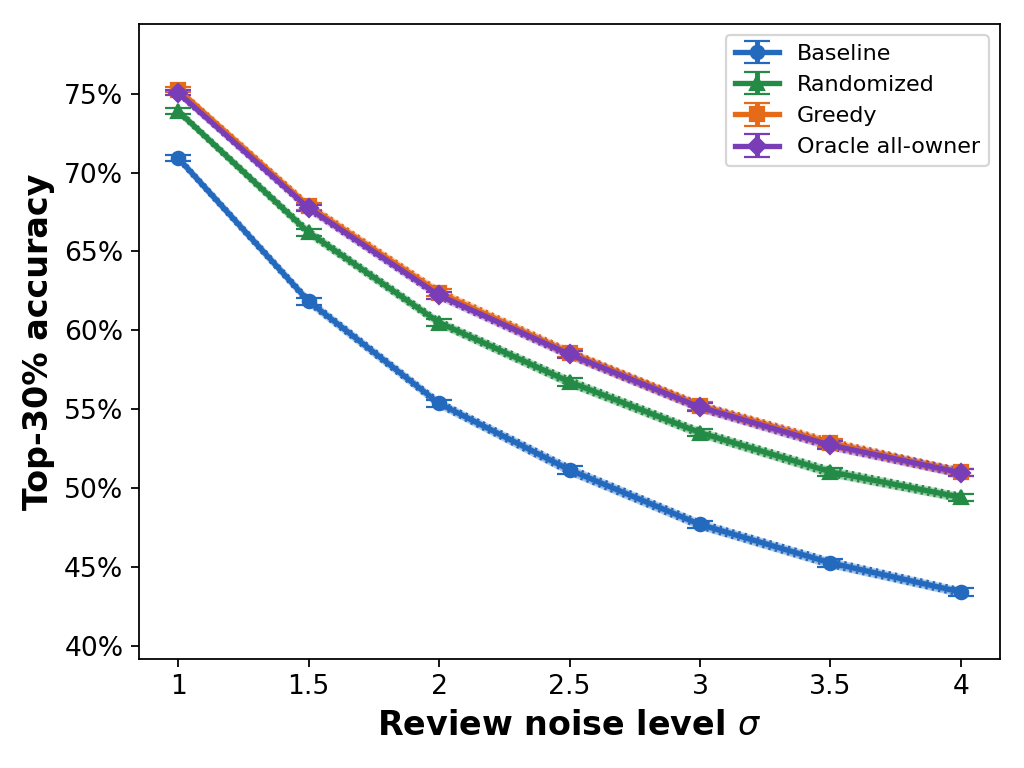}}
    \subfigure[Top-$30\%$, ICLR 2023]{\includegraphics[width=0.32\textwidth]{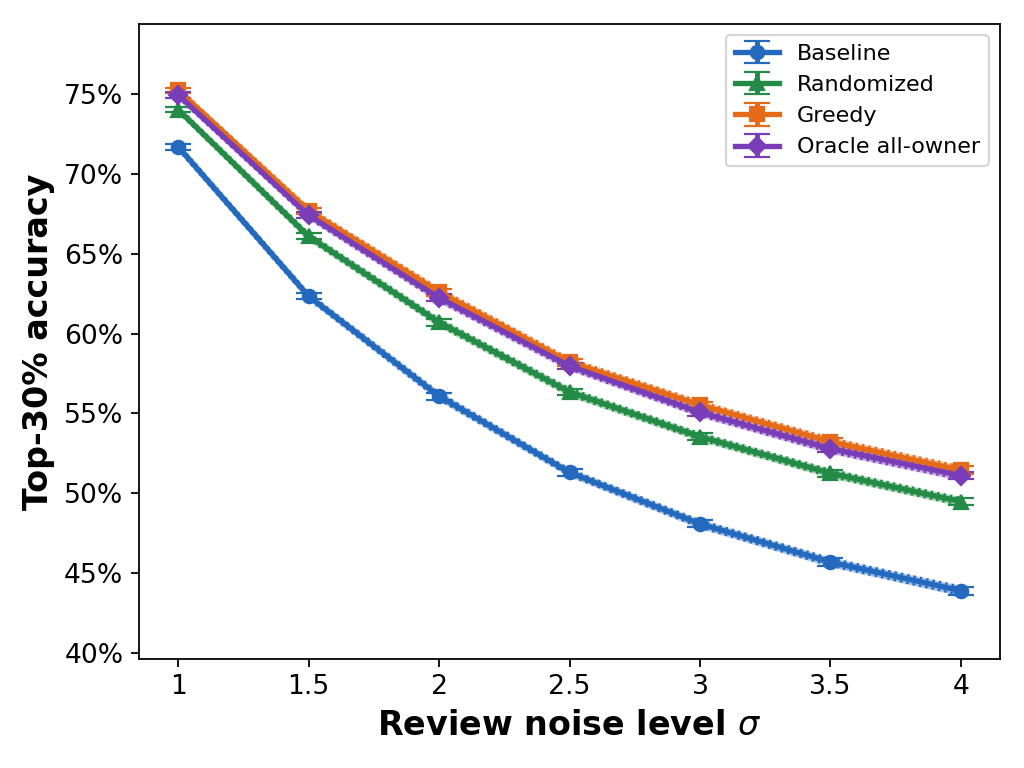}}
\caption{Performance across review-noise levels. \textbf{Upper}: MSE change relative to raw scores. \textbf{Lower}: Top-$30\%$ accuracy. Points show Monte Carlo means; shaded bands, dotted boundaries, and capped error bars show $95\%$ Monte Carlo confidence intervals.}
\label{fig:main}
\end{figure}

\vspace{2mm}
\noindent\textbf{Q1: How does the mechanism's calibration quality change under different levels of noise?} {Here we use the ICLR datasets, where we control the noise level in the simulation.
Figure~\ref{fig:main} shows how performance changes with noise. Across all three years and seven noise levels, the greedy method reduces MSE by $21$--$33\%$ relative to the raw scores. It also recovers $16$--$21\%$ of the surrogate top-$30\%$ papers missed by the raw-score selection. This pattern is stable across all three years.}
{The greedy partition also consistently outperforms its randomized counterpart and remains close to the oracle all-owner benchmark.}

\vspace{2mm}
\noindent\textbf{Q2: How does the Isotonic Mechanism balance informativeness and robustness under noisy author perceptions?} Our Isotonic Mechanism allows practitioners to choose the number of authors from whom to elicit ranking information (i.e., the $L$-strongness of the mechanism). Eliciting from fewer authors can permit larger blocks and improve informativeness. However, it leaves calibration more exposed to errors in authors' perceived rankings. Our second experiment examines this tradeoff. {Table~\ref{tab:iclr-stats} shows that imposing $L=2$ on the real ICLR graphs sharply reduces the number of calibrated papers and available comparisons. To study a wider range of $L$ without this data limitation, we use synthetic ownership relations with controlled overlap.}

\begin{figure}[tbh]
    \centering
    \includegraphics[width=0.28\textwidth]{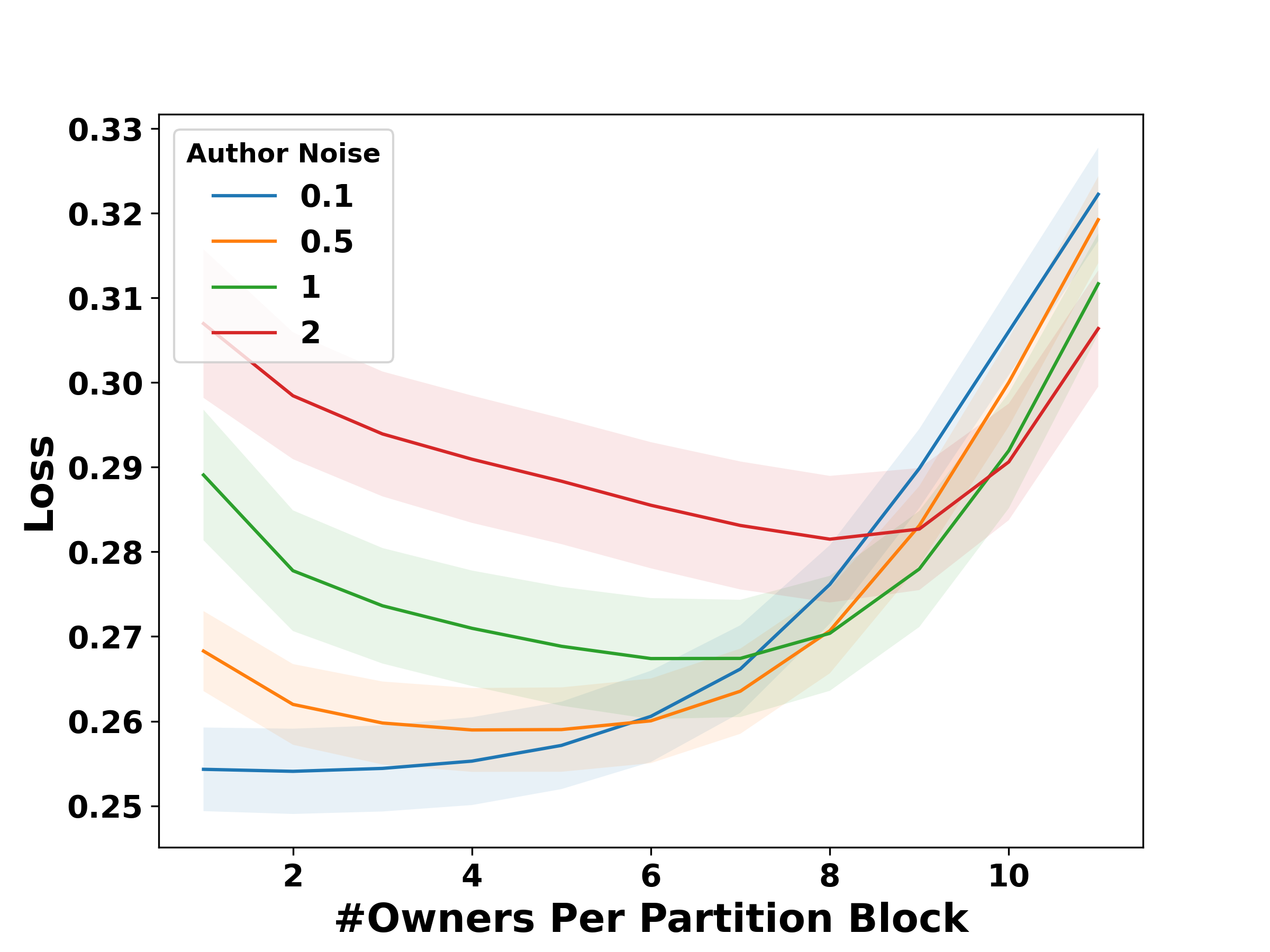}
    \includegraphics[width=0.28\textwidth]{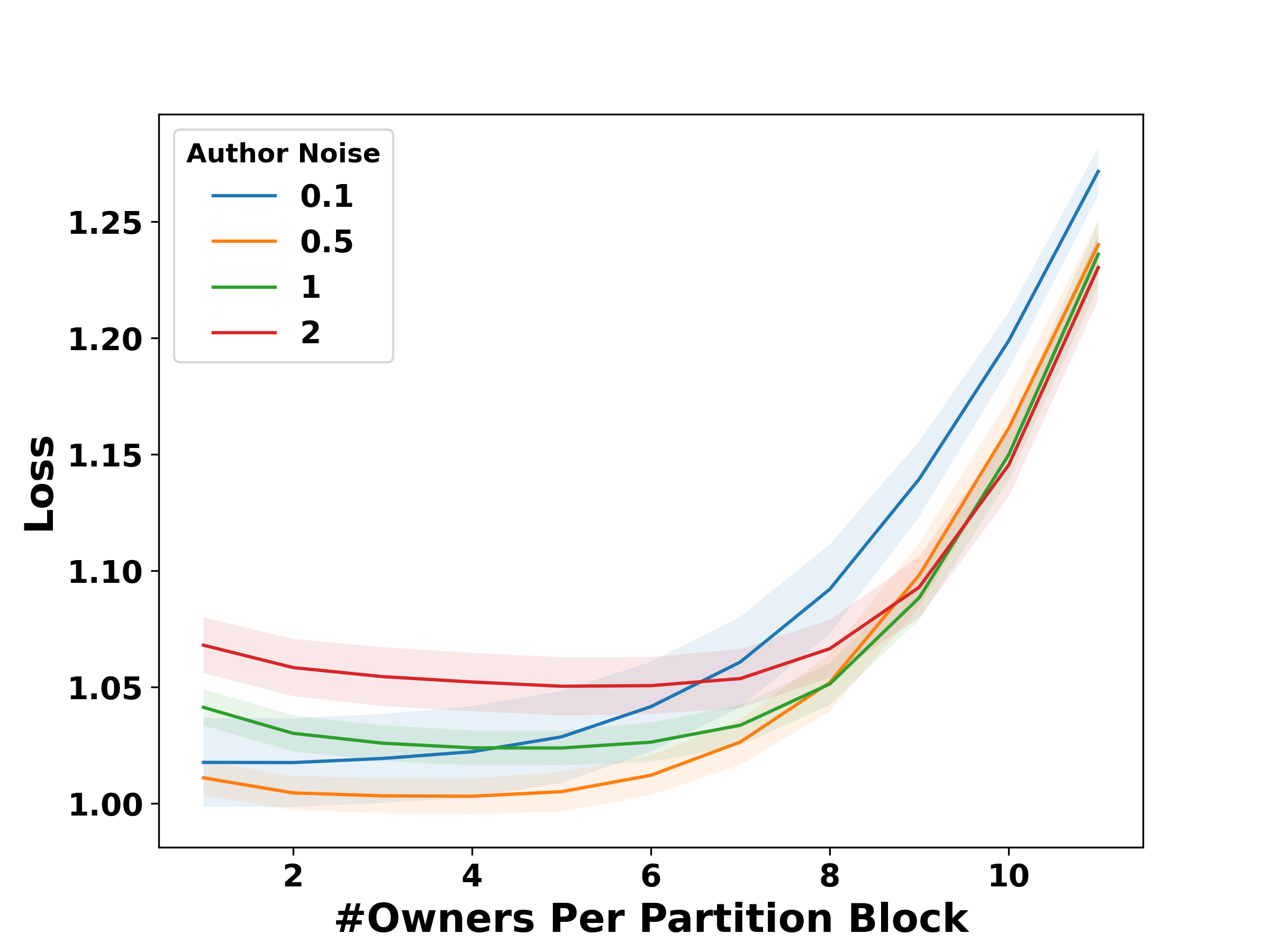}
    \includegraphics[width=0.28\textwidth]{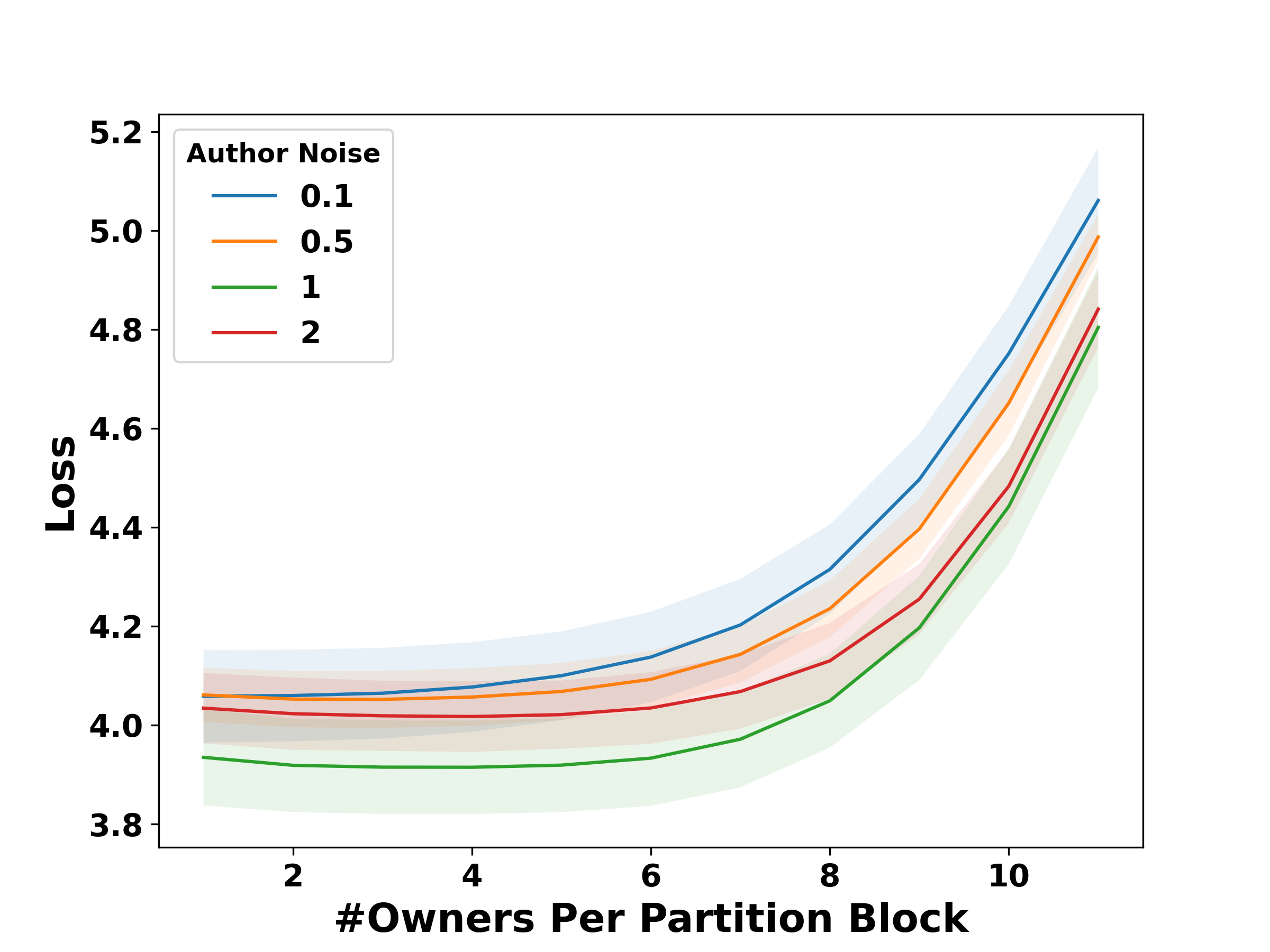}
    \caption{MSE of the Isotonic Mechanism on the $11$-level binary-tree instance as the number of authors per partition block and the level of author-perception noise vary. The review-noise levels are $\sigma=1$ (left), $2$ (middle), and $4$ (right).}
    \label{fig:tree}
\end{figure}

Specifically, in our simulation, author $j$ perceives paper $i$'s score as $R_i+\zeta_i^j$. We vary the standard deviation of the Gaussian perception noise $\zeta_i^j$ over $\{0.1,0.5,1,2\}$. {These values are smaller than the review-noise levels because we expect authors to perceive their own papers more accurately than reviewers do.}
Meanwhile, to vary the partition structure, we construct a class of instances characterized by an $11$-level binary tree with $2^{11}$ items and $2^{11}-1$ authors. At one extreme, the partition has a single block containing all items owned by the root author, so only one author can provide a ranking. At the other extreme, each leaf author's items form a block, and rankings can be elicited from that leaf author and all of her ancestors. Hence, there are $L$-strong partitions with $L$ ranging from 1 to 11.
Figure~\ref{fig:tree} shows a clear tradeoff. Under low perception noise (blue curve), larger blocks perform better because they provide more pairwise information. Under high perception noise (red curve), smaller {(though not necessarily the smallest)} blocks perform better because they have more authors per block (larger $L$). At intermediate noise levels (green and orange curves), the optimal block size balances block size and $L$-strongness. Thus, the choice of $L$ should be tailored to the perception-noise level in real-world applications.

\vspace{2mm}
\noindent\textbf{Q3: Evaluations on the real ICML 2023 data.} {The ICML experiment is a separate one-shot evaluation. For each paper, we use the least-confident review as the raw score and the mean of the remaining, higher-confidence reviews as the surrogate ground truth. We then apply the actual rankings reported by survey participants. Unlike the replicated ICLR simulation, this experiment cannot support a complete all-owner benchmark because rankings are observed only for participating authors. Participation was voluntary, so the sample may be selected. In addition, the reported rankings did not affect the final paper outcomes.}
The performance of the Isotonic Mechanism is presented in Table~\ref{tab:isotonic_and_distance}.
Both partition schemes reduce MSE relative to the baseline in this one-shot evaluation.

\begin{table}[tbh]
    \centering
    \begin{tabular}{ c c c c  }
    \toprule
     & Baseline &  Greedy &  Random   \\
    \midrule
    MSE & $1.4267$ & $1.3012$ & $1.3210$ \\
    F-statistic & - & $15.828$ & $9.911$   \\
    $p$-value & - & $7.032 \times 10^{-5}$ & $1.652 \times 10^{-3}$  \\
    \bottomrule
    \end{tabular}
    \caption{
    MSE under different partition schemes in the ICML dataset, together with the F-statistics and p-values for testing whether the loss decreases.
    }
    \label{tab:isotonic_and_distance}
\end{table}

%% file: discussion.tex
\section{Conclusion and Additional Discussion}
\label{sec:discussion}

In this work, we develop the Isotonic Mechanism to assist peer review at machine learning conferences with overlapping paper authorship.
We establish its truthfulness under our stated assumptions and design a polynomial-time algorithm for partitioning the ownership sets. The algorithm achieves a constant-factor approximation to the optimal partition, whose exact computation is hard. This facilitates the practical implementation of our mechanism.
Our mechanism also has limitations that suggest avenues for future work. We first discuss participation fairness. We then examine the applicability of the technical assumptions in our analysis and possible extensions and relaxations.

{
\vspace{2mm}
\noindent\textbf{Participation Fairness. }
A fixed partition may use rankings from some owners while leaving others unused. Call an owner \emph{eligible} if she owns at least two items, since only such an owner can provide a within-block comparison. We use \emph{owner reach} as a participation-based fairness measure: the fraction of eligible owners whose ranking is used for at least one multi-item block. For this comparison, we measure the \emph{informativeness} of a partition $\mathfrak{S}$ by the comparison-focused objective
\[
\obj(\mathfrak{S}):=\sum_{\cS_k\in\mathfrak{S}}|\cS_k|^2
=n+2\sum_{\cS_k\in\mathfrak{S}}\binom{|\cS_k|}{2};
\]
see Equation~\eqref{eq:obj-pairwise}.

To improve owner reach, the designer can randomize the global partition before eliciting rankings. At each step, the designer uniformly selects an owner with at least two unassigned items. That owner's unassigned items form the next block. This process continues until no owner has at least two unassigned items, after which all remaining items form singleton blocks. Each multi-item block is fully owned by the selected owner, so every realized partition is feasible. Moreover, each eligible owner has positive probability of being selected first and thus of having her ranking used. Truthfulness is preserved because every realized partition is truthful by Corollary~\ref{prop:truthfulness-partition}.

Table~\ref{tab:participation} compares this randomized construction with the deterministic greedy partition on the real ICLR 2021--2023 ownership graphs. Randomization uses rankings from more eligible owners but produces less informative partitions. Thus, improving owner reach comes at a cost. Optimizing this tradeoff remains an important direction for future work, and Example~\ref{ex:fairness-tradeoff} shows that the tension can be unbounded.

\begin{table}[tbh]
\centering
\small
\begin{tabular}{lcccc}
\toprule
& \multicolumn{2}{c}{Greedy} & \multicolumn{2}{c}{Randomized} \\
\cmidrule(lr){2-3}\cmidrule(lr){4-5}
Dataset & Providers (reach) & Informativeness & Providers (reach) & Informativeness \\
\midrule
ICLR 2021 & 743 ($40\%$) & $100\%$ & 1,130 ($61\%$) & $64\%$ \\
ICLR 2022 & 802 ($36\%$) & $100\%$ & 1,253 ($57\%$) & $64\%$ \\
ICLR 2023 & 966 ($36\%$) & $100\%$ & 1,528 ($56\%$) & $64\%$ \\
\bottomrule
\end{tabular}
\caption{Ranking providers, owner reach, and informativeness on the ICLR 2021--2023 ownership graphs. Parentheses report owner reach. Randomized results are averages over the same sampled partitions used in Section~\ref{sec:exp}, and greedy informativeness is normalized to $100\%$.}
\label{tab:participation}
\end{table}
\FloatBarrier

\begin{example}[An Extreme Informativeness--Owner-Reach Tradeoff]
\label{ex:fairness-tradeoff}
Fix an even $n=2K$ and consider $K+1$ owners over $n$ items. One senior owner owns all $n$ items, and for each $k\in[K]$ a junior owner $k$ owns exactly the pair $\{2k-1,2k\}$. All $K+1$ owners are eligible under the definition above.

The partition into the single block $[n]$ is the unique informativeness maximizer, with $\obj(\{[n]\})=n^2$, but its owner reach is only $1/(K+1)$. Conversely, the partition into the $K$ pairs $\{2k-1,2k\}$ uniquely maximizes owner reach at $1$, because using junior owner $k$ requires $\{2k-1,2k\}$ to be a block. Its informativeness is only $4K=2n$. Thus, maximizing owner reach reduces informativeness by a factor of $n/2$. Equivalently, the number of within-block pairwise comparisons falls from $\binom{n}{2}$ to $K=n/2$, a factor of
\[
\frac{\binom{n}{2}}{n/2}=n-1,
\]
while the informativeness optimum uses only one of the $K+1$ owners. This example demonstrates the intrinsic tradeoff between informativeness and owner reach.
\end{example}

We next discuss the applicability of our technical assumptions and possible relaxations.
}

\vspace{2mm}
\noindent\textbf{Owner's Private Information. }
{Assumption~\ref{assum:informed-owners} is an idealized benchmark. It requires each owner to know the relative ranking of her own items under the ground-truth scores. This requirement is stronger than assuming that she knows only the scores that she herself would assign. The assumption isolates a regime in which the mechanism is exactly truthful, and we consider a belief-based single-owner relaxation below.} Here we assume owners have accurate knowledge of the information we elicit, i.e., {relative ranking of their own items}. This assumption does not require owners to know the precise ground-truth scores of their items; agreement on the coarser ranking information suffices. We remark on the intrinsic tradeoff when eliciting information from owners. On the one hand, eliciting very fine-grained score information can be risky because owners may not have accurate information. On the other hand, eliciting overly coarse information may no longer be useful. As a compromise, we believe that eliciting ranking information is a reasonable ``middle ground'' to begin with. Exploring other forms of information is an interesting direction for future work.

Additionally, Section~\ref{sec:exp} evaluates the algorithm when authors have noisy perceptions of the true paper rankings. The Isotonic Mechanism still improves score estimation when owners supply additional, though imperfect, information not included in the review scores.
{A natural question is whether reporting an owner's private perceived ranking is optimal when it need not agree with the ground-truth ranking. We give an exact result for a single owner under a condition on her predictive beliefs. The condition requires the errors in future review scores, measured relative to her perceived scores, to be exchangeable conditional on her private information.

\begin{proposition}[Optimality of the Perceived Ranking]
\label{prop:ic-noisy}
Consider a single owner with private information $\mathcal{F}$ and an $\mathcal{F}$-measurable perceived-score vector $\bP\in\RR^n$. Let
\[
\boldsymbol{\eta}:=\by-\bP
\]
be the difference between the future review scores and her perceived scores. The owner reports before observing $\by$, and all relevant expected utilities are finite. Suppose that, conditional on $\mathcal{F}$, the distribution of $\boldsymbol{\eta}$ is exchangeable; that is, for every permutation $\rho$,
\[
\boldsymbol{\eta}\mid\mathcal{F}
\ \stackrel{d}{=}\
\rho\circ\boldsymbol{\eta}\mid\mathcal{F}
\quad\text{almost surely}.
\]
Under Assumption~\ref{assum:convex-utility}, any $\mathcal{F}$-measurable ranking consistent with $\bP$ maximizes the owner's conditional expected utility among all $\mathcal{F}$-measurable ranking reports.
\end{proposition}

The proof is in Appendix~\ref{append:ic-noisy}. The condition holds, for example, if $\by=\bP+\boldsymbol{\eta}$ and $\boldsymbol{\eta}$ is exchangeable and independent of $\mathcal{F}$. It is not implied merely by writing $\bP=\bR+\bvarepsilon$ with exchangeable perception noise independent of the review noise: conditioning on $\bP$ may destroy exchangeability of the residual $\by-\bP$. We therefore separate the incentive condition in Proposition~\ref{prop:ic-noisy} from the additive perception-noise model used for the statistical result in Proposition~\ref{prop:more-owners}.

That said, Proposition~\ref{prop:ic-noisy} addresses only a single owner's incentive. When multiple co-owners hold conflicting perceptions, the mechanism cannot guarantee an equilibrium in which every owner reports her perceived ranking. As demonstrated in Remark~\ref{rm:ne-noisy}, if one owner reports a conflicting perceived ranking, it can become suboptimal for another owner to report her own perceived ranking.
The result can also fail when an owner has ex-ante information about review noise that violates conditional exchangeability. Therefore, ensuring incentive compatibility when authors hold such information or have conflicting perceptions remains a fundamental design challenge.
Nevertheless, we believe our baseline model and its truthfulness result remain valuable. The mechanism may encourage coauthors to discuss and refine their perceptions before reporting a ranking. We also conjecture that truthful behavior may be the most robust strategy when authors are uncertain about others' reporting behavior, and we leave a formal analysis to future work.

\begin{remark}[On the Strategyproofness of the Isotonic Mechanism]
\label{rm:ne-noisy}
Consider an instance  with $2$ owners and $3$ items. The items have ground-truth scores
$
R_1 = 7, R_2 = 4, R_3 = 3,
$
i.e., one high-quality item and two lower-quality items. The ownership is completely overlapping, and we suppose both owners have equal influence ($\alpha_1 = \alpha_2 = \tfrac12$).
Assume the noise vector is exchangeable, taking values $z=(2,2,4)$ with
$
P(z=(2,2,4)) = P(z=(2,4,2)) = P(z=(4,2,2)) = \tfrac13.
$
Suppose Owner~1's perceived ranking is $\pi_1=(3,1,2)$, which differs from the ground-truth ranking, and she reports it. Suppose Owner~2's perceived ranking agrees with the ground-truth ranking, $\pi_{\mathrm{P},2}=(1,2,3)$.
Let Owner~2’s utility be a convex, non-decreasing function
$
\mathrm{U}(\hat{\bR})  =  \sum_{i=1}^{3}\max\{\hat{R}_i - 6.25,\,0\}.
$
Let $V(\pi_2)$ denote Owner~2's expected utility when Owner~1's report is fixed at $\pi_1=(3,1,2)$ and Owner~2 reports $\pi_2$. Evaluating all six possible reports, $V(\pi_2)$ is uniquely maximized at $\pi_2=(1,3,2)$ rather than at her perceived ranking $\pi_{\mathrm{P},2}=(1,2,3)$. In particular,
$$
V((1,3,2))=\frac{7}{2}
\quad>\quad
\frac{121}{36}=V((1,2,3)).
$$
Thus, when the owners' perceived rankings conflict, it can be suboptimal for one owner to report her own perceived ranking. The example shows that such reporting is not a dominant strategy; it does not establish a failure of group strategyproofness.
\end{remark}
}

\vspace{2mm}
\noindent\textbf{Noise Structures in Review Scores. }
The exchangeable noise structure in Assumption~\ref{assum:exchangeable-noise} is also a natural choice and has been widely adopted in mechanism design literature.
The rationale behind this assumption is as follows. An owner determines the ranking of her papers \emph{ex ante}, before the review outcomes are realized, and does not know how the papers will be reviewed. Symmetry of the review noise across her papers is therefore a natural benchmark, as captured by our exchangeability assumption. A similar justification for exchangeable noise appears in \citet{maskin2023borda}, though for a different mechanism-design problem. Assumption~\ref{assum:exchangeable-noise} strictly relaxes the i.i.d. noise assumption commonly used in the peer review literature; see, e.g., \citet{baba2013statistical, mackay2017calibration,tan2021least}.
Notice that the noise satisfying Assumption~\ref{assum:exchangeable-noise} does not even need to have a zero mean. In addition, if the review score of each paper is averaged over several reviews and the number of reviewers assigned to each paper is treated as an i.i.d. random variable, then the exchangeable noises allow the papers to have different variances.

Finally, an advantage of the exchangeable noise assumption is that it is non-parametric and does not require modeling a distributional family. If one is willing to assume a specific form for the noise distributions, then exchangeability may be further relaxed (e.g., see the recent work of \citet{yan2023isotonic}, which uses an exponential family to model review score distributions). However, without such distributional structure, we conjecture that it will be difficult to relax exchangeability in our setup.

\vspace{2mm}
\noindent\textbf{Owners' Utility Structures in Peer Review. }
{The convex utility in Assumption~\ref{assum:convex-utility} may be less familiar than the two assumptions above and is therefore worth discussing here. When review scores map to capped outcomes such as acceptance, spotlight, or oral decisions, utility is naturally convex over the score region below the relevant decision thresholds and tends to flatten or become concave near the top end. We formalize the convex part of this shape in Proposition~\ref{prop:assump3-threshold} in Appendix~\ref{append:assump3-threshold} and illustrate it in Example~\ref{ex:uniform-AC-noise} via a stylized model of bounded AC noise.
The concave-near-the-top behavior is acknowledged as a limitation and positioned as future work.
Accordingly, our goal is not to claim that realistic owner utilities are globally convex, but to explain why convexity is a reasonable modeling primitive on the score region most relevant for calibration.
We support this view both as a theoretical consequence of a stylized AC-noise model and as an empirical pattern in real conference data. Beyond this AC-noise microfoundation, the convex utility class also has standard economic motivations, which we briefly review as follows.}
To begin, it is important to distinguish two traits that affect utilities: \emph{quality} and \emph{quantity}. While utility theory often assumes concavity in the \emph{quantity} of products because of diminishing returns, convexity in \emph{quality} is also common, including in the seminal work of \citet{mussa1978monopoly} on pricing quality-differentiated goods and its numerous follow-up works.\footnote{More concretely, for tractability, these papers often normalize utility to be linear (i.e., weakly convex) in quality through reparameterization, while shifting the convexity into cost functions.}
In our setting, utility convexity in ratings reflects that items receiving higher ratings tend to generate significantly greater rewards. We believe such convex utility is particularly suitable for evaluating research (e.g., conference review), given the ``high-risk-high-reward'' nature of research recognized by various funding agencies~\citep{wagner2013evaluating, cao2022developing, franzoni2023encouraging}. A related setting arises in tournament design~\citep{lazear1981rank, rosen1985prizes}, where participants are incentivized to exert more effort or take more risk because potential rewards are significantly larger at the top. Participants' utilities are commonly assumed to be convex in their ratings, akin to paper scores in our setting.
In fact, this assumption of convexity in scarcity is naturally consistent with the classic assumption of concavity in quantity. This is because, if a product is hardly accessible (e.g., papers with high scores), or an award is conferred to only a few, we tend to derive more utility --- a phenomenon famously known as the ``Veblen effects'' in behavioral economics and social psychology~\citep{veblen2017theory, verhallen1982scarcity}.
In line with the ``Veblen effects'', most publication venues have indeed been implementing various metrics to promote (more scarce) high-quality research papers. For instance, with the massive number of accepted papers at AI/ML venues today, these conferences start to distinguish  accepted papers   with labels such as ``poster'', ``spotlight'' and ``oral''; oral papers are  promoted with extra resources such as longer presentations and more official media highlights, which naturally lead to much higher utility than an accepted poster paper.

{Appendix~\ref{append:assump3-threshold} provides the formal microfoundation and additional empirical discussion, including Figure~\ref{fig:convex-correlation} and Remark~\ref{rm:convex-rating}.}
Meanwhile, an author's value of a paper could depend on factors such as authorship order, the number of coauthors, and her relative contribution to the paper~\citep{demaine2023every}. This modeling perspective could more accurately reflect the fact that authors may assign different values to their coauthored submissions. Accounting for this heterogeneity in utility functions is an important challenge for future work.
{Another direction is to understand potential group manipulations or collusion under our mechanism. This is generally a challenging problem in mechanism design because it requires incentive analysis beyond unilateral deviations from a Nash equilibrium. Remark~\ref{rm:ne-noisy} shows only that reporting one's perceived ranking is not a dominant strategy under the current setup. For future work, one could investigate how defense techniques used in paper assignment~\citep[Section~4.2]{shah2022challenges} can be adopted to detect and mitigate group manipulations by paper authors. We also note that the ownership relation itself is a verified input in our applications rather than a self-reported quantity. Authorship is recorded at submission and known to the organizer, so owners cannot manipulate the partition by misreporting which items they own. Extending the analysis to settings where ownership is strategically reported is a further direction for future work.
}

Lastly, it is important to realize that theory might only take us so far in this field, due to the complex nature of human subjects. We view our work as a theoretical foundation for the initiatives aimed at eliciting self-evaluations and managing incentives in scientific reviews. While empirical results from the \emph{OpenRank} survey at ICML 2023 are promising \citep{su2024analysis}, further effort is required to implement this method in practice, where various challenges may arise. As an initial step, the Isotonic Mechanism can function as an independent layer atop existing peer review processes, with its calibrated scores serving as reference points to assist expert human decision-making (e.g., helping area chairs to detect anomalies and request additional reviewers).
We look forward to more real-world experiments to better understand the potential limitations of the Isotonic Mechanism in practice and drive iterative improvements towards solving the owner-assisted calibration problem.

%% file: proofs.tex
{
\section{Proof of Proposition~\ref{prop:ic-noisy}}\label{append:ic-noisy}

\begin{proof}[Proof of Proposition~\ref{prop:ic-noisy}]
Fix an $\mathcal{F}$-measurable ranking $\pi_{\bP}$ consistent with $\bP$. Let $K_{\boldsymbol{\eta}}(\omega,\cdot)$ be a regular conditional distribution of $\boldsymbol{\eta}$ given $\mathcal{F}$. For each fixed ranking $q$, define
\[
V_q(\omega)
:=\int_{\RR^n}\mathrm{U}\!\left(\hat{\mathrm{R}}\!\left(q;\bP(\omega)+\bx\right)\right)
K_{\boldsymbol{\eta}}(\omega,d\bx).
\]
This is a version of $\Ex[\mathrm{U}(\hat{\mathrm{R}}(q;\by))\mid\mathcal{F}]$. Fix $\omega$ outside the null set on which exchangeability may fail, and set $\bp=\bP(\omega)$. Under $K_{\boldsymbol{\eta}}(\omega,\cdot)$, the review-score vector is $\bp+\boldsymbol{\eta}$ with exchangeable $\boldsymbol{\eta}$. Theorem~\ref{thm:restate}, applied with $\bp$ in place of the ground-truth score vector, gives
\[
V_{\pi_{\bP}(\omega)}(\omega)\geq V_q(\omega)
\qquad\text{for every ranking }q.
\]
If $\pi$ is any $\mathcal{F}$-measurable ranking report, the finite report space gives
\[
\Ex\!\left[\mathrm{U}\!\left(\hat{\mathrm{R}}(\pi;\by)\right)\middle|\mathcal{F}\right](\omega)
=\sum_q \mathbf{1}\{\pi(\omega)=q\}V_q(\omega)
\leq V_{\pi_{\bP}(\omega)}(\omega)
\]
for almost every $\omega$. This proves the desired conditional optimality. A randomized deviation is a mixture of these reports and cannot improve conditional expected utility. Taking expectations gives the corresponding ex-ante inequality.
\end{proof}
}

{
\section{A Microfoundation for Assumption~\ref{assum:convex-utility}}
\label{append:assump3-threshold}

Consider a stylized two-stage decision pipeline. Raw review scores $\by$ are first calibrated into adjusted scores $\hat{\bR}$ by the Isotonic Mechanism, and the area chair (AC) then makes the final acceptance decision based on $\hat{\bR}$. If the AC's residual judgment is modeled as an additive post-calibration noise term $Z_i$ on the adjusted score, with the realized score capped to the rating range $[\underline{r},\bar{r}]$ for realism, then even a hard-threshold acceptance rule induces a convex ex ante utility in $\hat{R}_i$.

\begin{proposition}[A microfoundation for Assumption~\ref{assum:convex-utility} via clipped threshold outcomes]
\label{prop:assump3-threshold}
Fix an owner $j$ and consider any item $i\in \cI^j$. Suppose the adjusted score $\hat{R}_i$ lies in a bounded interval $[\underline{r},\bar{r}]$, and the realized score after AC clipping is
\[
\tilde{R}_i := \mathrm{clip}\bigl(\hat{R}_i+Z_i,\,\underline{r},\,\bar{r}\bigr) = \min\!\bigl(\max(\hat{R}_i+Z_i,\,\underline{r}),\,\bar{r}\bigr),
\]
where $\{Z_i\}_{i\in\cI^j}$ are i.i.d.\ with CDF $F$ and independent of the review noise. Let the owner's realized per-item payoff be $\mathbf{1}\{\tilde{R}_i \ge \tau\}$ for some decision threshold $\tau\in(\underline{r},\bar{r}]$. If $F$ is concave on $[\tau-\bar{r},\,\tau-\underline{r}]$, then the induced ex ante per-item utility
\[
u(x):=\Pr\bigl[\tilde{R}_i\ge\tau\,\big|\,\hat{R}_i=x\bigr]=1-F(\tau-x)
\]
is non-decreasing and convex on $[\underline{r},\bar{r}]$. Consequently, the owner's expected total payoff, viewed as a function of the adjusted-score vector, is non-decreasing and convex.
The same result extends when the realized per-item payoff is a non-decreasing step function $\sum_{k=1}^{K}a_k\,\mathbf{1}\{\tilde{R}_i\ge\tau_k\}$ with non-negative weights $a_k\ge 0$ and thresholds $\tau_1,\ldots,\tau_K\in(\underline{r},\bar{r}]$: the induced ex ante utility $u_{\mathrm{step}}(x):=\E\bigl[\sum_{k=1}^{K}a_k\,\mathbf{1}\{\tilde{R}_i\ge\tau_k\}\,\big|\,\hat{R}_i=x\bigr]$ is non-decreasing and convex on $[\underline{r},\bar{r}]$, provided $F$ is concave on each interval $[\tau_k-\bar{r},\,\tau_k-\underline{r}]$.
\end{proposition}

\begin{proof}[Proof of Proposition~\ref{prop:assump3-threshold}]
For any $\tau\in(\underline{r},\bar{r}]$, the clipped indicator satisfies $\mathbf{1}\{\tilde{R}_i\ge\tau\}=\mathbf{1}\{\hat{R}_i+Z_i\ge\tau\}$. Indeed, if $\hat{R}_i+Z_i<\underline{r}$ then $\tilde{R}_i=\underline{r}<\tau$ so both indicators equal $0$; if $\hat{R}_i+Z_i\in[\underline{r},\bar{r}]$ then $\tilde{R}_i=\hat{R}_i+Z_i$ and both indicators agree; and if $\hat{R}_i+Z_i>\bar{r}$ then $\tilde{R}_i=\bar{r}\ge\tau$ so both indicators equal $1$. Hence $u(x)=\Pr[x+Z_i\ge\tau]=1-F(\tau-x)$.

For any $x_1\le x_2$, we have $\tau-x_1\ge\tau-x_2$. Since $F$ is non-decreasing,
\[
u(x_1)=1-F(\tau-x_1)\le 1-F(\tau-x_2)=u(x_2),
\]
so $u$ is non-decreasing. For convexity, take any $x_1,x_2\in[\underline{r},\bar{r}]$ and $\lambda\in[0,1]$, and let $t_k=\tau-x_k$ for $k\in\{1,2\}$. By concavity of $F$ on $[\tau-\bar{r},\,\tau-\underline{r}]$,
$
F\bigl(\lambda t_1+(1-\lambda)t_2\bigr) \ge \lambda F(t_1)+(1-\lambda)F(t_2).
$
Multiplying by $-1$ and adding $1$ yields
\[
u\bigl(\lambda x_1+(1-\lambda)x_2\bigr) \le \lambda u(x_1)+(1-\lambda)u(x_2),
\]
so $u$ is convex on $[\underline{r},\bar{r}]$. The separable form follows from $\Pr[\tilde{R}_i\ge\tau\mid\hat{R}_i]=u(\hat{R}_i)$ together with linearity of expectation across items. Finally, the extension to $u_{\mathrm{step}}$ follows because a non-negative linear combination of non-decreasing convex functions is non-decreasing and convex.
\end{proof}

Under Proposition~\ref{prop:assump3-threshold}, the owner's separable total payoff $\sum_{i\in\cI^j}u(\hat{R}_i)$ is non-decreasing and convex on $[\underline{r},\bar{r}]^{n_j}$, hence satisfies Assumption~\ref{assum:convex-utility}, and the truthful Nash equilibrium result of Theorem~\ref{thm:ne-same-item} applies to this clipped threshold model.

Proposition~\ref{prop:assump3-threshold} convexifies threshold-style payoffs, which are not themselves convex, under a concavity condition on the AC-noise CDF. A complementary and unconditional observation is that if the base payoff is {already} non-decreasing and convex, then AC-noise smoothing preserves both properties without any restriction on the noise distribution.

\begin{corollary}[AC noise preserves non-decreasing convex utilities]
\label{cor:ac-noise-preserve}
Fix an owner $j$ and item $i\in\cI^j$ with adjusted score $\hat{R}_i\in[\underline{r},\bar{r}]$. Suppose the realized per-item payoff is $g(\hat{R}_i+Z_i)$, where $g:\RR\to\RR$ is non-decreasing and convex, $\E|g(x+Z_i)|<\infty$ for all $x\in[\underline{r},\bar{r}]$, and the AC noise terms $\{Z_i\}_{i\in\cI^j}$ are i.i.d.\ and independent of the review noise. Then, for any noise distribution, the induced ex ante per-item utility
\[
u(x):=\E\bigl[g(x+Z_i)\bigr]
\]
is non-decreasing and convex on $[\underline{r},\bar{r}]$, and the owner's expected total payoff is convex.
\end{corollary}

\begin{proof}[Proof of Corollary~\ref{cor:ac-noise-preserve}]
For any $x_1\le x_2$, monotonicity of $g$ gives $g(x_1+Z_i)\le g(x_2+Z_i)$ pointwise, so taking expectations yields $u(x_1)\le u(x_2)$. For any $\lambda\in[0,1]$, convexity of $g$ gives
\[
g\bigl(\lambda x_1+(1-\lambda)x_2+Z_i\bigr)\le \lambda\,g(x_1+Z_i)+(1-\lambda)\,g(x_2+Z_i),
\]
and taking expectations yields $u(\lambda x_1+(1-\lambda)x_2)\le \lambda u(x_1)+(1-\lambda)u(x_2)$. Summing over $i\in\cI^j$ preserves convexity of the total payoff.
\end{proof}

\paragraph{When does the concavity hypothesis hold?}
The condition that $F$ is concave on $[\tau-\bar{r},\,\tau-\underline{r}]$ is equivalent to the AC-noise density being non-increasing on that interval. We allow heterogeneity across decision tiers or owners, with thresholds spread over a range $[\tau_{\min},\tau_{\max}]$. A sufficient condition is then that $F$ is concave on $[\tau_{\min}-\bar{r},\,\tau_{\max}-\underline{r}]$, since this set contains every per-tier interval $[\tau_k-\bar{r},\,\tau_k-\underline{r}]$. We illustrate the condition with a stylized bounded uniform AC-noise model on the ICLR/ICML scale.

\begin{example}[A stylized model with bounded uniform AC noise]
\label{ex:uniform-AC-noise}
Take adjusted scores in the full ICLR/ICML range $[\underline{r},\bar{r}]=[0,10]$, with item-level decision thresholds in $[\tau_{\min},\tau_{\max}]=[5,7]$ spanning acceptance to oral selection. As a stylized model of bounded AC discretion, take
\[
Z_i \overset{\mathrm{i.i.d.}}{\sim} \mathrm{Uniform}[-c,c],
\qquad c  \ge  \max\!\bigl(\bar{r}-\tau_{\min},\,\tau_{\max}-\underline{r}\bigr)  =  7,
\]
where $c$ bounds the magnitude of the AC's override of the adjusted score, and uniformity within that bound reflects maximum entropy in the absence of further structure.

With $c\ge 7$, $[\tau-\bar{r},\,\tau-\underline{r}]\subseteq[-5,7]\subseteq[-c,c]$ for every $\tau\in[5,7]$. The CDF $F(z)=(z+c)/(2c)$ is linear and hence weakly concave on $[-c,c]$, so the hypothesis of Proposition~\ref{prop:assump3-threshold} is met for every threshold simultaneously. The induced ex ante per-item utility takes the closed form
\[
u(x) = \Pr[x+Z_i\ge\tau] = \frac{x-\tau+c}{2c}, \qquad x\in[0,10],
\]
which is non-decreasing and weakly convex on the full score range. Aggregating across decision tiers $\tau_k\in[\tau_{\min},\tau_{\max}]$ with non-negative weights $a_k$ yields the step utility $u_{\mathrm{step}}(x)=\sum_k a_k(x-\tau_k+c)/(2c)$, also non-decreasing and convex on $[0,10]$. The bound $c\ge 7$ is a convenient sufficient choice that keeps $u$ exactly linear on the whole range $[0,10]$, and it is not necessary for convexity. For smaller noise bounds $u$ is flat below a cutoff and linear above, which remains non-decreasing and convex on the calibration-relevant region.
\end{example}

The hypothesis of Proposition~\ref{prop:assump3-threshold} is also satisfied by standard noise distributions. Most relevantly, any zero-mean symmetric unimodal noise including Gaussian, Laplace, logistic, and Student-$t$ has CDF concave on $[0,\infty)$, so applying Proposition~\ref{prop:assump3-threshold} with $[\underline{r},\bar{r}]\subseteq[0,\tau_{\min}]$ delivers strict convexity of $u$ on the below-threshold sub-interval. A left-shifted exponential $Z_i=-\mu+W_i$ with $W_i\sim\mathrm{Exp}(\lambda)$ and $\mu\ge\bar{r}-\tau_{\min}$ also satisfies the hypothesis on $[-\mu,\infty)$.

\paragraph{Additional Discussion and Empirical Evidence.}
A natural question arises as to whether it would be more appropriate to directly model utilities based on the final review outcomes (e.g., ``oral'', ``spotlight'', ``poster''). This is a natural alternative. Our choice to model utility as a function of adjusted review scores is best viewed as a reduced-form approximation, motivated by the fact that the mechanism directly changes scores and that outcome-based rewards induce ex ante score-based utilities once the outcome probabilities conditional on scores are specified. This interpretation does not imply that the induced utility is always globally convex. Rather, it provides a bridge between outcome-based rewards and the score-based utility model studied in our theorem.

\begin{figure}[tbh]
    \centering
    \includegraphics[width=0.32\textwidth]{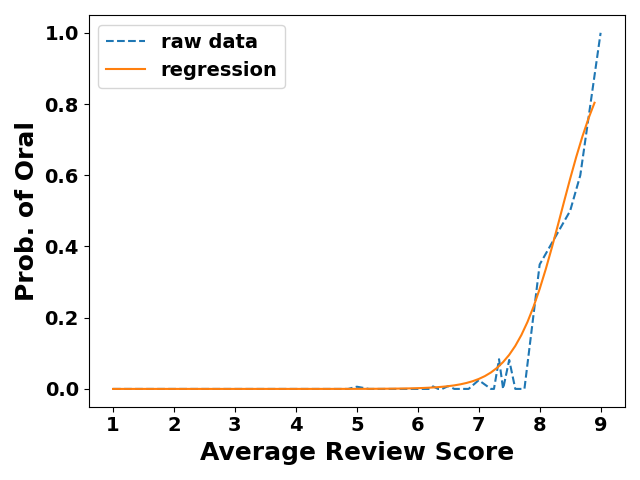}
    \includegraphics[width=0.32\textwidth]{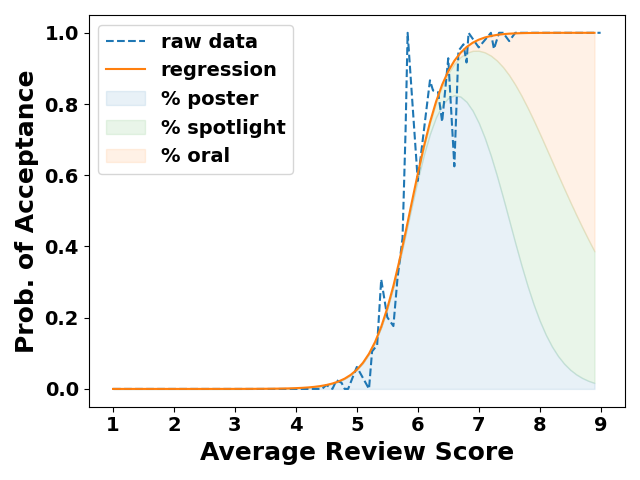}
    \includegraphics[width=0.32\textwidth]{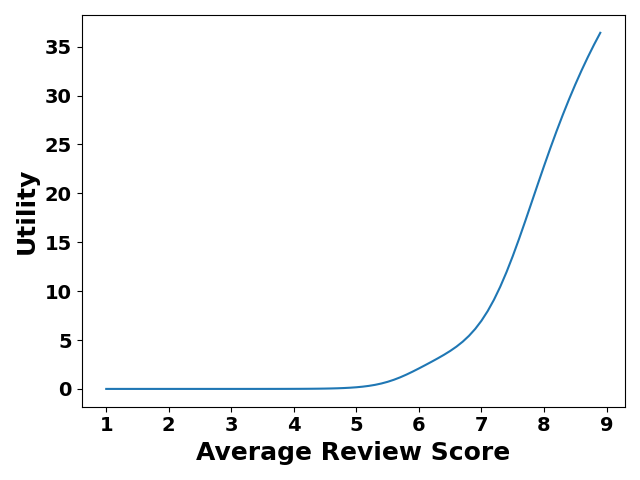}
    \caption{The first two plots, based on the ICLR 2022 dataset, show a paper's probabilities of acceptance as a poster, spotlight, or oral presentation as functions of its average review score.
    The dashed lines denote probabilities estimated from the raw data, and the smooth lines denote probabilities predicted by logistic regression.
    The last plot illustrates the expected utility curve based on a paper's chance of receiving different acceptance labels.
    }
    \label{fig:convex-correlation}
\end{figure}

As illustrated in the left panel of Figure~\ref{fig:convex-correlation}, the probability of a paper receiving an oral presentation at ICLR 2022 is approximately convex in its averaged review score. The middle panel plots the overall acceptance probability, summing over acceptances with different labels $\sigma \in \Sigma = \{ \text{``poster''}, \text{``spotlight''}, \text{``oral''} \}$, as a function of the review score. This curve is not globally convex.
Suppose an author derives utility $u(\sigma)$ for label-$\sigma$ acceptance for $\sigma \in \Sigma$.
The right panel of Figure~\ref{fig:convex-correlation} plots the expected utility as a function of score using $u(\sigma) \propto 1/\Pr[\sigma]$, where $\Pr[\sigma]$ is the ratio of accepted label-$\sigma$ papers among all papers with a given score (i.e., the utility $u(\sigma)$ of a label-$\sigma$ paper is assumed to be {proportional to} its scarcity).
In practice, authors with more papers may derive higher utility $u(\sigma)$ from scarcer paper labels, making the convex utility assumption easier to satisfy on the calibration-relevant score region. These authors also supply more ranking information to the mechanism. In addition, Remark~\ref{rm:convex-rating} illustrates how the design of rating metrics could encourage stronger convexity of author utility.
{Taken together, these panels exhibit the convex-then-concave pattern, with convexity over the score region below the acceptance threshold and saturation or concavity above. We therefore read Figure~\ref{fig:convex-correlation} as empirical support for Assumption~\ref{assum:convex-utility} on the score region most relevant for calibration, while recognizing that richer utility shapes remain important for future work.}

\begin{remark}[Rating Design and Convex Utility in Peer Review]\label{rm:convex-rating}
Convex utility can arise not only from the high-risk-high-reward nature of research but also from the design of peer review systems. Existing measures include best-paper awards and selective labels such as ``oral'' and ``spotlight.''
We observe that redesigning the review rating scale can further encourage convexity in owners' utility. One approach is to use fewer rating levels for excellent papers (e.g., merging ``7: very good'' and ``8: excellent'' into a single ``8: excellent'' rating) while retaining more fine-grained ratings for regular papers.
Under exchangeable review noise, such a design widens score differences among regular papers while narrowing them among excellent papers. It can therefore better distinguish papers near the acceptance margin. High-rated papers typically require less deliberation; for example, most papers with the highest ratings were accepted at ICLR 2022. Intuitively, the design can encourage convexity in the acceptance curve by stretching its middle region and compressing its high-score region.
A related redesign was adopted at ICLR 2024, where reviewers selected ratings from the non-linear scale $\{1,3,5,6,8,10\}$.
\end{remark}

{To relax the current utility assumption, an important direction is to study utility functions that are monotone in adjusted scores but convex over an intermediate score range and concave near upper score thresholds, as suggested by Figure~\ref{fig:convex-correlation} and by capped outcome spaces such as conference decisions. A complementary direction is to study such monotone utilities in the design of sequential-review mechanisms, despite the downside of prolonged peer review procedures~\citep{zhang2024eliciting}.}
}

\section{Proof of Theorem \ref{thm:ne-same-item}}\label{append:ne-same-item}

{Before we dive into the proof, we will introduce several definitions and lemmas.
}

\begin{definition}[Mixed Strategy Nash equilibrium]  \label{append:mixed-ne}
In a mixed strategy Nash equilibrium, each owner $j$ adopts a mixed strategy $x^j$ as a distribution supported on all possible rankings. We say a profile of owners' report $\{x^j\}_{j=1}^{m}$ forms a \emph{mixed strategy} Bayes-Nash Equilibrium (NE) under mechanism~$\cM$
if for any owner $j \in [m]$, given others' randomized report profile $x^{-j} = \{x^{j'}\}_{j'\neq j}$, the expected utility under $x^j$ is no worse than any other possible randomized report $\tilde{x}^j$, i.e.,
$$
\Ex_{\by} \left[ \Ex_{\pi^j\sim x^j} \Ex_{\pi^{-j}\sim x^{-j}} \mathrm{U}^{j}\left( \hat{\mathrm{R}}_{\cM}(x^j, x^{-j}; \by) \right)\right] \geq \Ex_{\by} \left[ \Ex_{\tilde{\pi}^j\sim \tilde{x}^j} \Ex_{\pi^{-j}\sim x^{-j}} \mathrm{U}^{j}\left( \hat{\mathrm{R}}_{\cM}(\tilde{x}^j, x^{-j}; \by) \right) \right].
$$
\end{definition}

\begin{definition}[Majorization]
We say $\ba$ majorizes $\bb$, denoted $\ba \succeq \bb$, if  $\sum_{i=1}^k a_{(i)} \ge \sum_{i=1}^k b_{(i)}, \forall k < n$ and $\sum_{i=1}^n a_{(i)}= \sum_{i=1}^n b_{(i)}$, where $a_{(1)} \geq \cdots \geq a_{(n)}$ and $b_{(1)} \geq \cdots \geq b_{(n)}$ are sorted in descending order from $\ba$ and $\bb$, respectively.
\end{definition}

\begin{lemma}[Hardy–Littlewood–Pólya inequality] \label{lm:majorize}
 For any vector $\ba, \bb \in \RR^n$,  the inequality
 $$ \sum_{i=1}^n h(a_i) \geq  \sum_{i=1}^n h(b_i)$$
 holds for all convex functions $h: \RR\to \RR$ if and only if $\ba \succeq \bb$.
\end{lemma}

\begin{lemma}\label{lm:decreasing}
  For any vector $\ba, \bb \in \RR^n$,   $\ba^+ + \bb^+ \succeq \ba^+ + \rho \circ \bb^+$ for any permutation $\rho$ on $\bb^+$, where $\ba^+ = (a_{(1)},a_{(2)},\dots,a_{(n)})$ and $\bb^+ = (b_{(1)},b_{(2)},\dots,b_{(n)})$
\end{lemma}
\begin{proof}[Proof of Lemma~\ref{lm:decreasing}]
It is clear that $(\ba^+ + \bb^+)_{(i)} = \ba^+_{i} + \bb^+_{i}=a_{(i)}+b_{(i)}$. And for any permutation $\rho$, we have $(\rho \circ \bb^+)_{(i)} = \bb^+_{(i)} = b_{(i)}$.
Next, we prove the lemma by definition.
The equality constraint holds since $\rho$ is a permutation. For any $k < n$, we have
$$
\sum_{i=1}^k (a_{(i)}+b_{(i)}) = \sum_{i=1}^k a_{(i)} + \sum_{i=1}^k (\rho \circ \bb^+)_{(i)} \ge \sum_{i=1}^k (\ba^+ + \rho \circ \bb^+)_{(i)}.
$$
We have proved that $\ba^+ + \bb^+ \succeq \ba^+ + \rho \circ \bb^+$ by definition.
\end{proof}

\begin{lemma}\label{lm:sum_order}
Suppose $\ba, \ba', \bb$ are in the same descending order and $\ba \succeq \ba'$, then $\ba + \bb \succeq \ba' + \bb.$
\end{lemma}

\begin{proof}[Proof of Lemma~\ref{lm:sum_order}]
We first verify that equality holds. Since $\sum_{i=1}^n a_i=\sum_{i=1}^n a_i'$,
$$
\sum_{i=1}^n (a_i+b_i)=\sum_{i=1}^n (a'_i+b_i).
$$

For the inequality at any $k<n$, we have $\sum_{i=1}^k a_i \ge \sum_{i=1}^k a'_{i} $. This readily implies that
$$
\sum_{i=1}^k (a_i+b_i) \ge \sum_{i=1}^k (a'_i+b_i).
$$
Therefore, by definition, we have $\ba + \bb \succeq \ba' + \bb$.
\end{proof}

\begin{thm}[Restatement of {\cite[Theorem~1]{su2021you}}]
\label{thm:restate}
Under Assumption \ref{assum:informed-owners}, \ref{assum:exchangeable-noise}, \ref{assum:convex-utility}, it is optimal for a single owner to truthfully report the ranking of its items $\pi^\star$, $$\Ex_{\bz}\left[ \mathrm{U}( \widehat\bR(\pi^\star) ) \right] = \max_{\pi} \Ex_{\bz}\left[ \mathrm{U}( \hat{\bR}(\pi) ) \right]. $$
\end{thm}

\begin{proof}[Proof of Theorem \ref{thm:ne-same-item}]
We first show that truthful reporting forms a Bayes-Nash equilibrium.
Pick any owner and let its utility function be $\mathrm{U}(\hat{\bR}) = \sum_{i=1}^{n} {U}(\hat{R}_i)$.
Suppose that the other $m-1$ owners are already truth-telling and without loss of generality we can index them from $1$ to $m-1$.
With slight abuse of notation, we let $\widehat\bR^j(\pi)$ be the random variable for the score estimates using owner $j$'s reported ranking $\pi$ w.r.t. the randomness of review noise $\bz$, i.e, $ \Pr[\hat{\bR}^j(\pi) = \hat{\mathrm{R}}^j(\pi; \bR + \bz)] = \Pr[\bz] $. Suppose that the first $m-1$ estimates use the true ranking and the true ranking $\pi^\star$ is the identity permutation for simplicity. It is clear that the item score estimates with all $m-1$ owners' truthfully reported rankings are all identical,
\[
\widehat\bR^1(\pi^\star) = \widehat\bR^2(\pi^\star) = \cdots = \widehat\bR^{m-1}(\pi^\star)  := \widehat\bR(\pi^\star).
\]

Suppose the $m$-th owner reports $\pi$.
Pick any $\{\alpha^j \}_{j=1}^m$ such that $\sum_{j=1}^m \alpha^j = 1$.
Then, her expected utility is
\[
\Ex_{\bz} \left[ \mathrm{U} ( \alpha^m \widehat\bR^{m}(\pi) + \sum_{j=1}^{m-1} \alpha^j \widehat\bR^j(\pi^\star)  ) \right] = \Ex_{\bz} \left[ \mathrm{U} (\alpha^m \widehat\bR^{m}(\pi) + (1-\alpha^m) \widehat\bR(\pi^\star) ) \right].
\]

Since the noise terms are exchangeable in distribution, the expectation can be replaced by averaging the term above with $n!$ permuted noise terms. Let $\widehat\bR(\pi; \bz) := \widehat\bR(\pi; \bR + \bz) $ and  $\rho \circ \bz$ be the noise vector $\bz$ after permutation $\rho$. By the linearity of expectation, we have

\[
\Ex_{\bz } \left[ \mathrm{U} (\alpha^m \widehat\bR^{m}(\pi) + (1-\alpha^m)\widehat\bR(\pi^\star) ) \right] = \Ex_{\bz } \left[ \frac{1}{n!}\sum_{\rho} \mathrm{U} \left( \alpha^m\widehat{\mathrm{R}}^{m}(\pi; \rho\circ \bz) + (1-\alpha^m)\widehat{\mathrm{R}}(\pi^\star; \rho\circ \bz)  \right) \right].
\]

Note that $\widehat{\mathrm{R}}^{m}(\pi; \rho\circ \bz) = \widehat{\mathrm{R}}(\pi; \rho\circ \bz)$ for any $\pi$.
Then, it suffices to show that, for any vector $\bz \in \RR^n$,
\begin{equation}\label{eq:what_we_prove}
 \sum_{\rho} \mathrm{U} \left( \alpha^m \widehat{\mathrm{R}}(\pi; \rho\circ \bz) + (1-\alpha^m)\widehat{\mathrm{R}}(\pi^\star; \rho\circ \bz)  \right) \le \sum_{\rho} \mathrm{U} \left( \alpha^m\widehat{\mathrm{R}}(\pi^\star; \rho\circ \bz) + (1-\alpha^m)\widehat{\mathrm{R}}(\pi^\star; \rho\circ \bz)  \right) .
\end{equation}

Let $\ba^+$ be the projection of $\ba$ onto the standard isotonic cone $\{ \ba | a_1 \geq a_2 \geq \dots \geq a_n \}$. Recall that we assume without loss of generality that $R_1 \geq R_2\dots \geq R_n$. Following from the coupling argument in the proof of Theorem 1 in \cite{su2021you}, we have {$\widehat{\mathrm{R}}(\pi^\star; \bz)  = (\bR + \bz)^+ $ and $\widehat{\mathrm{R}}(\pi; \bz)  = \pi^{-1}  \circ (\pi \circ \bR + \pi  \circ \bz)^+, \forall \bz \in \RR^n $.} So we can rewrite the inequality \eqref{eq:what_we_prove} as,
\begin{equation*}
\sum_{\rho} \mathrm{U} \left( (1-\alpha^m)  (\bR + \rho \circ \bz)^+ + \alpha^m \pi^{-1}  \circ (\pi \circ \bR + \pi  \circ \rho \circ \bz)^+  \right) \leq
\sum_{\rho} \mathrm{U} \left( (\bR + \rho \circ \bz)^+  \right) .
\end{equation*}

We can prove this inequality with the following steps:
\begin{align*}
 & \sum_{\rho} \mathrm{U} \left( (1-\alpha^m) (\bR + \rho \circ \bz)^+ + \alpha^m \pi^{-1}  \circ (\pi \circ \bR + \pi  \circ \rho \circ \bz)^+  \right)   \\
 \leq & \sum_{\rho} \mathrm{U} \left( (1-\alpha^m)  (\bR + \rho \circ \bz)^+ + \alpha^m (\pi \circ \bR + \pi  \circ \rho \circ \bz)^+  \right) \\
 \leq & \sum_{\rho} \mathrm{U} \left( (1-\alpha^m)  (\bR + \rho \circ \bz)^+ + \alpha^m (\bR + \pi  \circ \rho \circ \bz)^+  \right) \\
  = & \sum_{i=1}^{n} \sum_{\rho} U \left( (1-\alpha^m)  (\bR + \rho \circ \bz)^+_i + \alpha^m (\bR + \pi  \circ \rho \circ \bz)^+_i  \right) \\
  \leq & \sum_{i=1}^{n} \sum_{\rho} U \left( (1-\alpha^m)  (\bR + \rho \circ \bz)^+_i + \alpha^m (\bR + \rho \circ \bz)^+_i  \right) \\
  = & \sum_{\rho} \mathrm{U} \left( (\bR + \rho \circ \bz)^+  \right)
\end{align*}
The first inequality is due to Lemmas~\ref{lm:majorize} and \ref{lm:decreasing}, as $ \pi^{-1}$ is some permutation on $\bb^+ = (\pi \circ \bR + \pi  \circ \rho \circ \bz)^+$.
The second inequality follows from Lemmas~\ref{lm:majorize}, \ref{lm:sum_order} as well as Lemma 2.4 of \citep{su2021you}, which shows that $(\bR + \pi  \circ \rho \circ \bz)^+ \succeq (\pi \circ \bR + \pi  \circ \rho \circ \bz )^+, \forall \rho $. The third inequality follows from Lemma~\ref{lm:majorize} and \ref{lm:decreasing} where for each distinct permutation $\rho_j$, $ (1-\alpha^m) (\bR + \rho_j \circ \bz)^+_i, \alpha^m (\bR + \rho_j \circ \bz)^+_i$ are the $j$-th entry of $\ba^+, \bb^+ \in \RR^{n!}$ and $\pi$ is some permutation on $\bb^+$. The equalities are by the definition of the utility function and its linearity.

{
It now remains to prove that truthful NE gives every owner the highest equilibrium utility among all possible equilibria of the game.
Again, we pick arbitrary owner $i$.
Under truthful NE, the owner's utility is $\Ex_{\bz} \left[\mathrm{U}^i(\hat{\bR}(\pi^\star))\right]$, where $\widehat\bR(\pi^\star)$ is the item score estimate under the truthful ranking. We compare it to any (possibly mixed) Nash equilibrium (see Definition~\ref{append:mixed-ne}). Consider any pure strategy profile in this mixed Nash equilibrium, $\boldsymbol{\pi} = (\pi^1, \pi^2, \dots, \pi^m) \sim \bx = (x^1, x^2, \dots, x^m)$. This sampled profile of reported ranking corresponds to the weighted average of score estimates, $\sum_{j=1}^{m} \alpha^j \hat{\bR}(\pi^j) $.
We can see that the owner's expected utility under this pure strategy profile is no larger than that under the truthful strategy profile,
\begin{equation} \label{eq:jensen}
\begin{aligned}
    \Ex_{\bz}\left[\mathrm{U}^i \left( \sum_{j=1}^{m} \alpha^j \hat{\bR}(\pi^j) \right) \right] &\le \sum_{j=1}^{m} \alpha^j \Ex_{\bz}\left[ \mathrm{U}^i( \hat{\bR}(\pi^j) ) \right] \\
    &\leq \max_{j\in [m]} \Ex_{\bz}\left[ \mathrm{U}^i( \hat{\bR}(\pi^j) ) \right] \\
    &\leq \Ex_{\bz}\left[ \mathrm{U}^i( \widehat\bR(\pi^\star) ) \right],
    \end{aligned}
\end{equation}
where the first inequality is due to Jensen's inequality and linearity of expectation, the second inequality is by the property of arithmetic mean, last inequality is due to Theorem~\ref{thm:restate}.
Hence, the owner's expected utility under this Nash equilibrium, taking an expectation over the distribution of pure strategy profile, $\Ex_{\boldsymbol{\pi}\sim \bx} \Ex_{\bz} \left[ \mathrm{U}^i(  \sum_{j=1}^{m} \alpha^j \hat{\bR}^j ) \right] \leq  \Ex_{\bz} \left[ \mathrm{U}^i( \widehat\bR(\pi^\star) ) \right]$, is no larger than the expected utility under the truthful NE.
}
\end{proof}
\begin{remark}
In fact, one may have noticed that the proof of payoff dominance above does not rely on whether the non-truthful strategy profile forms an equilibrium or not. Hence, a slightly stronger result holds, that is, the truthful NE gives every owner the highest utility among all possible strategy profiles of the game. However, we would like to also point out that this additional observation does not imply that the truthful report is necessarily a dominant strategy for any owner, i.e., the best response to any other strategy profile of the other owners. Nor is the truthful NE necessarily unique in this game. That said, given its payoff-dominant property, there is other strong evidence, especially from empirical human preference and behavioral theory, to expect the truthful equilibrium outcome, as pointed out in Remark~\ref{rm:truthful-nash}.
\end{remark}

{
\section{Proof of Proposition~\ref{prop:coarser-better}}\label{append:coarser-better}

\begin{proof}[Proof of Proposition~\ref{prop:coarser-better}]
We first establish the per-block risk identity. For a block of size $s$, let
$C_s = \{\br \in \RR^s : r_1 \geq \cdots \geq r_s\}$.
When $R_1 = \cdots = R_s$, write $\bR = c\mathbf{1}_s$. Since $\operatorname{span}\{\mathbf{1}_s\}$ is the lineality space of $C_s$, projection onto $C_s$ is translation-equivariant along $\bR$, so
$$
\mathrm{proj}_{C_s}(\bR+\bz)-\bR = \mathrm{proj}_{C_s}(\bz).
$$
For $\bz \sim \mathcal{N}(0,\sigma^2\bI_s)$, writing $\bz=\sigma \boldsymbol{g}$ with $\boldsymbol{g}\sim\mathcal{N}(0,\bI_s)$ gives
$$
\Ex\norm{\mathrm{proj}_{C_s}(\bR+\bz)-\bR}^2
= \sigma^2 \Ex\norm{\mathrm{proj}_{C_s}(\boldsymbol{g})}^2
= \sigma^2 \delta(C_s),
$$
where the last equality is the metric characterization of statistical dimension~\citep[Proposition~2.4]{amelunxen2014living}.
The cone $C_s$ is isometric to the chamber $\{x\in\RR^s:x_1\leq\cdots\leq x_s\}$, whose statistical dimension is $H_s=\sum_{k=1}^s 1/k$~\citep[Proposition~3.5]{amelunxen2014living}. Therefore the risk of isotonic regression on this block equals $\sigma^2 H_s$.

For a partition with block sizes $(s_1,\ldots,s_K)$, isotonic regression is applied block by block and the squared-error risk is additive across blocks, so the total risk is $\sigma^2\sum_{k=1}^K H_{s_k}$. Hence $w(0)=0$ by convention and $w(s) = \sigma^2 s - \sigma^2 H_s = \sigma^2(s - H_s)$ for $s\geq1$.
Direct evaluation gives $w(0) = 0$ and $w(1) = \sigma^2(1 - 1) = 0$, and strict monotonicity follows from
$$
w(s+1) - w(s) = \sigma^2\left(1 - \frac{1}{s+1}\right) = \sigma^2 \cdot \frac{s}{s+1} > 0 \qquad (s \geq 1).
$$
For strict convexity on the non-negative integers, the successive increments are strictly increasing:
\[
\bigl[w(s+1)-w(s)\bigr]-\bigl[w(s)-w(s-1)\bigr]
=\sigma^2\left(\frac{s}{s+1}-\frac{s-1}{s}\right)
=\frac{\sigma^2}{s(s+1)}>0 \qquad (s\geq1).
\]
The same increment calculation also shows that total wellness is Schur-convex in the block sizes. Fix any partition $(s_1, \ldots, s_K)$ of $n$ and consider a transfer that increases a block of size $s_i$ by $1$ and decreases another block of size $s_j \geq 2$ by the same amount, with $s_i \geq s_j$. The change in total wellness is
\[
w(s_i + 1) - w(s_i) - [w(s_j) - w(s_j - 1)]
= \sigma^2\left(\frac{s_i}{s_i + 1} - \frac{s_j - 1}{s_j}\right) > 0
\]
whenever $s_i \geq s_j$. Hence $\sum_k w(s_k)$ is Schur-convex in the block sizes, and is maximized by the most unequal split.

Finally, compare a single block of size $n$ with a two-block refinement of sizes $n_1$ and $n_2=n-n_1$. The corresponding risks are $\sigma^2 H_n$ and $\sigma^2(H_{n_1}+H_{n_2})$, respectively, so the risk increase from splitting is
$$
\sigma^2 (H_{n_1} + H_{n_2} - H_n)
= \sigma^2 \sum_{j=1}^{n_2} \left(\frac{1}{j} - \frac{1}{n_1 + j}\right)
= \sigma^2 \sum_{j=1}^{n_2} \frac{n_1}{j(n_1 + j)}
> 0.
$$
For an equal split, taking $n_1=n_2=n/2$ gives risk ratio $2H_{n/2}/H_n\to2$ as $n\to\infty$, since $H_s=\ln s+\gamma+o(1)$. Equivalently, inverse-risk statistical efficiency is asymptotically halved.
\end{proof}
}

{
\section{Proof of Proposition~\ref{prop:more-owners}}\label{append:more-owners}

\begin{proof}[Proof of Proposition~\ref{prop:more-owners}]
Fix the review noise realization $\bz$ (and hence $\by = \bR + \bz$). Conditional on $\by$, the random vectors $\hat{\bR}^j := \hat{\mathrm{R}}(\pi^j ; \by)$ for $j = 1, \ldots, L$ are i.i.d.\ because the perception noise vectors $\bepsilon^1, \ldots, \bepsilon^L$ are i.i.d.\ and independent of $\bz$.
Let $\bmu(\by) := \Ex_\pi[\hat{\mathrm{R}}(\pi; \by)]$ denote the conditional mean.
By the standard bias-variance decomposition for an i.i.d.\ average $\hat{\bR}_L = \frac{1}{L}\sum_{j=1}^L \hat{\bR}^j$,
\begin{align*}
\Ex_\pi \norm{\hat{\bR}_L - \bR}^2
&= \norm{\bmu(\by) - \bR}^2 + \Ex_\pi\norm{\hat{\bR}_L - \bmu(\by)}^2 \\
&= \norm{\bmu(\by) - \bR}^2 + \frac{1}{L}\Ex_\pi\norm{\hat{\mathrm{R}}(\pi;\by) - \bmu(\by)}^2.
\end{align*}
The second equality uses $\Ex\norm{\bar{X}_L - \mu}^2 = \frac{1}{L}\Ex\norm{X_1 - \mu}^2$ for i.i.d.\ $X_1, \ldots, X_L$ with mean $\mu$.
Taking the outer expectation over $\bz$ yields~\eqref{eq:owners-bv}.
The bias term $\norm{\bmu(\by) - \bR}^2$ does not depend on $L$, so the risk is non-increasing in $L$, with strict decrease whenever the variance term is strictly positive, i.e., whenever perception noise yields distinct isotonic estimates with positive probability over the joint draw of $\bz$ and $\pi$.
Under Assumption~\ref{assum:informed-owners}, every owner reports $\pi^\star$ with probability one, so $\hat{\mathrm{R}}(\pi; \by) = \hat{\mathrm{R}}(\pi^\star; \by)$ is deterministic conditional on $\by$, and the variance term is zero.
\end{proof}
}

\section{Proof of Proposition~\ref{prop:simple-graph}}\label{append:simple-graph}
We prove that for any ownership instance $\cO = \{  \cI^j \}_{j \in [m]}$, we can construct a different ownership instance $\cO' = \{  \bar{\cI}^j \}_{j \in [m]}$ such that any $L$-strong partition $\cS$ in $\cO$ is a 1-strong partition in $\cO'$.

Pick any bipartite graph $\cG = (\cA, \cP, \cE)$ from the ownership instance $\cO = \{ \cI^j \}_{j \in [m]}$. Suppose we were to find $L$-strong partition $\cS$ in $\cG$. We can construct a bipartite graph $\cG' = (\cA', \cP', \cE') $, where the item set remains the same $\cP' = \cP$, the owner set $\cA'$ corresponds to all the $L$ element subsets of $\cA$, $\cA' = \{ \ell | u^{\ell}\subseteq \cA, |u^{\ell}| = L\}$ and the edge set captures the {common items} of every $L$ owners, $\cE' = \{ (i,\ell) | i \in \bigcap_{j \in u^{\ell}} \cI^{j} \}$.
We show that for any $L$-strong partition $\cS = \{\cS_1, \cS_2, \dots, \cS_K \}$ of $\cG$, there is an equivalent $1$-strong greedy partition $\cS' = \{\cS'_1, \cS'_2, \dots, \cS'_K \}$ of $\cG'$ such that $\cS_k = \cS'_k, \forall k\in K$.

The proof is a straightforward verification of this bipartite graph construction. Pick any $\cS_k \in \cS$ with $|\cT_k| \geq L$ in $\cG$, there exists an $\cS'_k$ with $|\cT'_k| = 1$  in $\cG'$. That is, we have $\cT'_k = \{ \ell \}$ for some $u^\ell \subseteq \cT_k$, then $\cS'_k = \cS_k \subseteq \bigcap_{j \in u^{\ell}} \cI^{j} $ by construction. Therefore, $\cS_k = \cS'_k, \forall k\in [K]$.

\section{Proof of Proposition~\ref{prop:hardness}}\label{append:hardness}
We consider the weight function $w(x) = \max\{x, 1\}-1$. For any partition with $K$ blocks, the objective function is $\obj(\cS) = n - K$. We show that finding the optimal partition under such an objective function is at least as hard as finding the minimum set. We prove via a standard hardness reduction argument: for any set cover instance, we can construct a bipartite graph instance in which the partition that maximizes the objective function $\obj(\cS) = n - K$ can be turned into a minimum set cover.

Pick arbitrary set cover instance with $m$ subsets of $n$ item, $\cV_1, \cV_2, \dots, \cV_m$. We construct a bipartite graph $\cG = (\cA, \cP, \cE)$ where $\cA = [m], \cP=[n]$ and $\cE = \{(p_i,a_j) | p_i\in \cV_j, j\in [m]\}$. There exists an optimal partition $\cS^*$ with $\obj(\cS^*) = n - K^*$, if and only if the minimum set cover has size $K^*$. For the ``if'' direction, we construct a partition $\cS^*$ with $\obj(\cS^*) = n - K^*$ from the set cover of size $K^*$. For each subset selected by the set cover, we construct a partition block with items in the subset minus the items already in the partition. This partition must be valid and have $K^*$ blocks. For the ``only if'' direction, we construct the set cover of size $K^*$ from the optimal partition $\cS^*$. For every block $\cS_k^*$ of the optimal partition $\cS^*$, we pick any vertex $a_j \in \cT_k^*$ from the corresponding owner set. Based on the $K^*$ distinct vertices, we select the subsets with the same indices, and it forms a set cover of all $n$ items by construction.

\section{Proof of Theorem \ref{thm:partition-necessary}}\label{append:necessity-proof}

\begin{proof}[Proof of Theorem \ref{thm:partition-necessary}]
The proof is structured as follows. First, we characterize the necessary conditions under which Mechanism~\ref{algo:linear-isotonic} is truthful, according to Lemma~\ref{lm:truthful-equivalence}.
Second, we show that for any Mechanism~\ref{algo:linear-isotonic} under the necessary condition, a Mechanism~\ref{algo:deterministic-partition} can be constructed to elicit as least as much ranking information.

Fix any ownership relation $\{  \cI^j \}_{j = 1}^{m}$.
For notational convenience, we let $\beta_i^j=0, \forall i\not\in \cI^j$, as $\beta_i^j$ does not affect the mechanism output anyway.
Our characterization of truthful Mechanism~\ref{algo:linear-isotonic} is based on the following lemma (restatement of Lemma), which shows that truthful condition holds for Mechanism~\ref{algo:linear-isotonic} only if its parameters admit certain partition structure as detailed in Equation \eqref{eq:partition-almost}.

{
\begin{lemma}[Restatement of Lemma \ref{lm:truthful-equivalence}]
Under Assumptions \ref{assum:informed-owners}, \ref{assum:exchangeable-noise}, and \ref{assum:convex-utility}, if a Mechanism~\ref{algo:linear-isotonic} with parameters $\big\{\mathfrak{S}^j , \bbeta^j \big\}_{j=1}^{m}$ is truthful, then  the following two conditions must both hold:
\begin{enumerate}
\item[\textup{(I)}] Each owner has balanced influence on the items within each of its partition blocks for any input, i.e.,
$$\text{for any } j\in [m], \cS\in \mathfrak{S}^j, i,i'\in \cS, \qquad \text{we have } \,  \omega^j_i = \omega^j_{i'},$$
where $\omega_i^j =  \alpha^j \beta^j_i / \sum_{j'\in \cJ^i} \alpha^{j'} \beta^{j'}_{i}$ denotes owner $j$'s relative influence on the output score of item $i$.
\item[\textup{(II)}] The parameters $\big\{\mathfrak{S}^j , \bbeta^j \big\}_{j=1}^{m}$  has a valid partition structure in the following sense,
\begin{equation}
\text{for any } j,j'\in [m], \cS \in \mathfrak{S}^j, \cS' \in \mathfrak{S}^{j'},\text{ if }\cS\cap \cS' \neq \varnothing, \text{ then } \beta^j_i = \beta^j_{i'}, \text{ for all } i,i'\in \cS\cup \cS'.\footnote{We remark that  $\beta_i^j$ is set to $0$ for any  $ i\not\in \cI^j$ in this statement.}
\end{equation}
\end{enumerate}
Moreover, these two conditions are equivalent to each other.
\end{lemma}
}

We start by observing a few corollaries from Lemma \ref{lm:truthful-equivalence}, which will be useful for our proof of Theorem~\ref{thm:partition-necessary}. First,    Condition (II) of the lemma is applicable to the situation with $j = j'$, in which case $\cS = \cS'$ since other sets in $\mathfrak{S}^j$  does not overlap with $\cS$. Hence we have $\beta^j_i = \beta^j_{i'} $ for each $i, i' \in \cS$. Second, let us now consider the case $j \not = j'$ and let $\cS \in \mathfrak{S}^j, \cS' \in \mathfrak{S}^{j'}$ satisfying  $\cS\cap \cS' \neq \varnothing$. Then the lemma implies $\beta^j_i = \beta^j_{i'}, \text{ for all } i,i'\in \cS\cup \cS'$. A consequence of this is that either $\beta^j_i = 0$ (i.e., $j$'s report does not affect item $i$), or $j$ must own every item in $\cS\cup \cS'$ since $j$ has a non-zero value for every item in  $\cS\cup \cS'$. In the later case, if $\cS' \not \subseteq \cS$, then some element $\bar{i}$ in $\cS' \setminus \cS$ owned by $j$ must belong to another set $\bar{\cS} \in \mathfrak{S}^j$. Condition (II) of Lemma \ref{lm:truthful-equivalence} then implies that the weights in  $  \bar{\cS}, \cS$   must all be the same, i.e.,  $\beta^j_i = \beta^j_{i'}$ for any $i \in  \cS$ and $i' \in \bar{\cS}$.

Observations above hint at a constructive proof of the theorem by iteratively merging any two  ``unnecessarily isolated'' partition blocks such as the $\cS, \bar{\cS}$ as above in the parameters
 $\big\{\mathfrak{S}^j,\bbeta^j \big\}_{j=1}^{m}$, until it becomes  a Mechanism~\ref{algo:deterministic-partition}. Formally, we will show that for any  Mechanism~\ref{algo:linear-isotonic} with a valid partition structure specified in Equation~\eqref{eq:partition-almost}, we can construct a Mechanism~\ref{algo:deterministic-partition} with some item set partition that elicits no less ranking information and thus has no worse statistical efficiency.

Our construction hinges on a \texttt{MERGE} operator that merges all those ``unnecessarily isolated'' partition blocks like the $\cS, \bar{\cS}$ above. Specifically,  if  some owner $j \in [m]$ has   two distinct blocks $\cS, \bar{\cS} \in \mathfrak{S}^{j}$ that satisfy the following conditions ---  for any $i \in \cS, \bar{i} \in \bar{\cS}$:  (1) $\beta^j_i = \beta^j_{\bar{i}}$; and (2) $\beta^{j'}_i = \beta^{j'}_{\bar{i}} \neq 0$ for any other owner $j'$ who fully owns all items in  $\cS \cup \bar{\cS}$ ---  then   \texttt{MERGE}   combines $\cS, \bar{\cS}$ as one set in $j$'s $\mathfrak{S}^j$ with the same parameter $\beta^j_i$ for any $i \in \cS \cup \bar{\cS}$.
Notice that \texttt{MERGE} strictly increases the amount of ranking information elicited from the owners, as the orderings of items among merged blocks  $\cS, \bar{\cS}$ are now used in the calibration.

We exhaustively apply \texttt{MERGE} on the parameters for every owner's every pair of item sets until there is no more valid merge, and then remove any block $\cS$ with zero weights, i.e., $\beta^j_i = 0, \forall i\in \cS$. The last procedure does not affect the mechanism but makes our following analysis cleaner. It is also obvious that the merge process must terminate, since the merged block can   only be as large as each owner's item set. In addition, after each merge, the value of any ${\bbeta}^j$ remains unchanged; it can be verified that the condition in Equation \eqref{eq:partition-almost} holds for the updated partition blocks.
Let the resulting parameters be $\big\{\tilde{\mathfrak{S}}^j, {\bbeta}^j \big\}_{j=1}^{m}$.
We claim that $\big\{\tilde{\mathfrak{S}}^j, {\bbeta}^j \big\}_{j=1}^{m}$ satisfies that, for any $j,j'\in [m]$, $\cS \in \tilde{\mathfrak{S}}^j, \cS' \in \tilde{\mathfrak{S}}^{j'}$,  $\cS, \cS'$ are either identical or disjoint.

We prove by contradiction: if this is not the case, then there must exist at least one valid merge. Suppose $\cS \cap \cS' \neq \varnothing$. Since $\big\{\tilde{\mathfrak{S}}^j, {\bbeta}^j \big\}_{j=1}^{m}$ satisfies the condition in Equation~\eqref{eq:partition-almost}, we have shown above that for owner $j$, she must (1) either have $\beta^{j}_i = 0, \forall i \in \cS\cup \cS'$, (2) or $j$ must own every item in $\cS\cup \cS'$ since $j$ has a non-zero value for every item in  $\cS\cup \cS'$. The first case is eliminated as blocks with zero weights are removed. In the second case, some element $\bar{i}$ in $\cS' \setminus \cS$ owned by $j$ must belong to another set $\bar{\cS} \in \mathfrak{S}^j$, and $\cS, \bar{\cS}$ can be merged, which would reach the contradiction. To check that the two condition for a valid \emph{merge} is satisfied, we can observe the following, according to the condition in Equation~\eqref{eq:partition-almost}:
\begin{enumerate}
    \item For this owner $j$, we have $\beta^j_i = \beta^j_{i'} = \beta^j_{\bar{i}}$ for any $\bar{i} \in \bar{\cS} \setminus \cS' $. Hence, $\beta^j_i = \beta^j_{i'}, \forall i,i' \in \cS\cup \bar{\cS}$.
    \item For any other owner $j'$ who owns all items in $\cS\cup \bar{\cS}$, we pick two items $i_1 \in \cS \cap \cS'$ and  $i_2 \in (\cS' \setminus \cS )\cap \bar{\cS} $. Let $i_1\in {\cS}_1$ and $i_2 \in {\cS}_2$ from $\mathfrak{S}^{j'}$. Since $\cS_1 \cap \cS \neq \varnothing$, $\beta^{j'}_{i_1} = \beta^{j'}_{i}, \forall i\in \cS$. Since $\cS_2 \cap \bar{\cS} \neq \varnothing$, $\beta^{j'}_{i_2} = \beta^{j'}_{i}, \forall i\in \bar{\cS}$. Since $\cS_1 \cap \cS \neq \varnothing$ and $i_1, i_2\in \cS$, $\beta^{j'}_{i_1} = \beta^{j'}_{i_2}$.
     Hence, $\beta^{j'}_{i} = \beta^{j'}_{i'} = \beta^{j'}_{i}, \forall i, i'\in \bar{\cS} \cup \cS$.
\end{enumerate}

Finally, consider $\mathfrak{S} := \bigcup_{j=1}^{m} \tilde{\mathfrak{S}}^j$ (after removing repeated item sets).
One can verify that any two blocks $\cS, \cS' \in \mathfrak{S}$ are disjoint, though the union of all its blocks is not necessarily $[n]$ (recall that zero weight blocks have been removed).
We claim that Mechanism \ref{algo:deterministic-partition} with parameter $\mathfrak{S}$ is at least as statistically efficient as  Mechanism~\ref{algo:linear-isotonic} with parameter $\big\{\tilde{\mathfrak{S}}^j, {\bbeta}^j \big\}_{j=1}^{m}$, and thus,  Mechanism~\ref{algo:linear-isotonic} with parameter $\big\{{\mathfrak{S}}^j, {\bbeta}^j \big\}_{j=1}^{m}$ before \texttt{MERGE}.
To see this, let us first observe that Mechanism \ref{algo:linear-isotonic} under $\big\{\tilde{\mathfrak{S}}^j, {\bbeta}^j \big\}_{j=1}^{m}$ proceeds in the following steps: (1) pick each distinct block   $\cS \in \mathfrak{S} = \bigcup_{j=1}^{m} \tilde{\mathfrak{S}}^j$, (2) elicit rankings of items in $\cS$ from any owner $j$ such that $\cS \in \tilde{\mathfrak{S}}^j$, (3) determine a weighted average of adjusted review scores based on the rank-calibrated score from those owners with weight $\alpha^j \beta_i^j$. These procedures are identical to Mechanism \ref{algo:deterministic-partition} with parameter $\mathfrak{S}$, except that Mechanism \ref{algo:deterministic-partition} elicits from all owners with complete ownership of a block $\cS \in \mathfrak{S}$. This completes the proof of our theorem.

\end{proof}

\begin{proof}[Proof of Lemma \ref{lm:truthful-equivalence}]

We will prove that the truthfulness of Mechanism~\ref{algo:linear-isotonic} implies condition (I), and condition (I) implies condition (II). Finally, we will prove condition (II) also implies condition (I), showing their equivalence.

We start by proving the first step, i.e., truthfulness of Mechanism~\ref{algo:linear-isotonic}  implies    condition (I). {We prove by contradiction.}
Suppose that for some owner $j\in [m]$ and for one of her partition blocks $\cS\in \mathfrak{S}^j$, there exists $ i,i'\in \cS$ such that $\omega^j_i \neq \omega^j_{i'}$. Without loss of generality, let $\omega^j_i < \omega^j_{i'} $. We construct an instance with owner utility $\{ \mathrm{U}^j\}_{j=1}^{m}$, review scores and ground-truth scores $\{  y_i, R_i \}_{i=1}^{n} $ such that the unbalanced influence would give the owner $j$ incentive to untruthfully report the pairwise order between $i,i'$.
Let $y_i = R_i > R_{i'} = y_{i'} $.
For any other item in the block $\cS$, let its review score and ground-truth score be equal, and be smaller than the ground-truth scores of $i,i'$ --- it is now possible to only misreport the ranking between $i,i'$, since the pairwise order of $i,i'$ is independent of other items in this block. In addition, since the owner's report strategy in block $\cS$ does not affect the items outside block $\cS$, it suffices to compare the owner's report strategy in block $\cS$.
Let the utility of owner $j$ be ${U}^j(\hat{R}) = \max\{ \hat{R}- c, 0 \}$ for some constant $c$ to be specified below. The utility function is convex w.r.t. the item's adjusted review score $\hat{R}$, conforming to Assumption~\ref{assum:convex-utility}.

We now compare two reporting strategy of this owner: one is to truthfully report the ground-truth ranking, the other is to report that item $i'$ is better than $i$, and the ground-truthful ranking for the rest of the items in this block. It suffices to show that the owner has strictly higher utility in the latter reporting strategy. Recall that in this problem instance, the review scores and ground-truth scores of the other items are equal, and are smaller than the ground-truth scores of $i,i'$. Isotonic regression implies   that the adjusted score of all other items remain the same as the ground-truth score in both reporting strategy, so it suffices to analyze adjusted score of $i,i'$ and compare the owner $j$'s utility on these two items.

If the owner $j$ is truthful, the adjusted score is $\hat{R}_i=R_i, \hat{R}_{i'}= R_{i'}$.
If the owner $j$ chooses the misreporting strategy, the calibrated scores based on the owner $j$'s ranking are $\tilde{R}_i^j=\tilde{R}_{i'}^j=\frac{1}{2} (R_{i'} + R_i)$, resulting in the final adjusted scores,
$\tilde{R}_i = \frac{1}{2} \omega_i^j  (R_{i'} + R_i) + (1-\omega_i^j)R_{i} $ and $\tilde{R}_{i'} = \frac{1}{2} \omega_{i'}^j  (R_{i'} + R_i) + (1-\omega_{i'}^j)R_{i'}.$
Let $\varepsilon =  \frac{R_{i} - R_{i'}}{2}$; it can be verified that $\tilde{R}_i = {R}_{i} - \omega_i^j \varepsilon$ and $\tilde{R}_{i'} = {R}_{i'} + \omega_{i'}^j \varepsilon$.
Notice that since $0 \leq \omega_{i}^j < \omega_{i'}^j \leq 1$, we have ${R}_i \geq \tilde{R}_i > \tilde{R}_{i'} > R_{i'} $ and $\omega_{i'}^j \varepsilon-  \omega_i^j \varepsilon > 0$.
Pick any $c \in [R_{i'}, R_{i'}+\omega_{i'}^j \varepsilon-  \omega_i^j \varepsilon )$,
we can derive the owner $j$'s utility for the two items under the misreport and under the truthful report, respectively:
\begin{align*}
   U^j(\tilde{R}_i) + U^j(\tilde{R}_{i'}) =  & \max\{{R}_{i} -  \omega_i^j \varepsilon - c, 0 \} +  \max\{{R}_{i'} +  \omega_{i'}^j \varepsilon - c, 0 \} \\
    = & {R}_{i} -  \omega_i^j \varepsilon - c + {R}_{i'} + \omega_{i'}^j \varepsilon - c \\
   U^j(\hat{R}_i) + U^j(\hat{R}_{i'}) =  & \max\{{R}_{i} - c, 0 \} +  \max\{{R}_{i'} - c, 0 \} \\
    = & {R}_{i} -  c
\end{align*}
We can see that  $ U^j(\tilde{R}_i) + U^j(\tilde{R}_{i'}) - U^j(\hat{R}_i) - U^j(\hat{R}_{i'}) = {R}_{i'} + \omega_{i'}^j \varepsilon -  \omega_i^j \varepsilon - c  > 0$ (by our choice of $c$), so misreporting the ordering between $j$ and $j'$ is strictly better than truthful report.

\medskip
We now prove that the condition (I) and (II) are equivalent and, therefore, are both necessary condition for Mechanism \ref{algo:linear-isotonic} to be truthful.

(I)$\Rightarrow$(II): We prove by contradiction. Suppose that there exists $j,j'\in [m], \cS \in \mathfrak{S}^j, \cS' \in \mathfrak{S}^{j'}$ such that $\cS\cap \cS'\neq \varnothing $ and there exists $ i_1,i_2 \in \cS \cup \cS', \beta^j_{i_1} \neq \beta^j_{i_2}$. We consider two possible cases:

\begin{enumerate}
    \item If $i_1,i_2 \in \cS$, then we can construct the problem instance with $\alpha^j = 1$ being the only non-zero weight such that $\omega^j_{i_1} = \beta^j_{i_1} \neq \beta^j_{i_2} = \omega^j_{i_2}$, which is a contradiction to the given condition. Similar contradiction arises when   $i_1,i_2 \in \cS'$.
    \item Otherwise,  let $i_1 \in \cS$ and $i_2 \in \cS'$ without loss of generality. Since $\cS\cap \cS'\neq \varnothing$, let $i_0 \in \cS\cap \cS'$. Note that the argument above already shows items within each block must have the same $\beta_i^j$ values, i.e., $\beta_{i_0}^j= \beta_{i_1}^j$ and $\beta_{i_0}^{j'} = \beta_{i_2}^{j'}$. Consequently, we have $\beta_{i_0}^j= \beta_{i_1}^j \neq \beta_{i_2}^j$
and $\beta_{i_0}^{j'} = \beta_{i_2}^{j'}$. We  consider a problem instance with $\alpha^j = \alpha^{j'} = 1$ being only non-zero weight, thus $w_{i_0}^{j'} = \frac{\beta_{i_0}^{j'}}{\beta_{i_0}^{j}+\beta_{i_0}^{j'}}$ and $w_{i_2}^{j'} = \frac{\beta_{i_2}^{j'}}{\beta_{i_2}^{j}+\beta_{i_2}^{j'}}$. Since $i_0, i_2\in \cS'$, we have $\omega^{j'}_{i_0} \neq \omega^{j'}_{i_2}$ because $ \beta_{i_0}^{j'} = \beta_{i_2}^{j'}$ but $\beta_{i_0}^j \neq \beta_{i_2}^j$. Contradiction is reached.
\end{enumerate}

(II)$\Rightarrow$(I):
We consider arbitrary owner $j \in [m]$ and any partition block $\cS \in \mathfrak{S}^j$.
Without loss of generality, suppose $|\cS| \geq 2$ since singleton set does not have item rankings.
Pick any two distinct items $i,i'\in \cS$ in the block. We make the following observations from the condition in Equation \eqref{eq:partition-almost}:
\begin{enumerate}
    \item $\beta_i^{j} = \beta_{i'}^{j}$ for the owner $j$ herself. This is implied by instantiating Equation \eqref{eq:partition-almost} to   $j = j'$, in which case $\cS = \cS'$ since other sets in $\mathfrak{S}^j$ does not overlap with $\cS$. The condition applies to any two items in $\cS$.
    \item $\beta_i^{j'} = \beta_{i'}^{j'}$, for any owner $j'(\neq j)$ who own any of $i,i'$, i.e., $j' \in \cJ^i \union \cJ^{i'}$. This is implied by instantiating Equation \eqref{eq:partition-almost} to the $j', j$ and some $\cS' \in \mathfrak{S}^{j'}$ such that $\cS' \cap \cS \neq \varnothing$. Any items in $\cS' \union \cS$ such as $i,i'$ must have the same $\beta$ values, i.e.,  $\beta_i^{j'} = \beta_{i'}^{j'}$.
    \item $\beta_i^{j'} = \beta_{i'}^{j'} = 0$, for any owner $j'$ who own neither of $i,i'$, i.e., $j' \not\in \cJ^i \union \cJ^{i'}$. This follows from the mechanism construction where we set $\beta_i^{j'} = 0$ for any $i \not\in \cI^{j'}$.
\end{enumerate}
The three cases above implies that, for any input credentials $\{ \alpha^j\}_{j=1}^{m}$, we must always have $ \omega^j_i = \omega^j_{i'}$ by definition for any two items $i, i'$ in any of $j$'s partition block $\cS$, which proves the condition \textup{(I)}.

\end{proof}

\section{Proof of Theorem \ref{thm:greedy-approximation} } \label{append:greedy-approximation}
\textbf{A Convenient Bipartite Graph View of the Ownership Relation.} To ease our discussion in the formal proofs regarding the optimal partition, we start by representing the overlapping ownership model as a bipartite graph $ \cG = (\cP,\cA,\cE)$. The vertex sets $\cP,\cA,$ with $|\cP|=n, |\cA|=m$ correspond to the disjoint set of $n$ items (papers) and $m$ owners (authors), respectively. The edge set $\cE \subseteq \cP \times \cA $ corresponds to the ownership relation: the owner $j$ owns the item $i$, if and only if there is an edge $(p_i,a_j) \in \cE$ between the vertices $p_i\in \cP$ and $a_j\in \cA$. The number of edges $|E| = \sum_{j \in [m]} |\cI^j|$ is precisely the $N$ in the theorem statement, which  captures the order of the problem's input size.   Let $E = (e_{i}^{j})_{m\times n} $ be the $0-1$ bi-adjacency matrix of the ownership in which $e_{i}^{j} = 1$ for each edge $(p_i,a_j) \in \cE$. Hence, the set of items owned by $j$ is exactly the neighbor set of vertex $a_j$, i.e., $\cI^j = \{ p_i\in \cP | e_{i}^{j} = 1  \}$.

We now formalize the partition-based scheme in Mechanism \ref{algo:deterministic-partition} using the bipartite graph characterization. Specifically, we consider the partition $\mathfrak{S} = \{\cS_1, \ldots, \cS_K\}$ on   the vertex set $\cP$ of the bipartite graph. By definition of a partition, we have   $\bigcup_{k=1}^{K} \cS_k = \cP$ and $\cS_{k'} \intersect \cS_{k} = \varnothing, \forall k\neq k' \in [K]$. For each partition block $\cS_k$, we denote $\cT_k = \{j \in \cA | \cS_k \subseteq \cI^j \}$  as the set of owners who owns all items in $\cS_k$. For the blocks with $|\cS_k| \leq 1$ or $|\cT_k|\leq 0$, the mechanism simply uses the raw review score.
Otherwise, by construction, the sub-bipartite-graph of $\cG$ induced by each $(\cS_k, \cT_k)$ and their incidence edges $\{(p_i,a_j) \in \cE | p_i\in \cS_k, a_j\in \cT_k \}$  {form a} complete bipartite graph, i.e., a complete overlapping ownership. Therefore, we can also visualize the partition of the partially overlapping ownership using the bi-adjacency matrix.

We now utilize the notations above to formally prove Theorem \ref{thm:greedy-approximation}.

\begin{proof}[Proof of Theorem \ref{thm:greedy-approximation}]
Pick any ownership instance   represented   as a bipartite graph $\cG$ and any wellness function $w\in \cW$. Let $\OPT(\cG) := \obj(\mathfrak{S}^*) $, where $\mathfrak{S}^* = \{ \cS_k^* \}_{k=1}^{K}$ is the optimal $1$-strong partition of $\cG$.
Consider the greedy algorithm that keeps picking the largest residual item set owned by some {owner}, until all items are assigned to some block, as described in Algorithm \ref{algo:greedy}.  Let $\mathfrak{S} $ denote the partition obtained by Algorithm \ref{algo:greedy}. By construction, $\mathfrak{S} $ is 1-strong  --- i.e., there is at least one owner who owns all items of each block in $\mathfrak{S}$.  We let $\ALG(\cG):= \obj(\mathfrak{S} ) $ be the objective value of the partition obtained by Algorithm \ref{algo:greedy}. The remainder of the proof will argue that the output $\mathfrak{S} $ simultaneously satisfies,   $$\frac{\ALG(\cG)}{\OPT(\cG)} = \frac{\obj(\mathfrak{S})}{\obj(\mathfrak{S}^*)} \geq \inf \big\{ \frac{w(x)}{ w'_{-}(x) x} \big| w'_{-}(x) > 0, x\geq 2 \big\},$$
for every $w \in \cW$.

Let $\cG \setminus j$ denote the induced subgraph of $\cG$ by removing the vertex $j$ as well as all its adjacent item nodes and corresponding edges from $\cG$. Formally, in this induced subgraph $\cG \setminus j = ( \cP', \cA',\cE')$, we have $\cP' = \cP \setminus \cI^j, \cA' = \cA \setminus \{j\}$ and $\cE' = \{(i,j) | (i,j)\in \cE \text{ and }  i\in \cP', j\in \cA' \}$.

We now prove the  approximation ratio of the greedy algorithm by induction on the owner set's size $m=|\cA|$. For the base case, for any bipartite graph $\cG=(\cP,\cA, \cE)$ with $|\cA| = 1$, it is clear that the greedy approach finds the exact optimal partition. For the inductive case, we show that if the approximation guarantee holds for any bipartite graph $\cG=(\cP,\cA,\cE)$ with $|\cA| \leq k$, then it must hold for any bipartite graph $\cG$ with $|\cA| = k+1$ as well.

Let $j$ be the very first owner picked by Algorithm \ref{algo:greedy} on the bipartite graph instance $\cG$ with $|\cA| = k+1$. Thus, the first block (i.e., item set) picked by the algorithm must be $\cI^j$.  Let $\mathfrak{S}^*$ be the optimal partition of $\cG$ under weight $w$. Note that the algorithm output $\mathfrak{S}$ is independent of $w$, but  $\mathfrak{S}^*$   typically depends on $w$.
For each block $\cS_k^* \in \mathfrak{S}^*$, let $p_k^*:= \abs{\cS_k^*}$ be the number of items in this block and $ a_k:= \abs{ \cS^*_k \intersect \cI^j } $ be the number of items covered by the   greedy algorithm's first choice $\cI^j$. Hence, $\abs{\cI^j} = \sum_{k=1}^{K} a_k$. We can then bound the approximation ratio of the greedy algorithm by decomposing the objective value as follows,
\begin{align*}
    \frac{\ALG(\cG)}{\OPT(\cG)}
    = & \frac{ w( \cI^j ) + \ALG (\cG \setminus j) }{  \sum_{k=1}^{K} w(p^*_k) - \sum_{k=1}^{K} w(p^*_k - a_k) + \sum_{k=1}^{K} w(p^*_k - a_k) } \\
    \geq &  \frac{ w( \cI^j ) + \ALG (\cG \setminus j) }{  \sum_{k=1}^{K} w(p^*_k) - \sum_{k=1}^{K} w(p^*_k - a_k) + \OPT (\cG \setminus j) },
\end{align*}
where the inequality uses the suboptimality of the partition $\{ \cS^*_k \setminus \cI^j \}_{k=1}^{K}$ for the  graph $\cG \setminus j$.

We will bound the approximation ratio by considering the two parts $ \frac{ w\left(\sum_{k=1}^{K} a_k \right) }{ \sum_{k=1}^{K} w(p^*_k) - \sum_{k=1}^{K} w(p^*_k - a_k) }$ and $ \frac{ \ALG (\cG \setminus j ) }{ \OPT (\cG \setminus j )} $ separately, and then  apply the mediant inequality to derive the lower bound. The second part can be easily  bounded according to the induction hypothesis. However,  the first part could be unbounded in general (this may happen when $p^*_k$'s are all at most 1). Fortunately,   the analysis for such special case can be carried out separately.  Specifically, if $\max_{k\in [K]} p^*_k \leq 1$ in $\cG$, by convexity of $w$ function, this implies that any 1-strong partition of the items are at least as good as $\mathfrak{S}^*$,  thus   the greedy algorithm is optimal.

Next, we consider the more typical situation with $p := \max_{k\in [K]}p^*_k \geq 2$. Here, invoking the   mediant inequality and a technical Lemma~\ref{lm:difference-ratio-general} below with $p\geq 2$,
we can lower bound the approximation ratio as follows,
\begin{align*}
& \frac{ w( \cI^j ) + \ALG (\cG \setminus j) }{  \sum_{k=1}^{K} w(p^*_k) - \sum_{k=1}^{K} w(p^*_k - a_k) + \OPT (\cG \setminus j) } \\
    \geq & \min \bigg \{ \frac{ w\left(\sum_{k=1}^{K} a_k \right) }{ \sum_{k=1}^{K} w(p^*_k) - \sum_{k=1}^{K} w(p^*_k - a_k) }, \ \frac{ \ALG (\cG \setminus j ) }{ \OPT (\cG \setminus j ) }  \bigg \} \\
    \geq & \inf \left\{ \frac{w(x)}{ w'_{-}(x) x} \big| w'_{-}(x) > 0, x\geq 2 \right\}.
\end{align*}

{
Finally, for running time analysis,  Algorithm~\ref{algo:greedy} can be efficiently implemented in almost linear time using appropriately chosen data structures. Specifically,  we employ the red-black tree data structure which can retrieve  the maximum element   $j^* = \arg\max_{j \in [m] } | \cI^{j} \setminus \bar{\cI} |$, i.e., the maximal  residual item sets among all owners, using $O(\log m)$ time. We will then update the item sets of all the owners who have at least one item in $\cI^{j^*} \setminus \bar{\cI}$. The number of operations is at most the number of edges connected to removed items from $j^*$ in the bipartite graph. We then update the red-black tree structure using the new sizes of  the residual item sets of the involved items. Note that the update of each owner's item set size takes $O(\log m)$ time by first deleting it in the red-black tree and then re-inserting it. The above recursion takes at most $m$ rounds since each round at least one owner's residual item set will become empty and there are only $m$ owners. In total,  the total running time is thus $O( |\cE| \log m)$.
}

\end{proof}

\begin{lemma}\label{lm:difference-ratio-general}
Consider any $\{ p_k^* \}_{k \in [K]}$ and $\{ a_k  \}_{k \in [K]}$ such that $p^*_k \geq a_k \geq 0 $ for any $    k\in [K]$ and  $\sum_{k=1}^{K} a_k \geq p := \max_{k\in [K]}p^*_k$.
For any $w \in \cW$ with $w(p)>0$, we have
    $$
    \frac{ w\left(\sum_{k=1}^{K} a_k \right) }{\sum_{k=1}^{K} w(p^*_k) - \sum_{k=1}^{K} w(p^*_k - a_k) } \geq \inf_{x \ge p} \frac{w(x)}{ w'_{-}(x) x}.
    $$
\end{lemma}

\begin{proof}[Proof of Lemma~\ref{lm:difference-ratio-general}]
Any $w\in \cW$ is convex, so its left derivative $w'_{-}$ is non-deceasing such that $w(p^*_k) - w(p^*_k - a_k) \leq w'_{-}(p_k^*)a_k $. In addition, $w'_{-}(p^*_k) \geq  0$ since $w(0)=0, w(p^*_k) \geq  0$.
Let $\sum_{k=1}^{K} a_k = a$, we have
$$
\frac{ w\left(\sum_{k=1}^{K} a_k \right) }{\sum_{k=1}^{K} w(p^*_k) - \sum_{k=1}^{K} w(p^*_k - a_k) }
\geq  \frac{w(a)}{\sum_{k=1}^{K} w'_{-}(p^*_k)a_k}.
$$
Since $a \geq \max_{k\in [K]}p^*_k = p$,  thus  $w'_{-}(p_k^*) \le w'_{-}(a)$ for any $k$ by convexity of $w$. Since $w(p) > 0$ and $w(0) = 0$, we must have $0 < w'_{-}(p) \leq w'_{-}(x)$ for any $x \geq p$. Therefore,
$$
\frac{w(a)}{\sum_{k=1}^{K} w'_{-}(p^*_k)a_k}  \geq \frac{w(a)}{\sum_{k=1}^K w'_{-}(a)a_k} = \frac{w(a)}{w'_{-}(a)a} \geq  \inf_{x \geq p} \frac{w(x)}{ w'_{-}(x) x}.
$$

 \end{proof}

\subsection{Tightness of Greedy's Approximation Ratio in the Monomial Class}\label{sec:tight}
One implication of Theorem \ref{thm:greedy-approximation} is that for $\alpha$-th degree polynomial function $w=|\cS_k|^\alpha$, the greedy algorithm has $\frac{1}{\alpha}$-approximation. Next, we show that this approximation ratio is provably tight for every weight in the monomial class.

Consider an instance with $n=M N$ items and $m = M + L $ owners for some positive integer $L, M, N $ such that $M^{L-1} \mid N $ and $ N (1-1/M)^{L}  \geq  L$.
The ownership relations are as follows (see also Figure \ref{fig:lower-bound} for the visualization of the instance):
\begin{itemize}[leftmargin=*]
\item There are $M$ owners $1, \cdots, M$, each of whom owns a disjoint set of $N$ items. Thus, collectively, they own all the $MN$ items.  Let the items owned by these owners as $\cI^1, \dots, \cI^{M}$ respectively.
\item We denote the item owned by the remaining $L$ owners  as $\cI^{ \ell + M }$ for $\ell \in [1, \dots, L]$. Let  $\abs{\cI^{\ell + M } \intersect \cI^{\ell}} = (1-1/M)^{\ell-1} N/M + 1$ and $\abs{\cI^{\ell+ M} \intersect \cI^{\ell'} } =  (1-1/M)^{\ell- 1} N/M, \forall \ell' \neq  \ell \in [M]$. Hence, $|\cI^{\ell+M}| = (1-1/M)^{\ell-1} N + 1$.
\end{itemize}

Observe that $\cI^{M+1}$ is the largest item set and $\cI^{M+2}$ becomes the largest item set after $\cI^{M+1}$ is removed. Therefore, we can see that the greedy algorithm proceeds by picking the $\ell$-th owner of the last $L$ owners from $\ell\in [1, \dots, L]$ and lastly the first $M$ owners, i.e., when each of whom has remaining item less than $N - \sum_{\ell=1}^{L} (1-1/M)^{\ell-1} N/M = (1-1/M)^{L} N$. As such, the greedy partition achieves the objective of at least
$ \ALG(\cG) \leq  \sum_{\ell = 1}^{L} [(1-1/M)^{\ell-1} N + 1]^\alpha +  M (1-1/M)^{\alpha L} N^{\alpha}.$

On the other hand, consider a partition strategy that only picks the first $M$ owners, which achieves the objective $\OPT(\cG) = M N^\alpha$. Simply using it as a lower bound for the optimal objective, we can bound the approximation ratio as, $$
\begin{aligned}
\frac{\ALG(\cG)}{\OPT(\cG) } &\leq \frac{ \sum_{\ell =1}^{L} [(1-1/M)^{\ell -1} N + 1]^\alpha +  M (1-1/M)^{\alpha L} N^{\alpha} }{ M N^\alpha } \\
&= \frac{ \sum_{\ell =1}^{L} [(1-1/M)^{\ell -1} N + 1]^\alpha}{ M N^\alpha } +  (1-1/M)^{\alpha L}.
\end{aligned}
$$
For any constant $M > 1$, we have as $ N, L\to \infty$,
$$
\begin{aligned}
 \lim_{N, L\to \infty} \frac{ \sum_{\ell =1}^{L} [(1-1/M)^{\ell - 1} N + 1]^\alpha }{ M N^\alpha } +  (1-1/M)^{\alpha L}
&= \lim_{N, L\to \infty}  \sum_{\ell =1}^{L} (1-1/M)^{(\ell - 1)\alpha} / M\\
&=  \frac{1/M}{1-(1-1/M)^{\alpha}}.
\end{aligned}
$$
Finally, as $M \to \infty$, the ratio matches with the worst case approximation guarantee $1/\alpha$,
$$
\lim_{M \to \infty} \frac{1/M}{1-(1-1/M)^{\alpha}}
= \lim_{M \to \infty} \frac{1}{\alpha-1/M}
= 1/\alpha.
$$

\input{plots/lower-bound-plot}

Meanwhile, despite the hardness result for the size-focused objective in Proposition~\ref{prop:hardness}, we would like to point out that the efficient algorithm for optimal partition under the $\alpha$-th degree polynomial objective may exist. In particular, consider the following instance with $m=n+1$, where the optimal partition cannot realize the set-cover problem for the hardness reduction. The owner $i$ owns the items $\{2i-1,2i\}$,$\forall i \in [n]$ and owner $n+1$ owns the items $\{1,3,\cdots,2n-1\}$. In this case, the optimal partition is not the minimum set cover, since $$ \obj\left(\left\{ \{1,3,\cdots,2n-1\},\{2\},\cdots\{2n\} \right\}\right) = n^{\alpha}+n>\obj\left(\left\{\{1,2\},\cdots,\{2n-1,2n\} \right\}\right) = n\times 2^\alpha. $$
Hence, it remains an open question on the hardness of finding the optimal partition under the $\alpha$-th degree polynomial objective.

%% file: plots/lower-bound-plot.tex
\begin{figure}[t]
    \centering

    \tikzset{every picture/.style={line width=0.75pt}}

    \begin{tikzpicture}[x=0.75pt,y=0.75pt,yscale=-1,xscale=1]

\draw    (157,61) -- (241.67,61) ;
    \draw [shift={(241.67,61)}, rotate = 180] [color={rgb, 255:red, 0; green, 0; blue, 0 }  ][line width=0.75]    (0,5.59) -- (0,-5.59)   ;
    \draw [shift={(157,61)}, rotate = 180] [color={rgb, 255:red, 0; green, 0; blue, 0 }  ][line width=0.75]    (0,5.59) -- (0,-5.59)   ;
\draw    (241.83,73) -- (326.5,73) ;
    \draw [shift={(326.5,73)}, rotate = 180] [color={rgb, 255:red, 0; green, 0; blue, 0 }  ][line width=0.75]    (0,5.59) -- (0,-5.59)   ;
    \draw [shift={(241.83,73)}, rotate = 180] [color={rgb, 255:red, 0; green, 0; blue, 0 }  ][line width=0.75]    (0,5.59) -- (0,-5.59)   ;
\draw    (439.67,108.33) -- (524.33,108.33) ;
    \draw [shift={(524.33,108.33)}, rotate = 180] [color={rgb, 255:red, 0; green, 0; blue, 0 }  ][line width=0.75]    (0,5.59) -- (0,-5.59)   ;
    \draw [shift={(439.67,108.33)}, rotate = 180] [color={rgb, 255:red, 0; green, 0; blue, 0 }  ][line width=0.75]    (0,5.59) -- (0,-5.59)   ;
\draw    (156.33,156.67) -- (171.67,156.67) ;
    \draw [shift={(156.33,156.67)}, rotate = 180] [color={rgb, 255:red, 0; green, 0; blue, 0 }  ][line width=0.75]    (0,5.59) -- (0,-5.59)   ;
\draw    (243.67,156.67) -- (259,156.67) ;
\draw    (330.33,156.33) -- (345.67,156.33) ;
\draw    (440,155.33) -- (455.33,155.33) ;
    \draw [shift={(455.33,155.33)}, rotate = 180] [color={rgb, 255:red, 0; green, 0; blue, 0 }  ][line width=0.75]    (0,5.59) -- (0,-5.59)   ;
\draw    (326.67,85) -- (411.33,85) ;
    \draw [shift={(411.33,85)}, rotate = 180] [color={rgb, 255:red, 0; green, 0; blue, 0 }  ][line width=0.75]    (0,5.59) -- (0,-5.59)   ;
    \draw [shift={(326.67,85)}, rotate = 180] [color={rgb, 255:red, 0; green, 0; blue, 0 }  ][line width=0.75]    (0,5.59) -- (0,-5.59)   ;
\draw    (208.67,211.17) -- (218,211.17) ;
    \draw [shift={(208.67,211.17)}, rotate = 180] [color={rgb, 255:red, 0; green, 0; blue, 0 }  ][line width=0.75]    (0,5.59) -- (0,-5.59)   ;
\draw    (293.17,211.83) -- (302.5,211.83) ;
\draw    (498,209.5) -- (507.33,209.5) ;
    \draw [shift={(507.33,209.5)}, rotate = 180] [color={rgb, 255:red, 0; green, 0; blue, 0 }  ][line width=0.75]    (0,5.59) -- (0,-5.59)   ;
\draw    (382.67,211) -- (392,211) ;

\draw (30,48.67) node [anchor=north west][inner sep=0.75pt]   [align=left] {owner $\displaystyle 1$ $ $};
\draw (164.67,35) node [anchor=north west][inner sep=0.75pt]   [align=left] {$\displaystyle N$ items};
\draw (418.67,84.33) node [anchor=north west][inner sep=0.75pt]   [align=left] {...};
    \draw (448.67,87.5) node [anchor=north west][inner sep=0.75pt]   [align=left] {$\displaystyle N$ items};
\draw (30,147.5) node [anchor=north west][inner sep=0.75pt]   [align=left] {owner $\displaystyle M+1$ };
\draw (30,201.83) node [anchor=north west][inner sep=0.75pt]   [align=left] {owner $\displaystyle M+\ell $ };
\draw (417.33,129.33) node [anchor=north west][inner sep=0.75pt]   [align=left] {...};
\draw (208.33,180.33) node [anchor=north west][inner sep=0.75pt]   [align=left] {$\displaystyle N( 1-1/M)^{\ell -1} + 1$ items};
\draw (80.38,175.22) node [anchor=north west][inner sep=0.75pt]  [rotate=-90.36] [align=left] {...};
\draw (30,102.67) node [anchor=north west][inner sep=0.75pt]   [align=left] {owner $\displaystyle M$ $ $};
\draw (159.33,130.33) node [anchor=north west][inner sep=0.75pt]   [align=left] {$\displaystyle N+1$ items};
\draw (78.88,78.72) node [anchor=north west][inner sep=0.75pt]  [rotate=-90.36] [align=left] {...};

    \end{tikzpicture}

    \caption{Illustration of the worst case instance for the greedy algorithm. $l = 1, \cdots, L$. }
    \label{fig:lower-bound}
\end{figure}

%% file: exp-append.tex
\section{Additional Experiments}\label{append;exp}

\begin{figure}[tbh]
    \centering
    \subfigure[Top-$1.5\%$, ICLR 2021]{\includegraphics[width=0.32\textwidth]{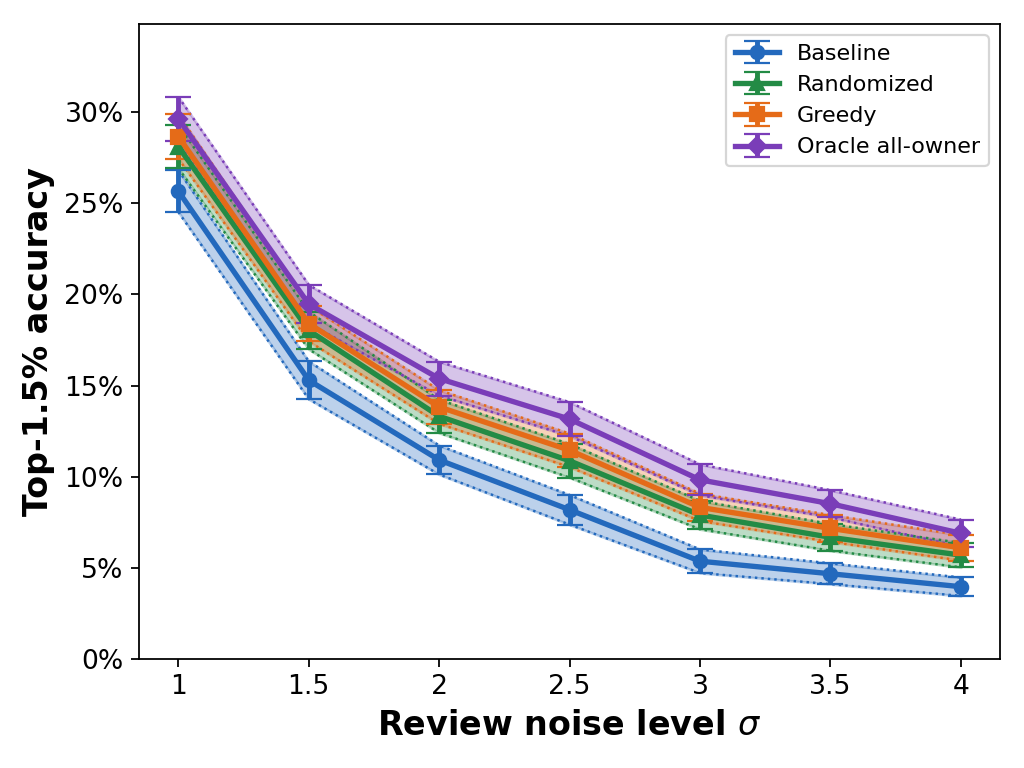}}
    \subfigure[Top-$1.5\%$, ICLR 2022]{\includegraphics[width=0.32\textwidth]{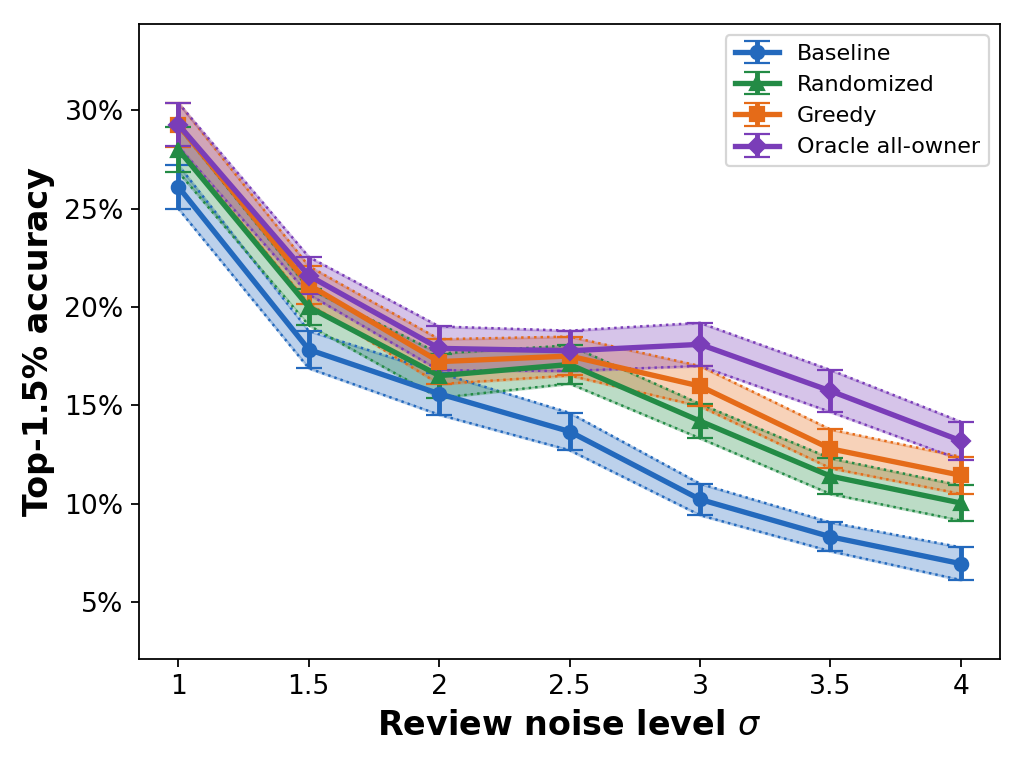}}
    \subfigure[Top-$1.5\%$, ICLR 2023]{\includegraphics[width=0.32\textwidth]{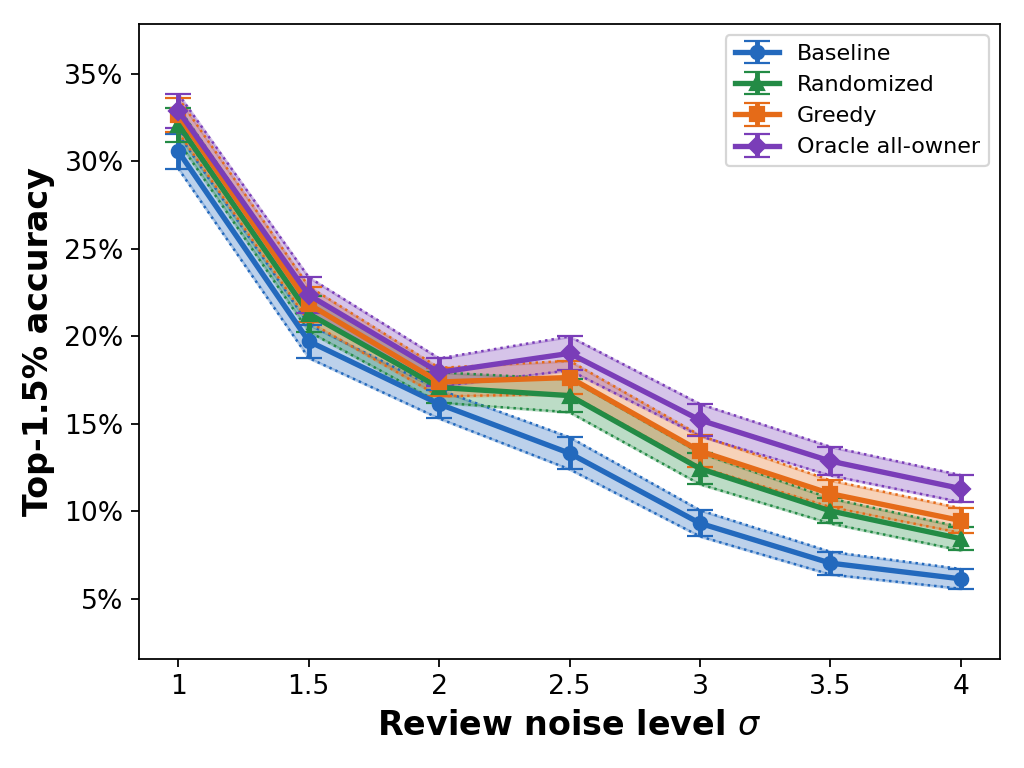}}
    \subfigure[Top-$5\%$, ICLR 2021]{\includegraphics[width=0.32\textwidth]{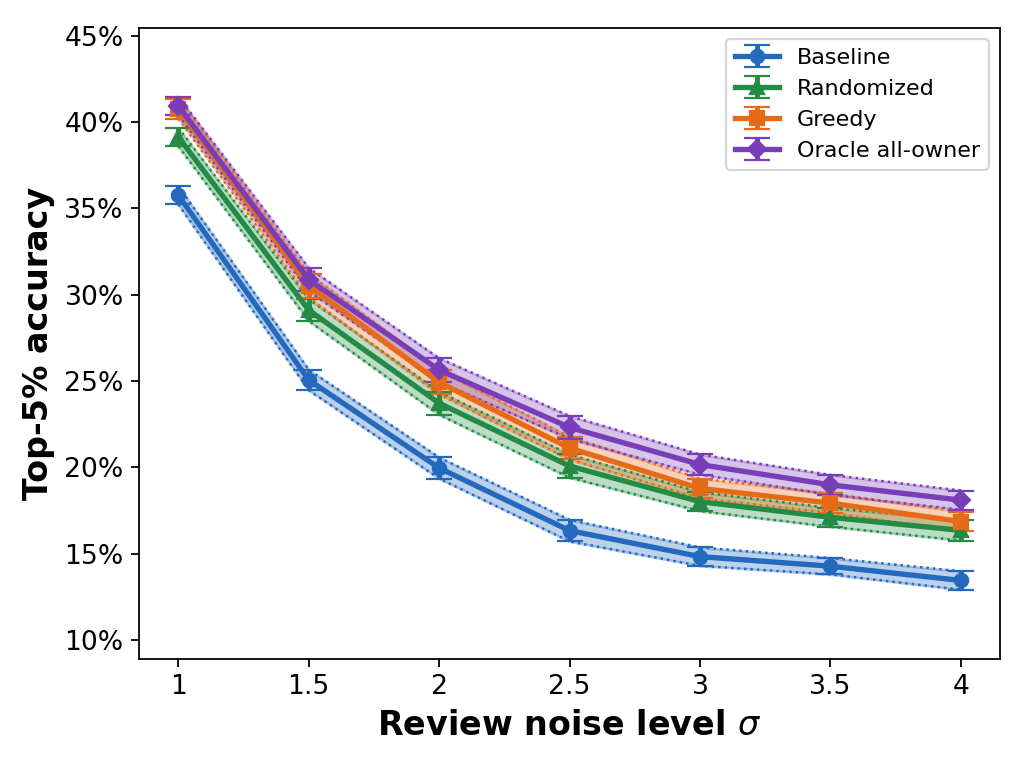}}
    \subfigure[Top-$5\%$, ICLR 2022]{\includegraphics[width=0.32\textwidth]{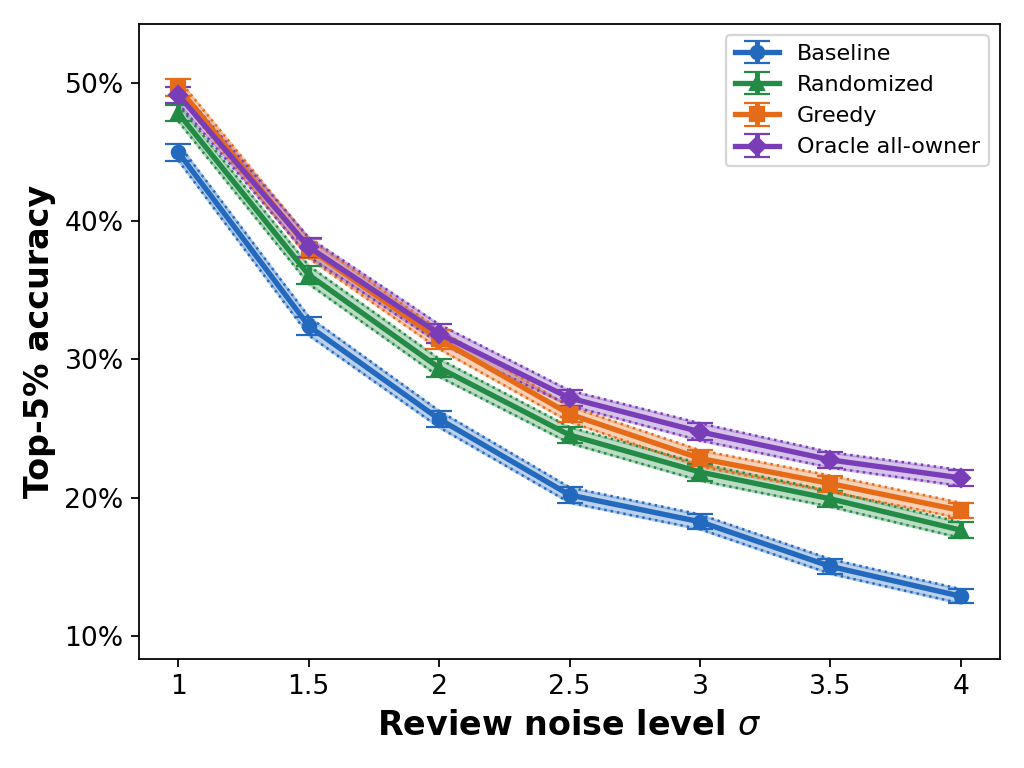}}
    \subfigure[Top-$5\%$, ICLR 2023]{\includegraphics[width=0.32\textwidth]{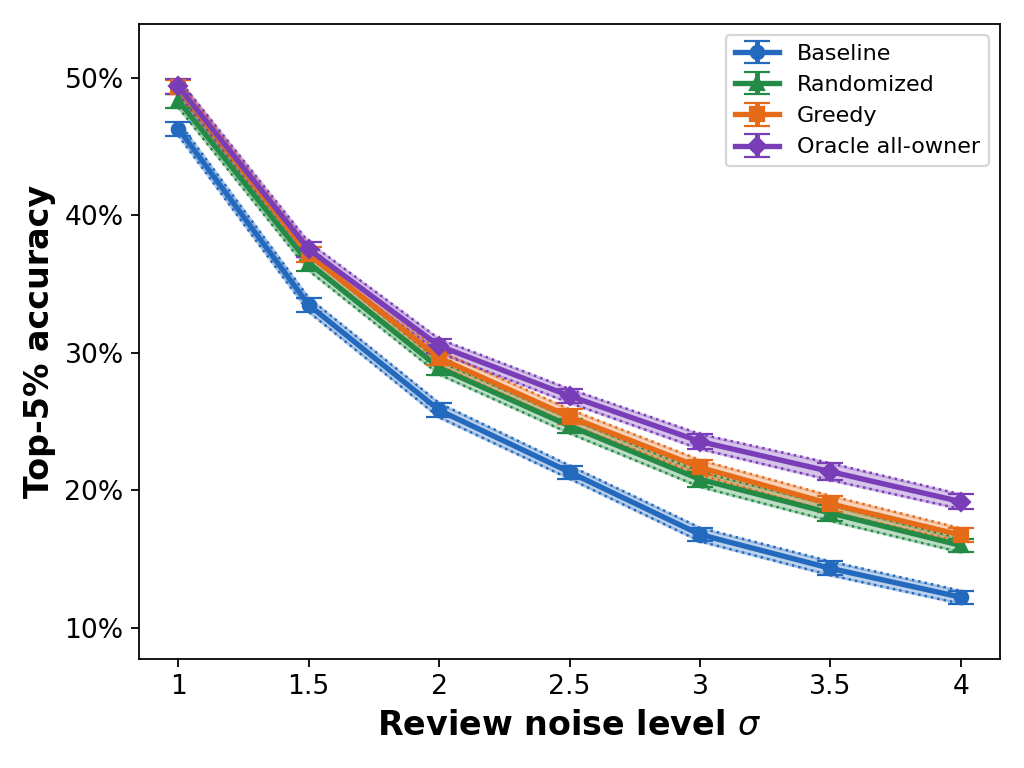}}
    \caption{Top-$1.5\%$ and top-$5\%$ accuracy under varying review noise in ICLR. Points show Monte Carlo means; shaded bands, dotted boundaries, and capped error bars show $95\%$ Monte Carlo confidence intervals.}
    \label{fig:precision-all}
\end{figure}

Figure~\ref{fig:precision-all} reports top-$1.5\%$ and top-$5\%$ accuracy under the same controlled-noise ICLR design. It includes all four methods from the main experiment. The greedy partition generally improves upon the baseline and randomized methods for these more stringent selections.

